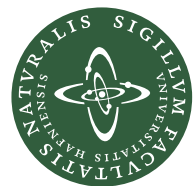

# Master's Thesis

Anders Christensen

# Inferential protein structure determination and refinement using fast, electronic structure based backbone amide chemical shift predictions

Academic supervisor: Professor Jan H. Jensen

Submitted: 24/10/2010

## Dansk resume

Denne rapport omhandler udviklingen af en ny, hurtig metode til udregning af backbone-amidproton kemiske skift i proteiner. Gennem kvantekemiske udregninger er den strukturbaserede forudsiglese det kemiske skift for amidprotonen i proteinet blevet parametriseret. Parametriseringen implementeres dernæst i et computerprogram kaldet `Padawan`. Programmet er efterfÃ¸lgende blevet implementeret i proteinfoldningsprogrammet `Phaistos`, hvori metoden andvendes til at *de novo* foldning af protein strukturer og til at forfine allerede eksisterende proteinstrukturer.

## Acknowledgments

This thesis represents the work I have done as a master student under supervision of Professor Jan. H. Jensen in his group of Biocomputational Chemistry. Thank you to all who have supported me during my work at the third floor of C-building at the H.C. Ørsteds Instituttet.

- Thank you to my supervisor, Jan H. Jensen for introducing me to the exciting fields of quantum chemistry and biocomputational chemistry.

- Thank you to the professors who have helped me towards a better understanding of quantum chemistry, Stephan P. A. Sauer and Sten Rettrup and Kurt V. Mikkelsen.

- Thank you to our partners at the Bio Informatics Group, Thomas Hamelryck and especially Mikael Borg for the massive support in introducing me to the `Phaistos` framework.

- Lastly, thank you to all the students, with whom I've had countless useful discussions on various topic. Especially Casper Steinmann and Kasper Thofte from the Jan H. Jensen group and also Janus Eriksen, Jacob Lykkebo and Jonas Elm of the Kurt V. Mikkelsen group.

# Contents

# 1   Introduction

In order to rationalize and optimize the chemical properties of proteins, based on the structure/funtion interplay, knowledge about the 3-dimensional structure is required. However, for many proteins the 3-dimensional structure cannot be obtained because they fail to crystallize, which a requirement for X-ray crystallography, or are too large for conventional NMR structure determination.

Traditional NMR structure determination relies on Nuclear Overhauser Effects (NOEs) and residual dipolar couplings (RDCs), which contain information that can be directly linked to distances and angles between atoms. These are difficult and time consuming to obtain, even for small proteins and can be impossible for larger proteins. Chemical shifts are more easily obtained than NOEs and RDCs, but the relationship between chemical shifts and the underlying structure is much less straightforward. [47] Chemical shift based structure determination thus relies on approximate methods to calculate the complicated relationship between structure and chemical shift.

Chemical shift based protein structure determination is an emerging area of research. The current state-of-the-art methods are the CS-ROSETTA [57] methods due to the Baker group at the University of Washington and the CHESHIRE [11] method due to the Vendruscolo group at Cambridge University. Briefly explained, these methods employ a Monte Carlo scheme which minimizes the differences between experimental chemical shifts and chemical shifts calculated from computer generated sample structures, in order to find the structure for which the calculated chemical shifts are in best agreement with experimental data. [42] Both methods are at the proof-of-concept stage and work well for proteins of up to about 120-150 residues.

For larger structures, however, the determination of the protein structure fails [62]. Current chemical shift based structure determination is limited by the accuracy of which chemical shifts can be predicted from protein structures [47]. Other uses, such as protein-ligand docking predictions are difficult to parametrize from empirical methods, because protein-ligand complexes generally are not available in the large emirical data sets needed for a parametrization. [68]

New approaches for the calculation of chemical shifts from protein structures are therefore needed in order to overcome this problem. Chemical shifts for the backbone $C^\alpha$ atom in proteins have recently been parametrized with high accuracy using electronic structure based methods. [67]. A similar approach is in the this work used for the backbbone amide proton, the chemical shift of which is much more difficult to predict. [15] [50]

The purpose of this work, is (1) to use electronic structure based methods to develop a new method to calculate chemical shifts from a protein structure, (2) to implement this method into a computationally fast protein structure prediction framework , (3) to validate the method by a comparison to quantum mechanical (QM) methods and the correlation with experimentally obtained chemical shifts as well as a comparison to existing chemical shift prediction methods and finally (4) to use method to predict protein structures.

To facilitate the rapid calculation of the chemical shift of proteins, a method will be developed, which directly relates structural information from a 3-dimensional protein structure to chemical shift. This is very analogous to a classical force field, in which a potential energy can be calculated using structural parameters such as bond distances and torsion angles. In this thesis, a method will be developed for the prediction of backbone amide proton chemical shifts as a proof-of-concept.

The method of predicting backbone amide proton chemical shifts will be interfaced to the protein structure determination software package, Phaistos [8] and used to include chemical shifts in protein folding simulations. Due to the relatively sparse amount of data included in backbone amide proton chemical shifts, these will be combined with existing energy terms which are known to predict correct folds. Finally a protein structure refinement is carried out using only chemical shift data.

This thesis is organized as follows:

- Section 2 contains a brief introduction to the quantum mechanical methods used in this work to calculate NMR shielding constants and chemical shifts.

- Section 3 is an introduction to the inferential structure determination principle employed in Phaistos and Phaistos in general.

- In section 4, the backbone amide proton chemical shift method by Parker, Houk and Jensen is presented and expanded to predict all backbone amide proton chemical shifts in the protein.

- Section 5 investigates the chemical shift perturbations due to ring current effects in aromatic side chains in the protein.

- Section 6 presents the computational implementation of the method and compare the method to QM methods and experimental data.

- In section 7, the results from protein folding simulations and structure refinements using the chemical shifts method are presented.

## 2 Calculation of the chemical shift

An external magnetic induction $\vec{B}^{(\text{External})}$ acting on a nucleus with a non-zero magnetic moment gives rise to a Zeeman splitting of the the energy levels of the spin states, which can be measured by NMR spectroscopy [29]. Due to electronic screening of the nucleus, the nucleus experiences a local magnetic field of

$$\vec{B}^{(\text{Local})} = (1 - \sigma)\,\vec{B}^{(\text{External})} \qquad (1)$$

where $\sigma$ is the shielding constant. This gives rise to variations of the Zeeman splitting of the energy levels as a function of the local molecular environment around the nucleus. This leads directly to the definition of the chemical shift. When the shielding constant is obtained for a reference nucleus, the chemical shift of of a nucleus with the shielding constant $\sigma$ is defined with respect to this reference as:

$$\delta \equiv \sigma_{\text{Ref}} - \sigma \qquad (2)$$

If we consider an external magnetic induction $\vec{B}$ acting on the $k^{th}$ nucleus of a molecule, we can write the NMR shielding tensor, the elements of which being the mixed derivatives of the energy with respect to $\vec{B}$ and the magnetic moment of the $k^{th}$ nucleus, $\vec{\mu}^{(k)}$: [60]

$$\underline{\underline{\sigma}}^{(k)} = \begin{pmatrix} \sigma_{xx}^{(k)} & \sigma_{xy}^{(k)} & \sigma_{xz}^{(k)} \\ \sigma_{yx}^{(k)} & \sigma_{yy}^{(k)} & \sigma_{yz}^{(k)} \\ \sigma_{zx}^{(k)} & \sigma_{zy}^{(k)} & \sigma_{zz}^{(k)} \end{pmatrix} \quad \text{in which} \quad \sigma_{\alpha\beta}^{(k)} = \frac{\partial^2 E\left(\vec{B}, \vec{\mu}^{(k)}\right)}{\partial B_\beta\, \partial \mu_\alpha^{(k)}}\bigg|_{|\vec{B}|=0,\ |\vec{\mu}^{(k)}|=0}$$

$$(3)$$

where the greek subscripts $\alpha, \beta$ represent the possible Cartesian components $(x,y,z)$ of the vector/matrix. The nuclear shielding constant is calculated as the average of the isotropic elements of the tensor, i.e. one third of the trace.

$$\sigma^{(k)} = \tfrac{1}{3}\text{tr}\left(\underline{\underline{\sigma}}^{(k)}\right) = \tfrac{1}{3}\left(\sigma_{xx}^{(k)} + \sigma_{yy}^{(k)} + \sigma_{zz}^{(k)}\right) \qquad (4)$$

A way to calculate derivatives of the energy, such as to calculate the shielding tensor (Eq. 3), is by perturbation theory. Once a proper Hamiltonian for the molecule in the external magnetic field is specified, Eq. 3 is in Rayleigh-Schrödinger perturbation theory the equivalent of the following sum-over-

states expression:

$$
\begin{aligned}
\sigma_{\alpha\beta}^{(k)} &= \left. \frac{\partial^2 E^{(2)}\left(\vec{B}, \vec{\mu}^{(k)}\right)}{\partial B_\beta \, \partial \mu_\alpha^{(k)}} \right|_{|\vec{B}|=0, \; |\vec{\mu}^{(k)}|=0} \\
&= \left. \left\langle \Psi_0 \left| \frac{\partial^2 H^{(2)}}{\partial B_\beta \, \partial \mu_\alpha^{(k)}} \right| \Psi_0 \right\rangle \right|_{|\vec{B}|=0, \; |\vec{\mu}^{(k)}|=0} \\
&+ \left. \sum_i \frac{\left\langle \Psi_0 \left| \frac{\partial H^{(1)}}{\partial B_\beta} \right| \Psi_i \right\rangle \left\langle \Psi_i \left| \frac{\partial H^{(1)}}{\partial \mu_\alpha^{(k)}} \right| \Psi_0 \right\rangle}{E_0 - E_i} \right|_{|\vec{B}|=0, \; |\vec{\mu}^{(k)}|=0}
\end{aligned}
\tag{5}
$$

where $\Psi_0$ and $E_0$ denote the wavefunction and energy of the unperturbed electronic ground state - $H^{(1)}$ and $H^{(2)}$ are the first and second order perturbing Hamiltonians, respectively. The summation runs over all excited states, $\Psi_i$, and corresponding energies, $E_i$. The first term in Eq. 5 is traditionally referred to as the diamagnetic term, while the second is refered to as the paramagnetic term [60].

Usually, derivatives of the energy are not obtained computationally by solving the above equation, but rather iteratively solving the coupled perturbed Hartree-Fock (CPHF) equations, in which the in which the external magnetic field is treated as a perturbation. [34] CPHF approaches are also implemented for the derivatives of non-variational wave functions such as MP2 and coupled-cluster. [27]

## 2.1 The gauge origin problem

When intoducing the field dependent Hamiltonians in Eq. 5 or the external magnetic field perturbation in the CPHF equations, a gauge origin dependence is introduced due to the external magnetic field

$$
\vec{B} = \nabla \times \vec{A}
\tag{6}
$$

which has the associated vector potential

$$
\vec{A}(\vec{r}) = \tfrac{1}{2} \vec{B} \times (\vec{r} - \vec{r}_{\mathrm{O}}) \, .
\tag{7}
$$

with an arbitrary gauge origin, $\vec{r}_{\mathrm{O}}$. The expression for the shielding of nucleus thus contains a gauge origin which can be arbitrarily chosen. Using a complete basis set, the arbitrary gauge origin dependence cancels out between the paramagnetic and diamagnetic terms. Using a finite basis set, however, the gauge origin dependence no longer cancels out and the value of the shielding constant depends explicitly on the arbitrary choice.

Although the arbitrary gauge origin dependence can in principle be circumvented by increasing the size of the used basis set towards infinity, this is computationally not a very useful approach and several methods exist to ensure unambiguous and unique results as well as to speed up the basis set convergence of NMR shielding calculations. The simplest methods make use of atom centered gauge origins, while other methods such as IGLO (Individual Gauge for Localized Orbitals) use localized occupied molecular orbitals together with individual gauge-origins chosen for each occupied orbital at its corresponding centroid. It is noteworthy, however, that the use of such local gauge origin approaches do not resolve the gauge origin problem - rather, unique results are enforced due to the choice of a well-defined set of local gauge origins and it can be shown that relatively rapid basis set convergence can be achieved.

In the GIAO-formulation (Gauge-Including Atom Orbitals) [12] used in all NMR calculations throughout this work, the molecular orbitals are expanded in London orbitals [36]. The concept of London orbitals is to introduce a field dependence complex phase onto the usual Gaussian-type basis functions:

$$
\begin{aligned}
\omega_\mu(\vec{B}, \vec{r}) &= \exp\left(-i\vec{A}\left(\vec{r}_\mu\right) \cdot \vec{r}\right) \chi_\mu(\vec{r}) \\
&= \exp\left(-\frac{i}{2}\left(\vec{B} \times (\vec{r}_\mu - \vec{r}_O) \cdot \vec{r}\right)\right) \chi_\mu(\vec{r}) \quad (8)
\end{aligned}
$$

where $\chi_\mu(\vec{r})$ are the usual field independent basis functions centered on $\vec{r}_\mu$. Thus the phase of the London orbitals depend parametrically on the field strength and gauge origin as well as the center of the basis function and the origin of the Cartesian coordinate system. The desired result is that integrals involving London orbitals only contain differences in vector potentials, thereby removing the reference to an absolute gauge origin. Several other methods made to achieve gauge origin independence based on reformulations of the diamagnetic term exist [60], such as for instance the CSGT (Continuous Set of Gauge Transformations) approach of Keith and Bader [30]. The GIAO formualtion is reported to have the fastest basis set convergence of methods presently available and has been implemented for for a variety of the methods, such as HF, DFT, MP2 and coupled cluster [60].

# 3   Inferential structure determination

## 3.1   The Bayesian inference principle

Experimental data are often insufficient to determine the structure of a protein on its own. As a consequence, energy minimizations based on experimental data are usually combined with a physical energy which is minimized at the same time. The disagreement between experimental data $D$ and a structural model $X$ is usually modeled in a cost function, $E_{data}$, which is also minimized during the run. [25] This approach thus effectively minimizes a hybrid energy:

$$E_{\text{hybrid}}(X) = E_{\text{phys}}(X) + w\,E_{\text{data}}(X) \qquad (9)$$

where $w$ is a weighting factor. It is, however, not obvious how to define a good cost function and optimal weight. [58] These have to be tweaked carefully by hand in order to do an optimal weighting between the included terms. Rieping, Habeck and Nigles describe an objective and rigorous method of weighting experimental data and physical energies in protein structures using a Bayesian inference principle [22]. Bayes theorem describes the relationship between one conditional probability and its inverse. [4] Since data can readily be obtained given a structure (through experiment), Bayesian inference can be used to determine a structure given experimental data. The Bayesian inference principle is thus given as:

$$P(X|D) \propto P(D|X)P(X) \qquad (10)$$

where $P(X|D)$ is the posterior probability of a structure, $X$, given experimental data, $D$; $P(D|X)$ is the likelihood of seeing the experimental data given a structure and finally $P(X)$ is the prior probability of the structure. The prior probability describes extra knowledge about structures. This can for instance be modeled as a force-field energy or via bias through conformational preferences in the Ramachandran map, such as done in Phaistos [8]. The task is thus to find structures $X$ with high likelihood given the set of experimental data $D$. When a prediction of the data is possible from a given trial structure, $P(D|X)$ can be calculated by assuming some distribution for the discrepancy between experimental and predicted data. In the following, an expression of the probability of a structure given a set of chemical shifts will be derived.

### 3.1.1   Chemical shift energy term

For a trial structure $X$ from which chemical shifts can be predicted, the probability of the experimental chemical shifts $D$ is given as the product of the probabilities for discrepancy between each set of experimental $d_i$ and predicted $d_i(X)$ values [22]. The discrepancy between experimental and predicted data is here assumed to follow a Gaussian distribution, i.e. a distribution of $p(\Delta x) = \left(\frac{1}{2\pi\sigma^2}\right)^{\frac{1}{2}} \exp\left(-\frac{(\Delta x)^2}{2\sigma^2}\right)$. The probability of the entire data set is equal to the product of the probability of each datapoint:

$$P(D|X) = \prod_{i=1}^{n} P(d_i|X) \tag{11}$$

Inserting the probability of $p(d_i - d_i(X))$ according to the Gaussian distribution into Eq. 11, the following result is obtained:

$$\prod_{i=1}^{n} P(d_i|X) = \left(\frac{1}{2\pi\sigma^2}\right)^{\frac{n}{2}} \exp\left(\frac{-1}{2\sigma^2} \sum_{i=1}^{n} [d_i - d_i(X)]^2\right) \tag{12}$$

where the standard deviation in the Gaussian distribution is assumed to be $\sigma$ and $n$ is the number of data points. From thermodynamics, a pseudo energy can then derived as:

$$E \propto -\log P(D|X) = \frac{1}{2\sigma^2} \sum_{i=1}^{n} [d_i - d_i(X)]^2 + \frac{n}{2} \log\left(2\pi\sigma^2\right) \tag{13}$$

Since all amide proton predictions derived in this work do not carry the same uncertainty, it is necessary to derive an expression for the probability and pseudo energy for a data set, where the data points have different uncertainties. Data points are grouped in $m$ bins according to their uncertainty. $\sigma_j$ is thus the uncertainty for every data point in the $j$'th bin. The likelihood of the data set is then the product of the likelihood of the data in each bin:

$$P(D|X) = \prod_{j}^{m} P(\{d_i\}_j, \sigma_j|X) \tag{14}$$

where $\{d_i\}_j$ is the data in the $j$'th bin with the associated standard deviation $\sigma_j$. Again assuming a Gaussian distribution, the following is obtained (similar to Eq. 12):

$$\prod_{j}^{m} P(\{d_i\}_j, \sigma_j|X) = \prod_{j}^{m} \left[\left(\frac{1}{2\pi\sigma_j^2}\right)^{\frac{n}{2}} \exp\left(\frac{-1}{2\sigma_j^2} \sum_{i=1}^{n} \{[d_i - d_i(X)]^2\}_j\right)\right] \tag{15}$$

The final expression for the chemical shifts based pseudo energy term used in this work is thus obtained:

$$E \propto -\log P(D|X) = \sum_{j=1}^{m} \left[\frac{1}{2\sigma_j^2} \sum_{i=1}^{n} \{[d_i - d_i(X)]^2\}_j + \frac{n}{2} \log\left(2\pi\sigma_j^2\right)\right] \tag{16}$$

### 3.2 Phaistos

The software framework `Phaistos` [8] constitutes an attempt to develop a fully probabilistic description of protein folding. The connection between the underlying physics and the probability of a protein in a conformation $j$ comes from the Boltzmann distribution and is proportional to $g_j \exp(-E_j)$, where $g_j$ and $E_j$ are the multiplicity and energy of the conformation. [8] Anfinsen's dogma [2] states that, the native structure of the protein is determined only by the protein's amino acid sequence and is the most probable state at thermodynamic equilibrium. `Phaistos` runs a Markov Chain Monte Carlo (MCMC) simulation to sample low energy structures [20] [40].

A protein is represented in `Phaistos` as a polymer with fixed bond angles and bond lengths. The dihedral angles of the backbone and side chains are the only allowed degrees of freedom. `Phaistos` samples protein-like backbone conformations from the generative model TorusDBN [7], which is a Ramachandran-map-like distribution of backbone conformations. To sample more efficiently from the allowed regions in the Ramachandran-map, the secondary structure must be assigned for the protein before a simulation. For an unknown structure, these can be obtained with high accuracy by various methods from sequence alone. In this work, the danger of erroneous secondary stucture assignment is eliminated because only proteins with known structures are used as examples. The secondary structures assignments used in this work are taken from known crystal structures.

When a trial backbone conformation has been generated from TorusDBN, the torsion angles of the side chains are sampled from the Basilisk model [24]. Basilisk is, in the same way as TorusDBN, a distribution over realistic torsion angles (see Fig. 1).

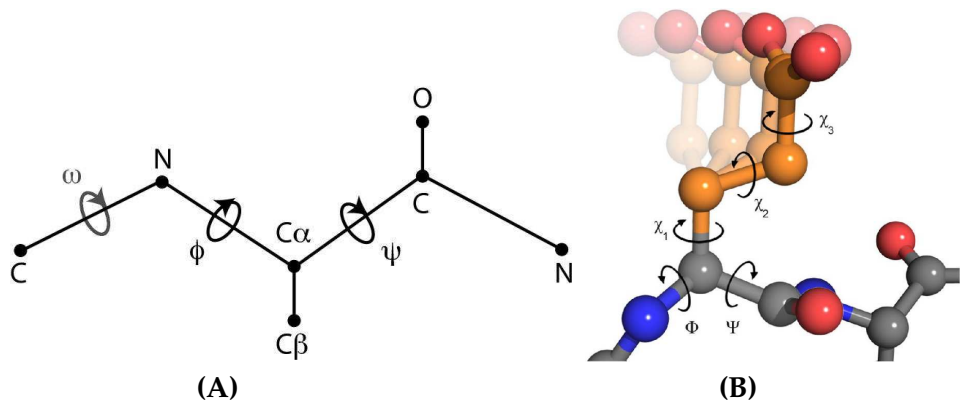

**Figure 1:** The allowed degrees of freedom in A) the protein backbone sampled from TorusDBN and B) a glutamate side chain sampled from Basilisk.

`Phaistos` samples conformational space through the distribution:

$$P \sim P(X|A, S) \prod_i P_i \qquad (17)$$

where $P_i$ is the probability of the structure according to the $i$'th energy function and $P(X|A, S)$ is probability of the structure according to TorusDBN and Basilisk. $A$ and $S$ denote the amino acid sequence and secondary structure assignment, respectively. The probability of the structure according to TorusDBN and Basilisk is included by using these as proposal distributions in the MCMC scheme.

The probabilities can be converted to a pseudo energy, through their Boltzmann factors $P_i \sim \exp(-\lambda_i E_i)$, where $\lambda_i$ is the Boltzmann factor for the $i$'th energy term. It is possible to calculate the energy $E$ associated with the probability $P$:

$$E = -\log P = -log \prod_i P_i = \sum_i \lambda_i E_i \qquad (18)$$

where $E_i$ is the energy according to the $i$'th energy term.

A number of different energy terms not based on chemical shifts are used in the protein folding simulations carried out in this work. These are described below.

### 3.2.1 Clash-Fast

The Clash-Fast energy term is a van der Waals contact term which ensures that no two atoms are closer than the sum of their van der Waal radii. If two atoms are closer than the sum, the energy of this function is returned as $\infty$, and the structure is rejected, otherwise the Clash-Fast energy is 0. This ensures an efficient discarding of many non-physical structures.

### 3.2.2 PP-Compactness

Folded proteins are in general compact, and the PP-Compactness energy term introduces a probabilistic bias towards compact structures. The probabilistic energy is given as the probability of observing the length of the protein ($N$) given the radius of gyration ($R_g$). Describing this as a spatial Poisson process, the probability is given as:

$$P(N|R_g) = \frac{1}{N} \left( \rho R_g^c \right)^N \exp \left( \rho R_g^c \right) \qquad (19)$$

The values for $\rho$ and $c$ are determined by a maximum likelihood estimation (resulting in $\rho = 0.334$ and $c = 2.274$ using a database of $\left( N, R_g \right)$ pairs derived from proteins structures in the Top500 database, which is a standard reference database of high quality structures.

### 3.2.3  MuCo and MuMu

The two empirical terms described here are probabilistic terms, which capture conformational preferences of proteins in the Top500 database. The MuCo term is based on a simple multinomial model that takes into account hydrogen bonding and compactness. Each amino acid falls into a number of bins depending on solvent exposure and hydrogen bonding. A probability is then assigned to the count vector based on the histogram observed in the Top500 database.

The MuMu term captures the probability of the number of neighbours expected for an amino acid, also based on a distribution obtained from the Top500 database. Both terms are under development and no publications are currently available for these. The cached version of MuMu is used in this work.

### 3.2.4  OPLS

The OPLS-aa (all atom) force field [28] is implemented in `Phaistos` and it is possible to specify the OPLS energy as an energy term.

## 3.3  Markov Chain Monte Carlo moves

Once the first trial protein structure has been generated by `Phaistos`, the next structure in the Markov Chain Monte Carlo simulation follows from the previos structure by randomly changing the backbone torsion angles according to the distribution of TorusDBN. Two different Monte Carlo move types are used in the presented work. (1) The DBN-Move randomly changes a backbone torsion angle according to TorusDBN and (2) the CRISP-Move, which randomly selects a 15 residue strand from the backbone and only changes backbone angles within the 15 residue strand.

The DBN-Move takes much larger conformational steps than the CRISP-Move, and is thus more efficient in exploring conformational space. However, the finer CRISP-Move is a better move-type when the structure is close to the minimum energy [Reference: Mikael Borg, personal communication].

For the *de novo* protein folding simulations carried out in this work, DBN-Moves and CRISP-Moves are used at ratio of 85 to 15, while in structure refinements, only CRISP-Moves are used.

# 4 Modeling the chemical shift of the backbone amide proton

Prediction of the chemical shift of the amide group atoms is very challenging, due to the high polarizability of the functional group. An amide group can be considered as a resulting average of two resonance structure (Fig. 2A and 2B). Pauling estimated, that an amide group in this resonance model has a preference of Fig. 2B of about 40%, while newer *ab inito* approaches estimate this preference to about 27% [51] [69]. Another model of the amide group is the polarization model (see Fig. 2C). In this model, the carbonyl oxygen is polarizing the C=O group by donating a positive charge to carbon, while the nitrogen lone pair is stabilized by the positive carbonyl carbon [69].

Regardless of how the amide group is described, it is easy to see that the electrons density around the amide group is easily perturbed. In a protein this includes torsion torsion angles of the nearby backbone bonds as well as the hydrogen bond strength of bonds directly at the hydrogen, and even hydrogen bonding at the carbonyl oxygen has a non-negligible effect [15]. Likewise, the chemical shift is very susceptible to nearby magnetic field caused by aromatic rings [10].

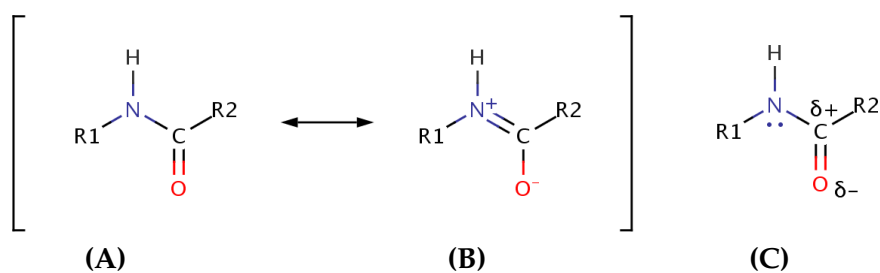

**Figure 2:** Graphic representation of the two models of the amide group, illustrating the easy polarization of the amide group.

Currently, the best methods for predicting the chemical shift of the amide protons correlates with experimental data with a linear correlation coefficient of only about 0.75-0.80 for the best methods and an RMSD of about 0.5 ppm when high resolution X-ray protein structures are used (for instance see table 15). This is in sharp contrast to the much less polarizable nuclei of the protein, such as $C^\alpha$ and $H^\alpha$, which are described with a correlation coefficient of up to 0.99 in favorable cases [66].

The amide proton and nitrogen are the backbone nuclei, the chemical shifts of which are the most difficult to predict [15]. As proof-of-concept for the use of quantum chemical methods to obtain accurate structural based prediction of chemical shifts in proteins, this work concerns the prediction of backbone amide hydrogen chemical shifts largely based on the work by Parker, Houk and Jensen [50].

Five programs capable of predicting amide proton chemical shifts currently exist, namely `PROSHIFT` [39], `SHIFTX` [48], `SPARTA` [61], `SHIFTS` [46] and `CamShift` [31]. Given a protein 3D-coordinate file, these programs predict the chemical shift of H, C and N nuclei in the protein based on empirical or semi-emprical parameters obtained from proteins with known structures and chemical shifts. The predictions of `SPARTA` and `CamShift` are not based on the underlying physics and chemistry of the system, but rather empirical correlations between structures and chemical shifts. `PROSHIFT`, `SHIFTX` and `SHIFTS` include physically founded terms in addition to this. The newest version of `SHIFTS` (Ver. 4.3) has further parameters based on quantum mechanical (QM) calculations. `CamShift` has the advantage that chemical shifts predictions are differentiable with respect to the cartesian coordinates of the atoms.

Another program, `CheShift` [67], is capable of predicting $C^\alpha$ chemical shifts, based exclusively on QM calculations. While the chemical shift predictions of `CheShift` are not as accurate (compared to matching experimental data) as the empirical programs mentioned when used on high quality X-ray crystal structures, `CheShift` has reported higher accuracy in discriminating between protein structures of varying quality based on chemical shifts than other methods [66].

Due to the restricted licensing of the `PROSHIFT` program, it is only available as an e-mail based web-service and is not included in most comparisons done throughout the rest of this work. However, we shall frequently benchmark and compare amide proton chemical shift predictions against `SHIFTX`, `SHIFTS` and `SPARTA` which all are freely available for download. Unfortunately `CamShift` was not released until August 2010, so a comparison to `CamShift` could not be made in this work due to time constraints.

In general the low accuracy of the amide proton chemical shift predictions is caused by two factors.

(1) Currently use crude simplification for the underlying physics which governs the chemical shift. For instance `SHIFTX` and `PROSHIFT` use very simple hydrogen bonding terms to describe the change in chemical shift due to hydrogen bonding effects in the protein, based solely on an $r^3$ dependence on the bond distance [50]. The empirical based programs thus use chemical shift terms which are largely based on structural features such as the secondary structure, rather than the actual chemical environment responsible for the chemical shift values.

(2) The chemical amide proton is very sensitive to small changes in geometries, such as hydrogen bonding distances, so even a small error

in the experimentally obtained atom coordinates will translate into large errors measured in ppm. This is further complicated by the fact that the positions of lighter atoms, such as hydrogen, are not very well defined from X-ray crystallography. A small number of known protein structures have the hydrogen atoms coordinates determined by neutron diffraction, which scatters on the nuclei. These are expected to give much more accurate results. However at the time of writing, only 39 protein structures from neutron diffraction experiments are currently deposited in the RCSB protein data bank (PDB - `www.pdb.org`). [5]

## 4.1 The Parker, Houk and Jensen model

Parker, Houk and Jensen [50] have developed a simple method to calculate the chemical shift of amide protons in proteins. The chemical shift is broken into five additive terms which are evaluated separately and added up to give the approximated chemical shift. The five terms depend on the local protein structure around the amide proton and quantifies the structural parameters as chemical shifts contributions. The terms consist of a backbone conformation term, three terms relating to the hydrogen bonding network of the amide group and finally a ring current term, which quantifies the effect of the magnetic field due to nearby aromatic side chains in the protein. This approach was found to correlate well with experimental data for 13 residues in human ubiquitin and the third IgG-binding domain of streptococcal protein G (Protein G). Using X-ray derived structures the correlation obtained by comparing to experimental data was $r = 0.83$ (RMSD = 0.7 ppm), while using the method on protein structures where the hydrogen bonding conformations were optimized, the correlation was improved to $r = 0.94$ (RMSD = 0.3 ppm) [50]. The functional form of the chemical shift of a backbone amide proton is in the Parker, Houk and Jensen model approximated as:

$$\delta_{H^N} = \Delta\delta_{BB} + \Delta\delta_{1^\circ HB} + \Delta\delta_{2^\circ HB} + \Delta\delta_{3^\circ HB} + \Delta\delta_{RC} \tag{20}$$

The backbone term, $\Delta\delta_{BB}$, is dependent on the internal geometry of the backbone, presumably especially on the proximity to the carbonyl oxygen within the residue. This term is given as a simple step function in the original work, but as it will be demonstrated, a better approximation is available. This is discussed in section 4.2.

The primary bond term, $\Delta\delta_{1^\circ HB}$, describe the contribution to the chemical shift from a hydrogen bond formed between the amide proton and a hydrogen bonding acceptor (these bonds are termed primary bonds). In the Parker, Houk and Jensen paper, only bonds to amide oxygen are described using an expression derived by Barfield [3]. In section 4.3 a similar method is used to describe bonds to the most common side chains, i.e. amides, alcohols and carboxylic acids. Later on, a simple approximation for solvent exposed amide protons as well as non-bonded amide protons is derived. Primary bonds are discussed in section 4.3.

The secondary bonding term, $\Delta\delta_{2°HB}$, describes the change in chemical shift of the amide proton due to the presence of a hydrogen bond to the carbonyl of the amide group. Further hydrogen bonding to primary and secondary hydrogen bonding partners such as amides, carboxylic acids and alcohols also have non-negligible effect on the chemical shift of the amide proton. These are termed tertiary hydrogen bonding partners and are described in the term $\Delta\delta_{3°HB}$. These are briefly discussed in section 4.4.

The last term, $\Delta\delta_{RC}$, is a ring current interaction term due to the magnetic field of nearby aromatic side chains. Parker, Houk and Jensen do not give an explicit functional form of this term, but use the equivalent term from the SHIFTS program [46]. In chapter 5, different approximations of this effect are compared in order to find the most accurate and computationally inexpensive approximation.

In Fig. 3 an overview of the bonding terms in the Parker, Houk and Jensen model is presented.

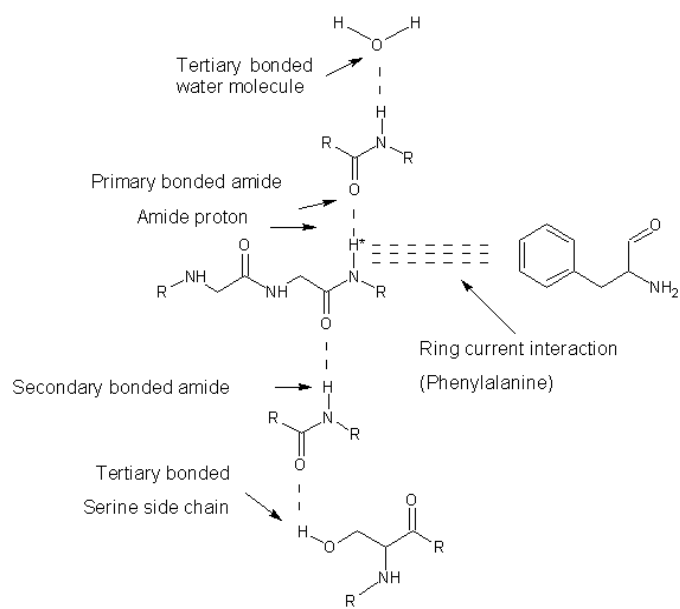

**Figure 3:** An example of the bonding partners that affect the chemical shift, illustrating the naming scheme. The amide proton depicted has a primary and a secondary hydrogen bond to an amide oxygen as well as tertiary hydrogen bonds to a water molecule and a charged side chain (serine/alcohol). A ring current interaction from a nearby phenylalanine side chain is also shown.

### 4.1.1 Hydrogen bond definition

We briefly define a hydrogen bond as the presence of a non-covalently bonded oxygen in a distance less than or equal to 2.5 Å from a hydrogen. This is not a completely arbitrary choice. Fig. 4 shows the distribution of the distance between every amide proton to the nearest possible oxygen hydrogen bonding partner. The protein structures used are taken from a data set of 15 high resolution X-ray crystal structures (see Appendix E). The shortest observed hydrogen bonding distance was 1.52 Åand only few bonds were found shorter than 1.7 Å. At a bond length of about 2.3 Å and longer, the bond lengths become increasingly rare, and after 2.4 Åonly few are present. In conclusion, the vast majority or hydrogen bonds found in a protein are described by this approximation.

Rare bonds, such as for instance hydrogen bonds to a neutral histidine nitrogen lone pair or sulfur, aromatic hydrogen bonding etc. are not considered in the following.

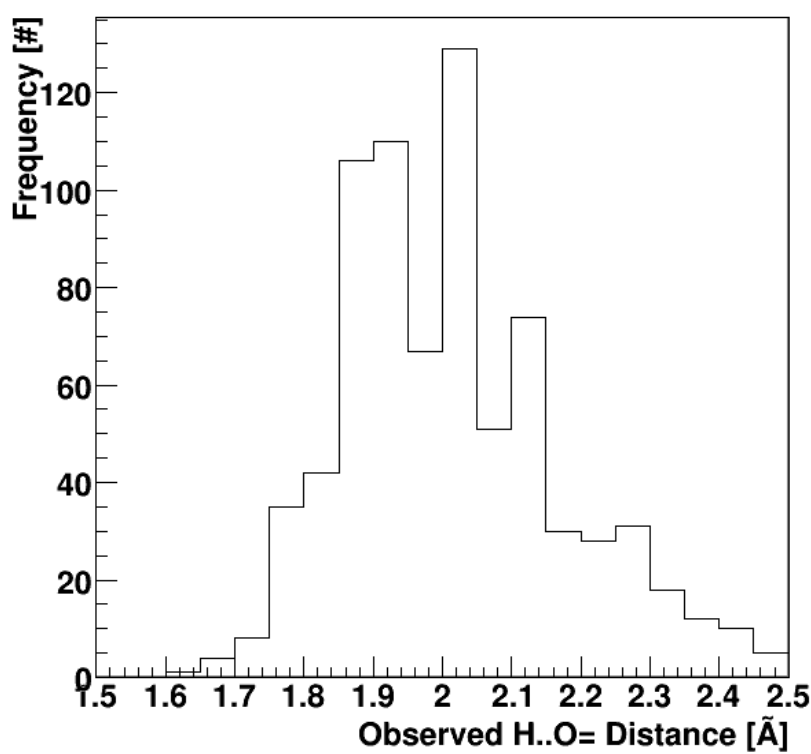

**Figure 4:** This figure represents the distribution of the distance from amide protons of 9 proteins to the nearest possible oxygen hydrogen bonding partner. Note also that no hydrogen bonding distances shorter than 1.5 Å were present and bonds shorter than 1.6 Å are extremely rare.

## 4.2 Backbone term

The form of the backbone contribution proposed by Parker, Houk, and Jensen is a three-interval step function which depends on the $r_\omega$ distance (the distance between the carbonyl oxygen and amide proton within the residue) and the dihedral angle between the N-H and C=O bond (see Fig. 5 for a graphic overview)

$$\Delta\delta_{BB}^{PHJ}(\omega) = \begin{cases} 4.5\,\text{ppm} & \text{if} & f(\omega, r_\omega) \leq 0.03 \\ 5.5\,\text{ppm} & \text{if} & 0.03 < \quad f(\omega, r_\omega) \leq 0.10 \\ 6.5\,\text{ppm} & \text{if} & 0.10 < \quad f(\omega, r_\omega) \end{cases} \tag{21}$$

where

$$f(\omega, r_\omega) = \frac{\cos^2(\omega)}{r_\omega^3}, \tag{22}$$

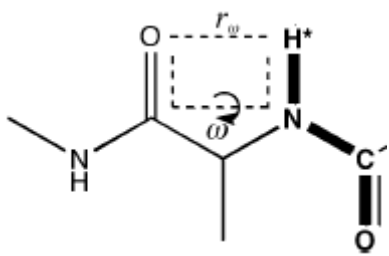

**Figure 5:** Sketch showing the structural parameters for $\omega$ and $r_\omega$ [50]

However, as it will be demonstrated in section 4.6.1, a functional form proposed by Czinki and Császár leads to significantly better results and is used instead [14]. This term is a 10'th-order cosine function fitted to *ab initio* chemical shifts computed on small peptides for a dense grid (1296 pairs) of $\phi$ and $\psi$ backbone angles. The coefficients obtained by Czinki and Császár can be found in Appendix B. The functional form is as follows:

$$\Delta\delta_{BB}^{CC}(\phi, \psi) = a + \sum_{n=1}^{N} \left[ b_n \cos(n\phi') + c_n \cos(n\psi') \right] + \sum_{n=1}^{N-1} \sum_{m=1}^{N-n} d_{nm} \cos(n\phi') \cos(m\psi') \tag{23}$$

where $\phi' = \frac{\phi + \pi}{2}$ and $\psi' = \frac{\psi + \pi}{2}$, where $\psi$ and $\phi$ are the usual backbone torsion angles.

Since the Czinki and Császár study was carried out at the B3LYP/TZ2P//B3LYP/6-31+G(d) level of theory relative to TMS at the B3LYP/TZ2P//B3LYP/6-311++G(d,p) level of theory, there is likely to be a linear offset between the chemical shifts obtained by Czinki and Császár and the empirically scaled DFT method used in this section (See methodology section). The Czinki and Császár study does not include any comparison to empirical data or higher order methods. DFT data obtained using a reasonably sized basis set often has a high linear correlation to well-behaved higher-order methods or experimental data, but the RMSD may not be low. To alleviate this problem, it is often enough to apply a simple linear scaling and much better results can be obtained. Here, a linear scaling to the Czinki and Császár backbone term is used in order to obtain better agreement with the DFT method used in this section, which is based on the parameters obtained by Rablen, Pearlman and Finkbiner [56]. A best fit gave the following values for $\Delta\delta_{BB}$, which is used as the backbone term in this work:

$$\Delta\delta_{BB} = (\Delta\delta_{BB}^{CC}(\phi, \psi) + 0.77 \text{ ppm}) \cdot 0.828 \qquad (24)$$

## 4.3 Primary hydrogen bonding terms

The Parker, Houk and Jensen paper only explicitly describes a functional form for the chemical shift due to amide-amide bonds by using an expression derived by Barfield [3]. In order to interface the chemical shift prediction to the protein structure prediction software package `Phaistos`, it is necessary to be able to predict the chemical shift of every amide proton in the protein. It is thus necessary to generate expressions for the chemical shift contributions due to other bonds than amide-amide bonds so that a meaningful chemical shift can be calculated for every amide proton.

### 4.3.1 Primary bonds to amide oxygen atoms

The chemical shift contribution due to a primary bond to an amide oxygen is described by Barfield. By studying conformations of formamide (FMA) dimers, mimicking amide-amide hydrogen bonding conformations, such as those found in $\alpha$-helices and $\beta$-sheet conformations, the effect of hydrogen bonding geometry was investigated. Based on QM calculations of the chemical shift of the dimer complexes the chemical shift contribution of the hydrogen bonds were parametrized as a best fit to the QM data. The best fit obtained was:

$$\delta_{1^\circ HB} = \left\{ 4.81 \cdot \cos^2(\theta) + \sin^2(\theta)[3.01 \cdot \cos^2(\rho) - 0.84 \cdot \cos(\rho) + 1.75] \right\}$$
$$\cdot e^{-2(r_{OH} - 1.760)} \qquad (25)$$

where $\rho$, $\theta$ and $r_{OH}$ are the dihedral H..O=C-N dihedral angle, H..O=C angle and the H..O hydrogen bond distance (see Fig. 6). The fit is in very good correlation with the QM data with a standard deviation of 0.09 ppm and $r = 0.993$.

The Barfield study has two important implications for the presented method of chemical shift prediction. (1) a ready-to-use formula of the chemical shift contribution from amide-amide bonds is given. And (2), that fact that the *only* important hydrogen bonding geometry parameters are $\rho$, $\theta$ and $r_{\mathrm{OH}}$ (see Fig. 6). Inclusion of more parameters did not increase the correlation or RMSD between calculated data and the fitted prediction. This is also confirmed in a study by Moon and Case, which features a thorough investigation of the N-H..O' and C-N-H..O'. Only in very unusual conformations of very low N-H..O' angles and close hydrogen bonding distances does the C-N-H..O' dihedral angle have some effect. The same formula is used in this work to describe the chemical shift contribution due to bonding to both backbone and side chain amide oxygens.

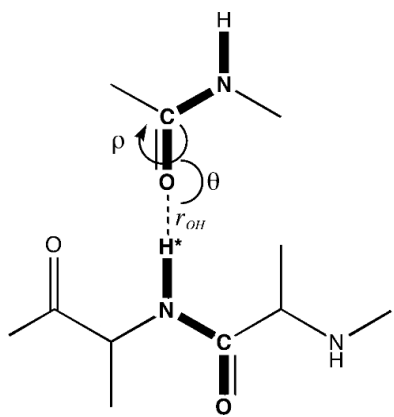

**Figure 6:** Sketch showing the geometric parameters used in the Barfield model of amide-amide interactions. The bonds of the amide groups are shown in bold for emphasis.

### 4.3.2 Bonds to carboxylic acids and alcohols

In this section, a model of the chemical shift contributions due to hydrogen bonding to carboxylic acids and alcohol side chains are described. These are common hydrogen bonding acceptors in proteins. Since the C-terminus and the aspartic acid and glutamic acid side chains are, in the current `Phaistos` framework, always given in the deprotonated state, only this state is considered in this section.

An approach similar to the amide-amide model due to Barfield is used. The model system consists of two molecules, modeling the amide hydrogen bonding donor and acceptor complex, and the geometric dependence is modeled by scanning conformations over relevant angles and distances.

As an approximation to the carboxylic acid functional groups found in the protein (aspartate, glutamate and the C-terminus), an acetate anion is used. The alcohol functional groups, threonine, serine and tyrosine are approximated by a methanol. It is, however, a fact, that the phenol -OH group of tyrosine side chain is not chemically completely similar to the -OH group of an alcohol, due to charge delocalization effects. For instance, the partial charge on tyrosine oxygen is -0.55 $e$, compared to -0.67 $e$ and -0.65 $e$ for serine and threonine, respectively [53]. Due to the rarity of backbone amide-tyrosine bonds, a less exact determination of these particular chemical shifts is unlikely to be critical for the overall efficiency of the method and no special action is taken to differentiate between different types of alcohol groups. As a backbone amide model, an *N*-methylacetamide (NMA) molecule is used.

Using minimal amide, carboxylic acid and alcohol models, a scan over a range of bond angles and distances is carried out. The hydrogen bonding distance is modeled in a range from 1.5 Å to 2.5 Å in 0.125 Å steps. From Fig. 4 it is evident, that shorter hydrogen bonds are unlikely and the cutoff used in determining the presence of a hydrogen bond in this work is 2.5 Å. The dihedral angles H..O=C-C and H..O-C(..)H$^O$ are scanned over the full 360° range in 15° intervals. Finally, the H..O-C angle was scanned from 180° to 90° in 10° intervals. To avoid steric clashes between the methyl groups of the NMA molecule and the methanol/acetate, the N-H..O bond angle was fixed at 180°. See Fig. 7 for an overview of the models. The size of both grids were 1944 model systems, for which the NMR shieldings was calculated.

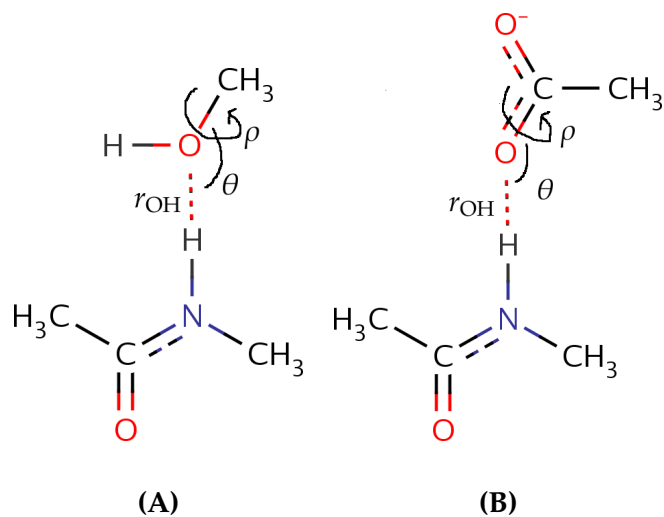

**Figure 7:** Sketches showing the geometric parameters and the systems used in the modeling of chemical shift contributions due to hydrogen bonding. In A), the $\theta$ angle is defined as the H..O-C angle, while $\rho$ is defined as the H..O-C(..)H$^O$ dihedral angle. In B), $\theta$ is the H..O-C angle, and $\rho$ is the H..O=C-C dihedral.

As in the case of amide-amide bonds, the effect of variations in the N-H..O and C-N-H..O angles were found negligible when calculating the resulting chemical shift due to bonds to carboxylic acids and alcohols. This is shown in section 4.6.2.

Barfield calculates the chemical shift of a number of model systems and uses the data to fit a formula. The decision was taken to use a much less laborious approach and store the data in look-up tables and do interpola-tion between data points. A look-up tables is a very efficient tool when the amount of data large and the fitting surface is complicated. This approach has recently been applied by Scheraga and coworkers in the `CheShift` chemical shift prediction server. [67] The `CheShift` server predicts $C^\alpha$ chemical shifts of proteins given an input structure by comparing the geometry to a QM-generated database of more than 500,000 conformations. Given the large amount of complicated data structures, fitting an analytical function or series to the `CheShift` database is impracticable.

The look-up table in this work consist of a complete 3D-grid of evenly spaced, discrete data points, each corresponding to a specific geometry described by the three geometric parameters, $\rho$, $\theta$ and $r_{OH}$. To allow for the prediction of any geometry between the grid points, an interpolation algorithm is required. Two different interpolation algorithms were compared: simple trilinear interpolation and tricubic interpolation, which enforces continuous first derivatives of the hyper surface. Tricubic interpolation relies on the partial derivatives at the grid points. For the equally spaced grid, these can be approximated as:

$$\frac{\partial f(q_i)}{\partial q} = \frac{f(q_{i+1}) - f(q_{i-1})}{2\Delta q}, \qquad (26)$$

where $\Delta q$ is the grid spacing in the $q$ direction. At the boundaries, the partial derivative were given based on the difference to the only neighbor in the given direction, e.g:

$$\frac{\partial f(q_0)}{\partial q} = \frac{f(q_1) - f(q_0)}{\Delta q} \qquad (27)$$

For dihedral angles, cyclic boundaries were used. For the rare geometries out of the range of the look-up table, the nearest value in the out-of-range direction is used; e.g for a hydrogen bond distance of 1.45 Å a distance of 1.5 Å would be used. Most of the code for the tricubic interpolation was taken from "Numerical recipes in C" [55].

The difference between the interpolation scheme proved to be of very little importance due to the grid density. In all cases, the difference was found to be less than 0.03 ppm, which is much less than about any other uncertainty in other approximations we have made in our prediction of the amide proton chemical shift. For a set of 210 randomly generated input geometries the linear correlation between the methods was 0.999 for the alcohol and the carboxylic acid look-up data.

Due to the diminutive difference between simple trilinear and smoother tricubic interpolation scheme, no other interpolation methods were tested, and the simplest and fastest interpolation scheme, trilinear interpolation, was subsequently used in the `C++` implementation of the method.

### 4.3.3 Solvent exposed amide protons

Accurate modelling of solvent around the protein is an extremely difficult problem in modern computational chemistry due to the complexity involved. Since `Phaistos` generate structures, that do not contain explicit solvents (e.g. crystallographic water molecules), the simplistic model used in this work implicitly assumes the presence of water molecules, when the amide proton is considered to be solvent exposed.

The chemical shift contribution from solvent exposure of the amide proton is here assumed to be equivalent to the contribution from a hydrogen bond to a water molecule. A typical contribution from a water molecule in a energy minimized hydrogen bonding conformation is found in section 4.6.3 to be +2.07 ppm. As a crude approximation solvent exposed amide protons are assigned with a fixed +2.07 ppm primary bond contribution.

### 4.3.4 Modeling of other bonds types

An explicit evaluation of the chemical shift due to other bond types than the previously mentioned is not considered, as these are extremely rare. These should in principle include hydrogen bonds to partners such as neutral histidine, aromatic rings and a few others. Another question is the amide protons not participating in hydrogen bonding, which are buried inside the protein. How these amide protons interact with the surrounding environment is a non-trivial problem, due to the number of possible interactions. It is likely, that such a buried amide proton will form some hydrogen bond-like interaction, thus increasing the chemical shift of the amide proton.

Since no model is available for the chemical shift of the amide protons described here, all such unlikely and complicated cases are for the sake of simplicity treated by the code as if they were solvent exposed. +2.07 ppm is a typical deshielding caused by a hydrogen bond, and this is again used as a crude estimation of the chemical shift due to these types of primary bonds.

| Bonding partner: | Non-bonded | Water | Amide | Charged side chain |
|---|---|---|---|---|
| $\Delta\delta_{2^{\circ}HB}$, secondary | +0 | +0.15 | +0.3 | +0.8 |
| $\Delta\delta_{3^{\circ}HB}$, tertiary (to primary) | +0 | +0.05 | +0.1 | +0.8 |
| $\Delta\delta_{3^{\circ}HB}$, tertiary (to secondary) | +0 | +0.1 | +0.2 | +0.8 |

**Table 1:** Small contributions to the chemical shift. Interactions to secondary and tertiary hydrogen bonding partners. All numbers are given as ppm.

## 4.4   Secondary and tertiary hydrogen bonding terms

One of the main findings by Parker, Houk and Jensen is that not only the hydrogen bond directly at the amide proton has an effect on the chemical shift, but also the hydrogen bonds to the carbonyl oxygen. Bonds at the carbonyl oxygen are here called secondary bonds. The effect of secondary bonds has previously been described for nitrogen [15], but was not shown for the amide proton before Parker, Houk and Jensen. Further hydrogen bonding by a third group to the primary and secondary hydrogen bonding partners, respectively, was also shown to give a non-negligible contribution, particularly if the bonding partner is a charged side chain. These are termed tertiary bonds (see Fig. 3 for an example). The chemical shifts due to secondary and tertiary bonds are described by the terms $\Delta\delta_{2°\mathrm{HB}}$ and $\Delta\delta_{3°\mathrm{HB}}$, respectively. See table 1 for an overview of these terms. In the `Phaistos` implementation, the chemical contribution due to tertiary bonds to a primary hydrogen bonding partner are only considered for amide protons bonded to another backbone amide group.

## 4.5   Ring current term

The aromatic side chains in the protein give rise to magnetic fields due to ring current effects. These usually have an effect in the range of -1.2 ppm to +0.3 ppm. This is described in the term $\Delta\delta_{\mathrm{RC}}$ in Eq. 20. This effect is described in several studies, but the parameters used in descriptions by different authors differ wildly (see for instance table 11 in the next chapter), and it is not obvious which study has the most accurate set of parameters. Chapter 5 reviews the parameters and formalisms used in other studies in order to find the most computational efficient and accurate description of the ring current effect by a comparison to parameters obtained using QM methods.

### 4.6 Results

#### 4.6.1 Backbone terms

As a benchmark between the backbone terms suggested by Czinki and Császár and Parker, Houk and Jensen, a set of 34 backbone fragments were constructed. The backbone conformations were sampled from a structure of the third IgG-binding domain from streptococcal protein G (Protein G) in the PDB database (code: 1IGD). The backbone fragments were constructed as tri-alanine peptides, with the ends capped as -CO-CH$_3$ and NH-CH$_3$ towards the N-terminus and C-terminus respectively, and side chains were truncated as alanine groups to minimize the computational effort. Fragments including proline or glycine residues were discarded to ensure uniformity of the sample set. Subsequently, all hydrogen positions were optimized as well as the amide O=C-N-H moiety at the N-H of the middle alanine reside. Lastly, the chemical shift of the amide proton of the central alanine (shown in the grey box on Fig. 8) was calculated quantum chemically. See the methodology section for a more detailed description of the methods used. The resulting geometric parameters can be found in table 18 in Appendix C. From the optimized structures the chemical shift was calculated by QM as well as using the two backbone terms.

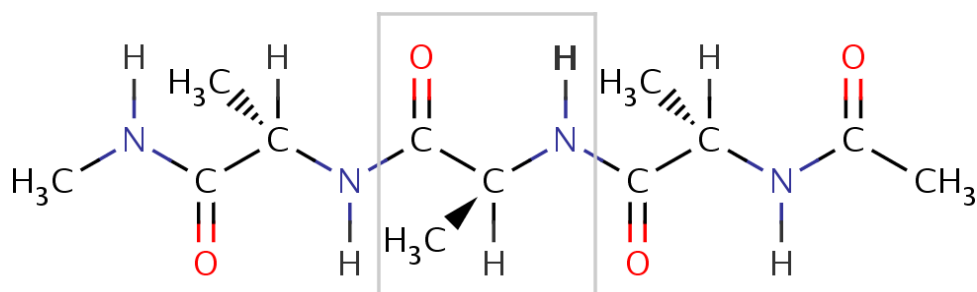

**Figure 8:** Conceptual sketch showing the extend of the systems used as backbone fragments. The $\phi$ and $\psi$ backbone angles are taken as the torsion angles of the central backbone residue shown in the grey box. The amide proton of the central rescue is furthermore show in bold font.

The linear correlation between QM and the approximate methods are 0.80 and 0.94 for the Parker, Houk and Jensen model and the Czinki and Császár model, respectively. Applying a linear correction to $\Delta\delta_{BB}^{CC}(\phi, \psi)$ will then give the lowest possible RMSD between the two methods. See table 2). The best fit obtained was:

$$\Delta\delta_{BB} = (\Delta\delta_{BB}^{CC}(\phi, \psi) + 0.77 \text{ ppm}) \cdot 0.828 \tag{28}$$

The RMSD for the PHJ model is 0.34 ppm and 0.37 ppm for the unscaled Czinki and Császár model. Most of the error in the Czinki and Császár

model, is due to a relatively large constant offset. After applying the linear correction, the RMSD is down to only 0.17 ppm, due to the good linear correlation of 0.94, which is of course preserved after a linear transformation.

|  | PHJ | CC | CC (scaled) |
|---|---|---|---|
| $r$ | 0.80 | 0.94 | 0.94 |
| RMSD [ppm] | 0.34 | 0.37 | 0.17 |
| Slope | 0.93 | 1.21 | (1) |
| Intersect [ppm] | 0.31 | -0.77 | (0) |

**Table 2:** The correlation between chemical shift of 34 tri-alanine backbone fragments obtained using DFT at the GIAO/B3LYP/6-311++g(d,p)//B3LYP/6-31g(d) level of theory and two approximations. PHJ is the approximate step function of Parker, Houk and Jensen, while CC is the approximate 10th-order cosine series of Czinki and Császár. The slope and intersect of the Czinki and Császár was used as a best-fit values to scale the Czinki and Császár backbone term and obtain a better agreement with the DFT data. The statistics of this best-fit is given in the last column, resulting in a significant decrease in RMSD.

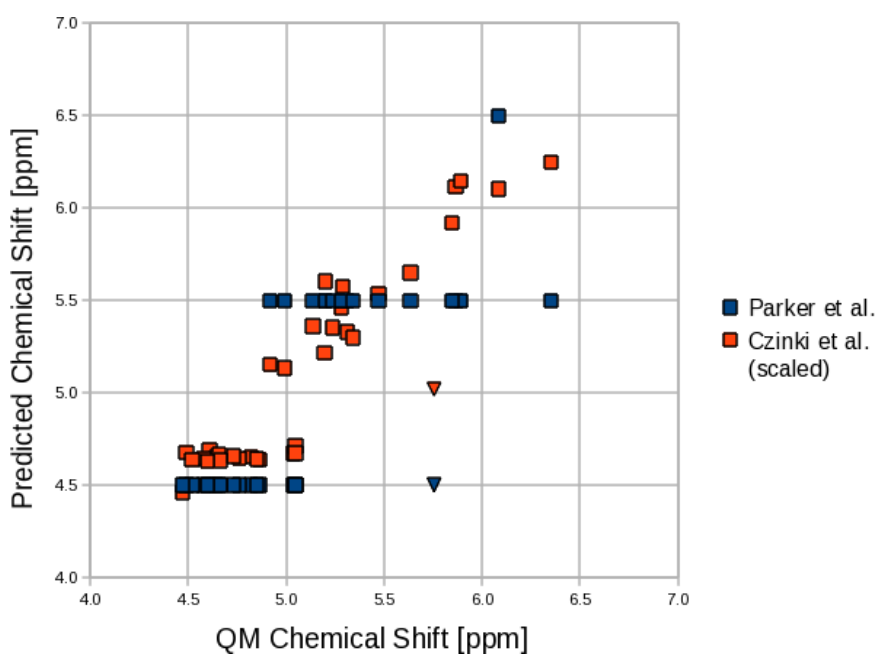

**Figure 9:** For at set of model backbone fragments from Protein G, the chemical shift of the amide proton is calculated quantum chemically (x-values). The predictions of model used by Parker, Houk and Jensen is shown in blue, while the best-fit scaled term derived by Czinki and Császár is shown in orange (see text). An outlier, not included in the fit is displayed marked with a triangular dot.

One backbone fragment was discarded in the fit data, due to it's very unusual conformation. The geometry is depicted in Fig. 10. This unusual conformation places the amide proton of residue 55 only 2.25 Å from the nitrogen of residue 54. In this position, the amide proton has a weak interaction to the nitrogen causing an extra deshielding of about 1 ppm, compared to the predictions of the backbone terms. This is of course a flaw in the backbone term approximations, which mostly predicts the interaction from the nearby carbonyl oxygen and to a lesser extend the side chain of the residue. For the majority of the data points, however, the approximation is in very good agreement with the QM data.

Fig. 9 shows a scatter plot of the data from original Parker, Houk and Jensen step function backbone term and the new proposed scaled backbone term against DFT data. It is very clear from the graphic representation, that much better agreement is obtained by using the Czinki and Császár term.

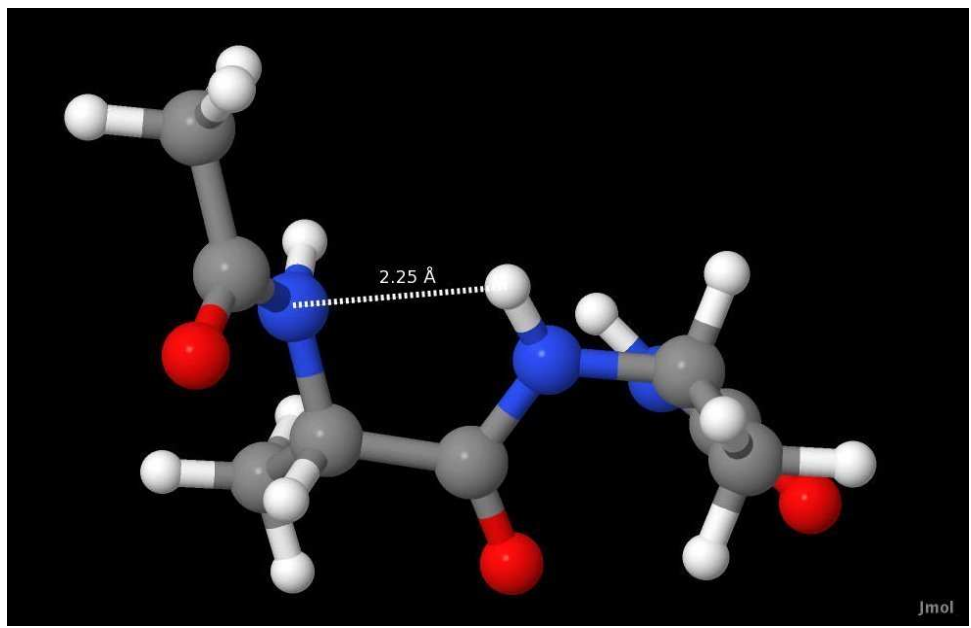

**Figure 10:** The very unusual backbone geometry around the amide proton of Protein G residue 55. The distance is between to the nitrogen of residue 54 is only 2.25 Å. In this position, the amide proton forms weak hydrogen bond-like interaction due to the nitrogen p-orbital. Such polarization effect are not accounted for in the approximate backbone terms, resulting in an underestimation of the predicted chemical shift compared to a QM calculated chemical shift.

### 4.6.2 Redundant bond angles

To test the hypothesis of the N-H..O and the C-N-H..O angles being redundant, a test system was set up using an methanol-NMA dimer and a acetate-NMA dimer. The H..O-C bond was fixed at 180° and the hydrogen bonding distance was set to 2.0 Å. The N-H..O angle was varied between 140° an 180°, while the C-N-H..O was varied between 60° to 120°. Further extending the scan over the C-N-H..O was not meaningful since it would cause a steric clash between the alcohol and the methyl groups of the NMA molecule.

The deviation between the lowest and the highest chemical shift in the tested geometries was 0.14 ppm in the methanol-NMA dimer systems and 0.28 ppm in the acetate-NMA dimer systems (see table 3 for the calculated data). These results largely agree with the results obtained by Barfield, where these angles were found to be much less significant than the $\rho$ and $\theta$ angles (see fig. 7). As an approximation the two bond angles are ignored when performing the scan over relevant angles and bonding distances. The approximation seems to be a bit better for bonds to alcohols, compared to bonds to carboxylic acids. A likely explanation is the stronger bond strength of carboxylic acids hydrogen bonds.

| 3A: NMA-Methanol dimer | | | |
|---|---|---|---|
| Chemical shift [ppm] | C-N-H..O = 60° | = 90° | = 120° |
| N-H$^N$..O = 140° | 6.26 | 6.29 | 6.32 |
| = 160° | - | 6.24 | - |
| = 180° | - | 6.19 | - |

| 3B: NMA-Acetate dimer | | | |
|---|---|---|---|
| Chemical shift [ppm] | C-N-H..O = 60° | = 90° | = 120° |
| N-H$^N$..O = 140° | 7.25 | 7.27 | 7.28 |
| = 160° | - | 7.49 | - |
| = 180° | - | 7.53 | - |

**Table 3:** Variation of other tested bond angles in calculation of the amide proton chemical shift in (A) an NMA-methanol dimer and (B) an NMA-acetate dimer.

### 4.6.3 Hydrogen bonding to a water molecule

A water molecule is placed near the amide hydrogen atom of a probe NMA molecule and a B3LYP/6-311++G(d,p) minimization is carried out. The resulting structure is a local minimum of an amide hydrogen bonded to a water molecule. From this geometry the chemical shift of the entire dimer is then calculated.

Using four different starting geometries, the water molecule was minimized into two different conformation (see Fig. 11A and 11B). The hydrogen bonding geometry of these two conformations had a few similarities. The water oxygen was in both cases aligned into the N-H bond axis. The hydrogen bonding distance was also similar at 2.04 Å and 2.07 Å, respectively. To separate the change in chemical shift due to change of the internal geometry of the NMA molecule, another NMR calculation was carried out using the optimized systems, but with the water molecule removed. One minimization done without the geometry restriction led to a 60° rotation of a methyl group. The subsequent analysis of the systems with the water molecule removed revealed that the rotation caused an extra shielding of about 0.2 ppm. However, by using the NMA geometries from the optimized dimer as a reference, this artifact was removed. The resulting chemical shift due to the water molecule turned out to be very similar, at +2.04 ppm and +2.09 ppm respectively. The chemical shift in these minima serve as rough figures for the chemical shift due to solvent exposure. We thus assign a contribution of +2.07 ppm to the total chemical shift of solvent exposed amide protons.

| Chemical Shift | Minimum A | Minimum B |
|---|---|---|
| Optimized NMA-Water Dimer | 6.23 ppm | 6.45 ppm |
| Optimized NMA alone | 4.19 ppm | 4.37 ppm |
| Difference | +2.04 ppm | +2.09 ppm |

**Table 4:** The chemical shift of the amide proton in two local energy minima. "NMA-Water Dimer" is the chemical shift of the NMA amide proton in the dimer. "NMA alone" is the amide proton chemical shift is the resulting chemical shift of the NMA amide proton in the optimized configuration when the water molecule is removed and no furter optimization is carried out. The resulting difference corresponds to the change in chemical shift due a hydrogen bond to a water molecule.

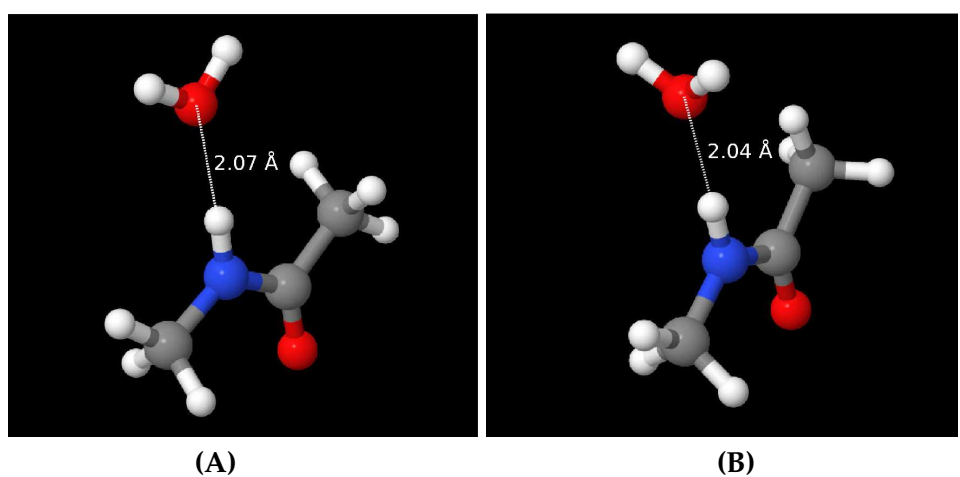

**(A)**                    **(B)**

**Figure 11:** Two different local energy minima of the NMA-water dimer. The hydrogen bonding distances are almost identical. The water molecule is rotated 90° between A and B, and one methyl group has a 60° difference in rotation between A and B.

## 4.7 The vailidity of an additive model

One key question in this work is the general validity of Eq. 20. The foundation of Eq. 20 is that the chemical shift contributions work in an additive manner as suggested by Parker, Houk and Jensen. Eq. 20 is valid, if the contribution from a primary bond has the same effect on amide protons with a high chemical shift due to the backbone conformation, as it has on amide protons with small backbone conformation contributions, or at least if the error is acceptably small.

To test this, a series of seven backbone fragments were constructed, and for each backbone, the isotropic shielding was calculated when the amide proton was in a hydrogen bonding conformation at varying hydrogen bonding distances. The seven backbone fragments were picked from the set used in section 4.2(see Fig. 8), such that the chemical shift due to the backbone would cover a range of 4.6-6.3 ppm, which is the lowest to highest observed in Protein G. The hydrogen bonding partner was the oxygen of a formamide molecule, as used in the Barfield study. The hydrogen bonding distance was varied between 1.5 Å and 2.5 Å, and the N-H..O′ and H..O′=C′ bond angles were restricted to 180°. No optimization was carried out on the dimer.

From the chemical shift of the backbone fragment amide proton, the chemical shift contribution of a primary bond was calculated (1) using the formula by Barfield and (2) via a QM calculation on the dimer, from which the contribution could be inferred. The difference between the two methods corresponds closely to the error caused by using the additive model of Eq. 20. The QM derived values are shown in table 5. By subtracting the non-bonded backbone chemical shift and the Barfield term for a given conformation from the QM calculated chemical shift of the dimer, we obtain a difference corresponding to the error caused by the additive approximation. In table 6, the Barfield term is given for each of the six hydrogen bonding distances and the error cause by the additive approximation is given for each dimer conformation. The error is thus effectively calculated as (see table 5 and 6):

$$\Delta\delta^{\text{Err}} = \delta_{\text{Conf}} - (\delta_{\text{Non-bonded}} + \Delta\delta_{\text{Barfield}}) \qquad (29)$$

### 4.7.1 Results - the additive model

The size of the error increases with the size of the chemical shift contribution due to hydrogen bonding. For the backbone conformations with low chemical shifts, the Barfield formula has a tendency to underestimate the chemical contribution, while the Barfield formula overestimates the chemical shift, when the chemical shift is large due to the backbone conformations.

Protein G is a representative protein for a distribution of backbone chemical shifts, since it features common types of conformations (i.e. $\alpha$-helix and $\beta$-sheet as well as a few irregular elements). The distribution of Protein G backbone chemical shifts can be seen on Fig. 9. It is noted that most chemical shifts fall into two main groups. One group is in a range of 4.5-5.0 ppm. These are mostly residues in the $\alpha$-helix, which is the common group with generally the lowest chemical shifts. The second group is that of $\beta$-sheet and irregular residues which have higher chemical shifts mostly due to the proximity of the carbonyl oxygen of the residue. These chemical shifts are mostly found in the range of 5.0-5.5 ppm, but also up to about 6.5 ppm.

Since hydrogen bonding distances shorter than 1.8 Å are extremely rare (See Fig. 4) and the backbone contribution to the chemical shift usually is found in the range of 4.5-5.5 ppm, the error caused by the additive approximation will for the vast majority of chemical shifts found in the protein be less than 0.2 ppm, often below 0.1 ppm. Note that table 5 covers a very wide range of chemical shifts from 5.8-13.0 ppm, which is wider than the range found in most proteins.

| HB dist [Å] | $\delta_{Conf1}$ | $\delta_{Conf2}$ | $\delta_{Conf3}$ | $\delta_{Conf4}$ | $\delta_{Conf5}$ | $\delta_{Conf6}$ | $\delta_{Conf7}$ |
|---|---|---|---|---|---|---|---|
| 2.5 | 7.21 | 7.22 | 7.07 | 6.79 | 6.11 | 5.97 | 5.80 |
| 2.3 | 7.52 | 7.53 | 7.40 | 7.15 | 6.49 | 6.35 | 6.18 |
| 2.1 | 8.02 | 8.04 | 7.93 | 7.70 | 7.08 | 6.94 | 6.77 |
| 1.9 | 8.89 | 8.92 | 8.82 | 8.61 | 8.03 | 7.89 | 7.73 |
| 1.7 | 10.41 | 10.43 | 10.35 | 10.17 | 9.64 | 9.49 | 9.34 |
| 1.5 | 12.99 | 13.00 | 12.93 | 12.77 | 12.29 | 12.14 | 12.00 |
| Non-bonded | 6.27 | 6.23 | 6.01 | 5.68 | 4.93 | 4.81 | 4.63 |
| Max diff [ppm] | 6.72 | 6.78 | 6.92 | 7.09 | 7.36 | 7.33 | 7.37 |

**Table 5:** The chemical shift of the amide proton i seven different backbone conformations ($\delta_{Conf1}$ to $\delta_{Conf7}$), while engaging in a hydrogen bond to a formamide molecule at six different bond-distances. The chemical shift of the non-bonded amide proton in the different backbone conformations is also given. All chemical shifts are given in ppm.

| HB dist [Å] | $\Delta\delta_{Barfield}$ | $\Delta\delta^{Err}_{Conf1}$ | $\Delta\delta^{Err}_{Conf2}$ | $\Delta\delta^{Err}_{Conf3}$ | $\Delta\delta^{Err}_{Conf4}$ | $\Delta\delta^{Err}_{Conf5}$ | $\Delta\delta^{Err}_{Conf6}$ | $\Delta\delta^{Err}_{Conf7}$ |
|---|---|---|---|---|---|---|---|---|
| 2.5 | 1.14 | 0.16 | 0.15 | 0.12 | 0.07 | -0.04 | -0.06 | -0.08 |
| 2.3 | 1.50 | 0.19 | 0.18 | 0.14 | 0.09 | -0.05 | -0.07 | -0.10 |
| 2.1 | 2.05 | 0.23 | 0.22 | 0.17 | 0.10 | -0.06 | -0.08 | -0.12 |
| 1.9 | 2.98 | 0.28 | 0.27 | 0.22 | 0.13 | -0.07 | -0.10 | -0.15 |
| 1.7 | 4.54 | 0.35 | 0.34 | 0.27 | 0.16 | -0.09 | -0.13 | -0.19 |
| 1.5 | 7.14 | 0.43 | 0.42 | 0.33 | 0.19 | -0.11 | -0.16 | -0.23 |

**Table 6:** The estimated error cause by the additive approximation of Eq. 20 as predicted by Eq. 29 on the dataset of table 5. The error shown is given in ppm. The error corresponds to the difference between the predicted chemical shift of the formula by Barfield and the QM calculated chemical shifts calculated for seven dimers of different backbone conformations.

# 5  Ring Current

In the Parker, Houk and Jensen [50] model of amide proton chemical shifts, the chemical shift difference caused by ring current effects ($\Delta\delta_{RC}$), is not described formally and an estimate of the term is taken from the equivalent term given by the SHIFTS [49] web interface[1]. The ring current effect stems from the fact, that in a cyclic aromatic moiety, the conjugated $\pi$-orbitals will form a ring, through which electrons density will flow and in effect induce a magnetic field as if it was a molecular electromagnetic coil. For an amide proton in a protein this magnetic effect usually has an effect of about -1.2 to +0.3 ppm [48].

This section compares three different approximations to the ring current chemical shift effect and by quantum chemical methods obtains an accurate estimate of the necessary parameters in order to find the most accurate and computationally efficient model.

## 5.1  Ring Current Formalisms

The ring current effect has previously been described mainly by the three following formalisms [23]: The Johnson-Bovey formalism which is based on classical electrostatics [18], the Hückel-based Haigh-Mallion theory [23] and the simple point-dipole model by Pople [54]. In general these methods describe the change in chemical shift due to a nearby aromatic ring, formally as:

$$\Delta\delta_{RC} = i \, B \, G, \tag{30}$$

where $G$ is a geometric factor, depending on the spatial oritentation and distance of the ring relative to the proton. $i$ is the ring current intensity relative to benzene, and $B$ is the ring current intensity for a benzene ring. So in principle $i_{benzene} \equiv 1$. The next three subsections contain descriptions of the three formalisms.

### 5.1.1  The classical point-dipole model by Pople

In the simplest approximation, the magnetic field caused by an aromatic ring is equvalent to a magnetic point dipole perpendicular to the molecular plane and centered in the ring center. Pople [54] [44] gives the geometric dependence of the chemical shift due to the presence of an aromatic ring as:

$$G_{PD}(\vec{r}, \theta) = \frac{1 - 3\cos^2\theta}{|\vec{r}|^3}, \tag{31}$$

where $\theta$ is the angle between the normal to the plane of the aromatic ring passing through its center and the vector connecting the center of the ring and the proton and $\vec{r}$ is the vector from the proton to the ring center. Despite the simplicity of the point-dipole model, it has not gained much popularity due to the existence of the seemingly more advanced Johnson-Bovey and Haigh-Maillon methods. [44] The only reported problem, however, with the point-dipole model is the underestimation of the ring current effect from macrocyclic rings, such as those found in porphyrins [52]. Since macrocyclic ring are not found in the proteins encountered in this work, this effect can safely be disregarded.

---

[1] http://casegroup.rutgers.edu/qshifts/qshifts.htm

### 5.1.2 The Johnson-Bovey model

This model, also sometimes referred to as the Waugh-Fessenden-Johnson-Bovey model, is an classical expansion of the point-dipole model. In this model, the electrons are appoximated as a current in two closed circular line loops, and the resulting magnetic field is the magnetic field of two such loops. In litterature, the loops are usually placed 0.64 Å above and below ring aromatic ring plane, and the loop radius is taken as the center-to-atom distance. The loop radii are here set to 1.39 Å and 1.182 Åconsistent with references [10] and [23]. A further generalization of the Waugh-Fessenden-Johnson-Bovey exist, which is the Farnum-Wilcox double-toroidal-shell model [19]. In this model, the current loop no longer consists of circular lines, but on series of toroidal surface shells with common centers, but different width. In that sense the Johnson-Bovey theory is the special case of the Farnum-Wilcox model, with just one shell with infinitely small width. However, the Farnum-Wilcox model is found to give worse prediction than the simpler Johnson-Bovey, and at least 24 shells are needed to get satisfactory results [23]. For this reason, the Farnum-Wilcox model is not considered in this work. The total ring current contribution from one loop is in the Johnson-Bovey model given with the geometric dependence:

$$G_{JB}(\rho', z') = \frac{1}{\sqrt{(1+\rho')^2 + (z')^2}} \left[ K(k') + \frac{1 - (\rho')^2 - (z')^2}{(1-\rho')^2 + (z')^2} E(k') \right], \qquad (32)$$

in which

$$k' = \sqrt{\frac{4\rho'}{(1+\rho')^2 + (z')^2}} \qquad (33)$$

and $\rho'$ and $z'$ are the radial and vertical cylindrical coordinates of the probe atom relative to the loop center and loop plane, respectively and defined in units of the loop radius, $a$. $K$ and $E$ are the complete elliptical integrals of the first and second kind, respectively. We apply the following expansions for the elliptical integrals of the first and second kind, respectively:

$$
\begin{aligned}
K(k') &= \int_0^{\frac{\pi}{2}} \frac{1}{\sqrt{1 - (k')^2 \sin^2 \theta}} \, d\theta \\
&= \frac{\pi}{2} \left[ 1 + \left(\frac{1}{2}\right)^2 (k')^2 + \left(\frac{1 \cdot 3}{2 \cdot 4}\right)^2 (k')^4 + \left(\frac{1 \cdot 3 \cdot 5}{2 \cdot 4 \cdot 6}\right)^2 (k')^6 + \ldots \right] \quad (34)
\end{aligned}
$$

$$
\begin{aligned}
E(k') &= \int_0^{\frac{\pi}{2}} \sqrt{1 - (k')^2 \sin^2 \theta} \, d\theta \\
&= \frac{\pi}{2} \left[ 1 - \left(\frac{1}{2}\right)^2 (k')^2 - \left(\frac{1 \cdot 3}{2 \cdot 4}\right)^2 \frac{(k')^4}{3} - \left(\frac{1 \cdot 3 \cdot 5}{2 \cdot 4 \cdot 6}\right)^2 \frac{(k')^6}{5} - \ldots \right] \quad (35)
\end{aligned}
$$

In the following sections this expansion is evaluated to the 20th order, which is accurate beyond the 10th decimal.

Johnson and Bovey also give an explicit expression of the $B$-factor as:

$$
B_{\text{JB}} = -\frac{\mu_0}{4\pi} \frac{ne^2}{6\pi m_e a} \quad (36)
$$

where $\mu_0$ is the permeability of free space, $n$ is the number of electrons in the loop (i.e. half the number of $\pi$-electrons in the ring), $m_e$ is the electron mass and $a$ is the radius of the loop.

It has been noted [23] [37] [44], that using the geometric intensity factor of the Johnson-Bovey model, $B_{\text{JB}}$, does not give an accurate estimation for $\Delta\delta_{\text{RC}}$, and differ by a scaling factor [37]. For this reason and for comparability to the data obtained by Case [10], we use the analytical value and scale the resulting differences into the relative intensity $i$-factors. The number of electrons in one of the two current loops is 3 in 5- and 6-membered aromatic rings, and Eq. 36 evaluates for 6-membered rings to -3.25 ppm and for 5-membered rings to -3.79 ppm.

### 5.1.3 The Haigh-Mallion model

The third model to be presented is the Hückel-based model by Haigh and Mallion [23]. This is today the most commonly used method of estimating ring current effects and the Haigh-Mallion model is employed in both the `SHIFTS` and the `SHIFTX` chemical shift models. In this case, the geometric term is given as:

$$
G_{\text{HM}} \left( \vec{r_i}, \vec{r_j} \right) = \sum_{ij} S_{ij} o_{ij} \left( \frac{1}{|\vec{r_i}|^3} + \frac{1}{|\vec{r_j}|^3} \right) \quad (37)
$$

where $\vec{r_n}$ is the distance from the $n$-th atom to the proton. Note that the summation is done over pairs of atom involved in mutual bonding. The sum is over the pairs is thus $ij \in \{12, 23, 34, 45, 56, 61\}$ in the case of a six membered ring. The numeration is done recursively around the ring. $o_{ij}$ is the area of the triangle formed by the $i$th atom, $j$th atom and projection of the the proton onto the ring plane, $\vec{o}$. Let vectors from $\vec{o}$ to the $i$th and $j$th atoms be $\vec{r_i}$ and $\vec{r_j}$ respectively. A normal vector to the plane, $\vec{n}$, is then defined as:

$$\vec{n} = (\vec{r_2} - \vec{r_1}) \times (\vec{r}_{\text{Last}} - \vec{r_1}) \tag{38}$$

where $\vec{r_n}$ is the vector of the $n$th atom of the ring. $S_{ij}$ is a term describing the sign of the triangular area, defined as:

$$S_{ij} = \begin{cases} 1 & \text{if} \quad \vec{r_i} \times \vec{r_j} \text{ is parallel to } \vec{n} \\ -1 & \text{if} \quad \vec{r_i} \times \vec{r_j} \text{ is antiparallel to } \vec{n} \end{cases} \tag{39}$$

## 5.2 QM modelling of the ring current effect

A probe molecule is used, for which the isotropic shielding is calculated using quantum chemical methods. The probe was chosen to be the *cis* amide hydrogen of a *trans-N*-methylacetylamide (NMA) (Fig.12a) in order to provide a small and relatively inflexible molecule with high degree of resemblance to the amide proton and the chemistry of an amide group of protein backbone. For more computationally demanding calculations, such as calculations using HF and MP2 with large basis sets as well as CCSD and CCSD(T) calculations, the two methyl groups are removed in order save considerable amounts of computational time, and the probe molecule used is then formamide (FMA) (Fig.12b). See the section on computational methodology for an detailed description of the exact methods used in the calculations.

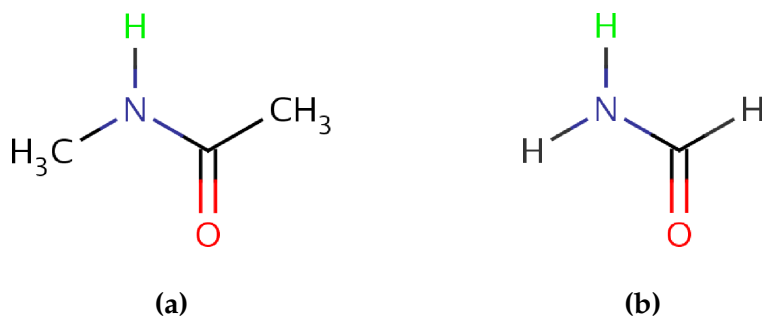

**(a)**                  **(b)**

**Figure 12:** The molecules used as probes for the ring current effects on amide protons. *N*-methylacetamid (a) and formamide (b). The probe hydrogen used through out this section is the *cis-N*-proton, which is marked in green color.

A large number of dimer systems, consisting of an amide probe molecule and an aromatic ring in different conformations is constructed (see the methodology section). This way, it is possible to calculate the chemical shift perturbation on the amide proton due to the aromatic ring. Following the general approach of Boyd and Skrynnikov [9], we thus write the absolute shielding of the probe hydrogen as:

$$\sigma_H^{\text{Dimer}} = \sigma_{\text{Conformation}} + \sigma_{\text{Local}} + \sigma_{\text{RC}} \qquad (40)$$

The chemical shift of the probe hydrogen atom in the dimer system will apart from ring current effect also be influenced by the conformation of the probe (described in the $\sigma_{\text{Conformation}}$-term) as well as any possible interactions with the aromatic moiety, such as electrostatic forces, possible hydrogen bonding, spin-spin repulsion and other effects which can be difficult to quantify and separate (described in the $\sigma_{\text{Local}}$-term). Finally, the chemical shift perturbation due to the aromatic ring is approximiated as $\sigma_{\text{RC}}$.

Reference systems for each dimer system which have approximately identical local interactions between the molecules, only excluding the ring current effect, are constructed in order to filter these hard-to-quantify effects. These are modeled as corresponding dimer system, where the aromatic ring has been replaced by an olefinic analogue. The definition of an olefinic analogue here is an aromatic ring, which by addition of two hydrogen atoms has lost its aromaticity. The protonation is done such that the planar geometry of the ring is still enforced, and the overall carbon/nitrogen chain structure is highly comparable. The olefinic analogue is placed, such that the ring center corresponds to the center of the aromatic ring, relative to the probe hydrogen and the coordinates of the carbon/nitrogen atoms are matched as closely as possible. This approach ensures that $\sigma_{\text{Conformation}}$ and $\sigma_{\text{Local}}$ are largely retained, while $\sigma_{\text{RC}}$ is filtered out. See Fig. 13 for an example of a dimer set. Using this substitution scheme, we model the absolute shielding of the hydrogen atom in the reference system as:

$$\sigma_H^{\text{Ref}} = \sigma_{\text{Conformation}} + \sigma_{\text{Local}}, \qquad (41)$$

thus enabling us to estimate the ring current contribution to the chemical shift due to the aromatic ring as:

$$\Delta\delta_{\text{RC}} = -\sigma_{\text{RC}} \approx \sigma_H^{\text{Ref}} - \sigma_H^{\text{Dimer}} \qquad (42)$$

Due to the large set of different dimer conformations (roughly 150-250 dimers per type of side chain - see the section on methodology on how these are constructed), we assume that the small errors and uncertainties caused by the approximation in Eq. 42 show up as Gaussian errors, which cancel out upon fitting the intensity parameters. The predictive power of the three ring current models thus lies in the linear correlation coefficient relative to the QM data.

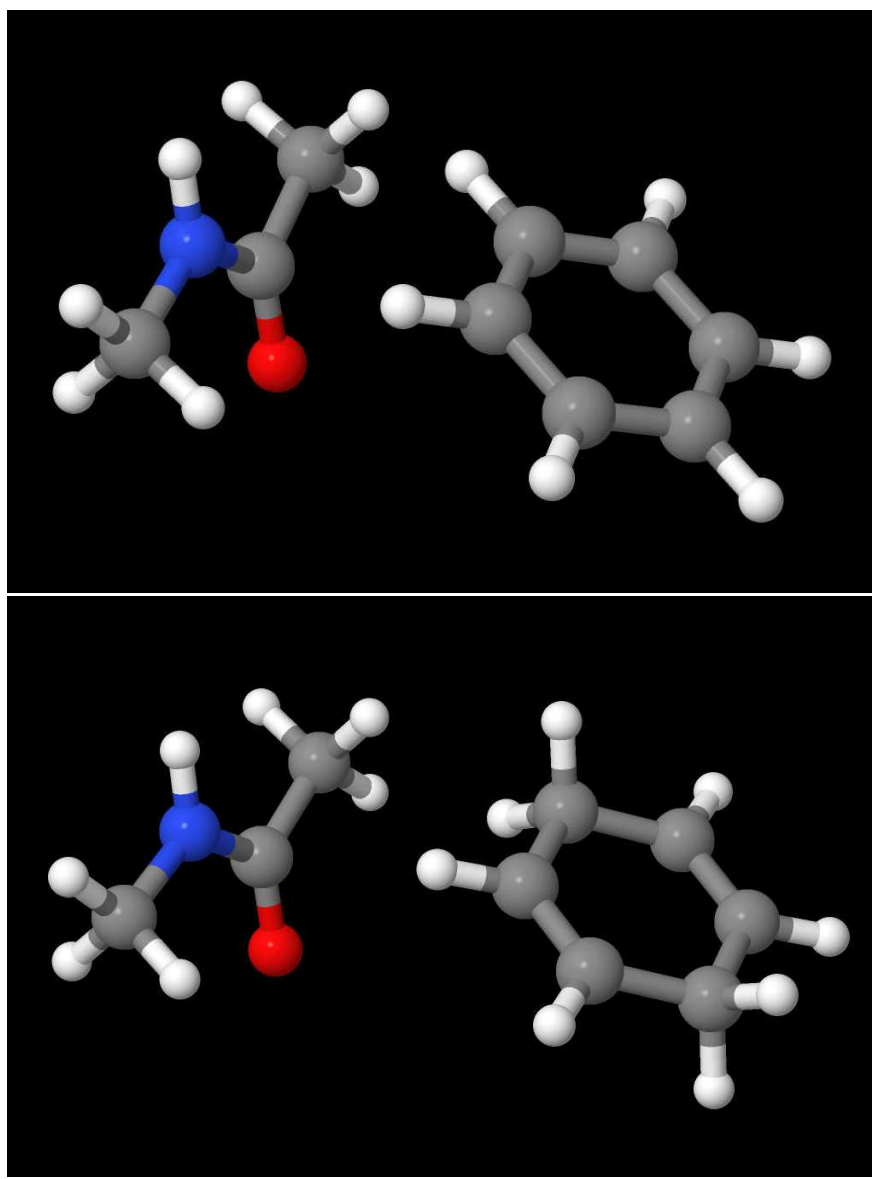

**Figure 13:** One set of dimers, with the top being the aromatic dimer, and the bottom being the corresponding reference dimer with the olefinic analogue.

### 5.2.1 Approximate side chains and olefinic analogues

In the dimer system, we have two different approximations for each aromatic side chain. The aromatic approximations of the side chains are cut off, so $C^\beta$ of the side chain is replaced by a hydrogen atom. The olefinic analogues to the aromatic side chains are constructed from their aromatic counterparts, but have two extra hydrogens added, in order to break the conjugated ring. The hydrogens are added such that the planar geometry is retained and the geometry of the heavier atoms is highly conserved. See table 7 for an overview as well as Fig. 35-39 in the Appendix E for sketches of the molecules used.

| Side chain | Analogue | Olefinc analogue |
|---|---|---|
| Phenylalanine | Benzene | 1,4-cyclohexadiene |
| Tyrosine | Phenol | cyclohexa-1,4-diene-1-ol |
| Tryptophan | Indole | 2,3,5,6-tetrahydroindole |
| Histidine | Imidazole | 4,5-dihydroimidazole |
| Histidine$^+$ | Imidazolium | 4,5-dihydroimidazolium |

**Table 7:** List of the side chain approximations used in this work and their olefinic analogues.

### 5.2.2 Basis set extrapolations at correlated levels of theory

In this work, DFT (and the very popular B3LYP functional) is largey used to obtain NMR shielding constants. Due to the slightly semi-empiric nature of the approximated exchange-correlation functionals used, B3LYP data can not in general be expected to show convergence towards experimental NMR values or values obtained at very accurate levels of theory when increasing the basis set size [45]. It is often the case, that a small error compared to correlated methods of high order can be obtained with DFT, if a simple linear correction or scaling factor is applied to DFT data. [32] In this work, a comparison of B3LYP to higher order correlated methods is used to obtain such a linear scaling factor.

In order to obtain very accurate shielding constants a series of calculations is done at the CCSD(T) level of theory with very large basis sets, in order to give an estimate of the CCSD(T) shielding at the complete basis set limit. However, since the dimers used in this work are too large for CCSD(T) calculations (70 electrons) with meaningful basis sets, we also compare CCSD(T) calculations to the cheaper methods, CCSD, MP2 and HF.

In the estimation of the complete basis set limit (CBS) for a method, we follow the approach of Moon and Case [45] and Kupka *et al.* [32] [33]. By utilizing Dunning's correlation consistent basis sets (cc-pV$x$Z; where $x \in \{D, T, Q, 5, 6\}$ is the valence orbital splitting in the basis set), we have a sequence of basis sets of a defined increasing quality. Kupka *et al.* describe a method to extrapolate calculated NMR shielding values towards the case of infinite basis set size. The shielding or chemical shift values at a certain basis set size $x$, where $x$ denotes the valence orbital splitting in the basis set, a monotone, asymptotic function is fitted. Kupka *et al.* suggest the use of a three parameter exponential decreasing function (Eqn. 43).

$$\sigma(x) = \sigma^{III}(\infty) + A^{III}\exp(-x/B^{III}) \tag{43}$$

where $\sigma^{III}(\infty)$, $A^{III}$ and $B^{III}$ are the fitting parameters, with $\sigma^{III}(\infty)$ being the estimated shielding at the complete basis.

Frank Jensen has constructed a set of basis sets for the purpose of DFT NMR shielding calculations with, the polarization consistent pcS basis sets. The pcS basis sets have been shown to faster basis set convergence than the Dunning-type basis sets for DFT shielding calculations. pcS-1 is of double zeta quality basis set, pcS-2 is triple zeta and so forth. For basis sets of similar valence orbital splitting, the pcS basis set contains more basis functions of low angular momentum, compared to the Dunning-type basis sets. For the shielding calculated at the pcS-$n$ level of theory, a value of $x = n + 1$ is used in Eq. 43.

A non-linear least-squares Marquardt-Levenberg algorithm [38] [35] is used to fit the parameters.

It is necessary to benchmark the quality of different sized basis set, at various levels of theory in order to give good estimates of shielding constants by extrapolating methods. Here, this is done by comparing the calculated absolute shielding value of the single FMA molecule to experimental values and high levels of theory. Inferring the experimental gas-phase [1]H shielding values from $CH_4$ ($\sigma_H = 30.61$ ppm [26]), an experimental value of $\sigma_H = 26.24$ ppm in the gas phase [65] at 483 K is obtained. At this temperature, thermal motion cause rapid switching of the *trans* and *cis* amide protons and the peaks are not separable, so this value has to be considered as an average over the two proton chemical shifts, so the value of $\sigma_H = 26.24$ ppm can only serve as a ballpark figure. It is noted by Ruud *et al.* [59], that a zero-point vibrational correction (ZPVC) has to be added to *ab initio* proton shielding constants in order to obtain close agreement to experimental data. The ZPVC is found to be in the range of -0.18 to -0.70 ppm for all hydrogens in the study (if alcohol hydrogens are disregarded). While the amide functional group is not specifically studied in the work by Ruud *et al.*, the functional groups for which the proton ZPVC was studied which are chemically most similar to the amide group are arguably the =CRH ene group and the RN-$H_2$ amine group for which the ZPVC was found to be $-0.46 \pm 0.13$ ppm and $-0.18 \pm 0.03$ ppm, respectively. A ZVPC thus suggests that the uncorrected amide proton shielding constant calculated at a high level of theory is expected to be overestimated by 0.18 to 0.70 ppm compared to the experimental value.

### 5.3 Results

#### 5.3.1 Basis set convergence

Since the number of dimers is large, it is not possible to do calculations at the correlated level of theory on every dimer. DFT calculations are cheap in this context, even with relatively good basis sets. In this subsection a complete basis set limit extrapolation of the amide proton isotropic shielding at the CCSD(T) level of theory is carried out in order to evaluate the quality of CCSD, MP2 and HF chemical shifts using smaller basis sets. An FMA molecule is minimzed at the B3LYP/aug-cc-pVTZ level of theory and an NMR shielding calculation is carried out. Given the computational resources, it was possible to carry out NMR calculations on FMA at the CCSD(T), CCSD and MP2 levels of theory using basis set no larger than cc-pV5Z. At the HF level of theory it was possible to carry out a calculation using the cc-pV6Z basis set. See Table 8 for an overview of the obtained shielding constants.

The amide proton isotropic shielding at the MP2 level of theory converges to a value 0.14 ppm from CCSD(T)/CBS, which is a very good benchmark for shielding constants. Since the calculation of chemical shift rely on the subtraction of a reference, a significant amount of error cancellation is expected, and the final error is expected to be much smaller. The convergence of the CCSD shielding towards the CCSD(T)/CBS shielding is very good, with CCSD/CBS being only 0.09 ppm off. The Hartree-Fock shielding converge to a shielding value 0.25 ppm higher than the CCSD(T)/CBS.

Due to the size of the benzene/FMA dimer system used in the next section (a total of 70 electrons), it was only possible to carry out CCSD and CCSD(T) NMR calculations on a dimer using much smaller Pople basis sets. The largest Pople basis set used was 6-311++G(d,p), but a smaller basis set, 6-31+G(d), was also used. These, however are far from converged size and thus do not serve as good measures of the isotropic shielding, not even at the CCSD or CCSD(T) level of theory. The highest level of theory in which it was possible to carry out a dimer NMR calculation was MP2/cc-pVQZ. At this particular level of theory, the shielding is 0.16 ppm off compared to CCSD(T)/CBS.

Seemingly all the MP2, CCSD and CCSD(T) methods show very good convergence towards an amide proton shielding constant close to the experimental gas phase value. All three methods, however, converge towards a somewhat larger value. As mentioned, this can be attributed to the vibrational averaging of the experimental value, which must be calibrated by approximately +0.18 to +0.46 ppm in order to obtain an empirical equilibrium value corresponding to the calculated shielding constant. A factor not investigated was the dependence of the used geometry, which is known to cause deviations in calculated $^1$H shielding constants of around $\pm$ 0.1 ppm between MP2 and B3LYP geometries when larger basis sets are used - see supplementary material of Ref. [56]. This uncertainty is thus expected to be smaller than the ZVPC and is left out.

In conclusion, the MP2/cc-pVQZ level of theory serves as an accurate level of theory for the calculation of chemical shifts and shows very good agreement with CCSD(T)/CBS extrapolated values.

### 5.3.2 Fitting B3LYP data to correlated levels of theory.

The previous section compared the quality of various levels of theory for which we can calculate the chemical shift contribution from an aromatic ring. In this section we use an appropriate level of theory to obtain a linear scaling correction to chemical shift contribution due to ring current effects, obtained at the B3LYP/6-311++G(d,p)//B3LYP/aug-cc-pVTZ level of theory.

Four dimer systems consisting of an FMA molecule and a benzene ring were selected from the large data set of benzene/FMA dimers, in such a way that the ring current contribution varied over a range of -0.72 ppm to +0.15 ppm, at the B3LYP/6-311++G(d,p) level of theory, in even sized steps. For these four dimer systems, the isotropic shielding was calculated using various methods and basis sets. Here the chemical shift is modeled by

$$\Delta \delta_{\text{RC}}^{(\text{uncorrected})} = \sigma_{\text{H}}^{\text{Probe}} - \sigma_{\text{H}}^{\text{Dimer}} \tag{44}$$

where $\sigma_{\text{H}}^{\text{Probe}}$ is the shielding of the probe atom in the probe molecule in vacuum and $\sigma_{\text{H}}^{\text{Dimer}}$ is the shielding of the probe atom in the probe molecule in the dimer. Note that an NMR calculation for a reference dimer is not carried out. In section 5.3.3 this is shown to give a negligible small error for benzene. The linear scaling factor is unaffected, whether a reference calculation is carried out, since this would also have to be scaled. If the B3LYP chemical shifts and the chemical shifts obtained at the correlated level of theory differ not only by a scaling factor but also by a non-negligible constant offset, this will show up in these fits too.

| Basis set | | Method | | | | |
|---|---|---|---|---|---|---|
| | Size | CCSD(T) | CCSD | MP2 | HF | B3LYP |
| 6-31+G(d) | 60 | 28.23 | 28.30 | 28.07 | 25.52 | 28.19 |
| 6-311++G(d,p) | 87 | 27.85 | 27.90 | 27.70 | 27.76 | 27.64 |
| cc-pVDZ | 57 | 28.06 | 28.09 | 27.90 | 27.89 | 27.67 |
| cc-pVTZ | 132 | 27.29 | 27.35 | 27.16 | 27.34 | 27.17 |
| cc-pVQZ | 255 | 26.92 | 27.00 | 26.80 | 27.13 | 26.94 |
| cc-pV5Z | 438 | 26.78 | 26.86 | 26.65 | 27.04 | 26.83 |
| cc-pV6Z | 693 | - | - | - | 27.00 | 26.78 |
| $\sigma^{III}_{cc\text{-}pVxZ}(\infty)(Q)$ | | 25.58 | 26.68 | 26.45 | 27.00 | 26.73 |
| $\sigma^{III}_{cc\text{-}pVxZ}(\infty)(5)$ | | 26.64 | 26.73 | 26.50 | 26.99 | 26.73 |
| $\sigma^{III}_{cc\text{-}pVxZ}(\infty)(6)$ | | - | - | - | 26.99 | 26.73 |
| pcS-0 | 44 | 29.32 | 29.36 | 29.31 | 29.37 | 28.88 |
| pcS-1 | 66 | 27.55 | 27.58 | 27.40 | 27.38 | 27.29 |
| pcS-2 | 141 | 27.02 | 27.09 | 26.89 | 27.10 | 26.91 |
| pcS-3 | 321 | 26.75 | 26.83 | 26.62 | 26.99 | 26.77 |
| pcS-4 | 543 | - | - | 26.31 | 26.99 | 26.76 |
| $\sigma^{III}_{pcS\text{-}n}(\infty)$ (T) | | 26.79 | 26.90 | 26.70 | 27.05 | 26.80 |
| $\sigma^{III}_{pcS\text{-}n}(\infty)$ (Q) | | 26.67 | 26.78 | 26.57 | 27.00 | 26.75 |
| $\sigma^{III}_{pcS\text{-}n}(\infty)$ (5) | | - | - | 26.36 | 26.99 | 26.75 |
| aug-pcS-0 | 48 | - | - | - | - | 28.62 |
| aug-pcS-1 | 105 | - | - | - | - | 26.93 |
| aug-pcS-2 | 216 | - | - | - | - | 26.79 |
| aug-pcS-3 | 444 | - | - | - | - | 26.74 |
| aug-pcS-4 | 726 | - | - | - | - | 26.79 |
| $\sigma^{III}_{aug\text{-}pcS\text{-}n}(\infty)$ (T) | | - | - | - | - | 26.77 |
| $\sigma^{III}_{aug\text{-}pcS\text{-}n}(\infty)$ (Q) | | - | - | - | - | 26.75 |
| $\sigma^{III}_{aug\text{-}pcS\text{-}n}(\infty)$ (5) | | - | - | - | - | 26.76 |
| $\sigma_{Expt'l\ (gas)}$ | | 26.24 | | | | |

**Table 8:** The absolute isotropic chemical shielding of the *cis* amide proton in gas-phase formamide at the HF, MP2, CCSD and CCSD(T) level of theory using Dunning's correlation consistent basis sets. All values are given as ppm. The experimental value is obtained at 483 K [65]. $\sigma^{III}(\infty)$ is obtained using Eq. 43 and fitted over all values in the above column. The size indicates the number of basis functions in the system at the given basis set size. All shielding constants are given in ppm.

The effect of increased inclusion of correlation energy was investigated, by using 6-311++G(d,p) and the series of methods HF, MP2, CCSD and CCSD(T). Unfortunately, due to the size of the system (70 electrons), it was not possible to carry out calculations at the coupled cluster level of theory with larger basis sets than 6-311++G(d,p). An overview is found in table 9.

Regardless of basis set or method the linear correlation to B3LYP/6-311++G(d,p) was 0.992 or better. It is thus demonstrated, that applying a linear correction based on correlated methods is a very good approximation. No constant offset (intercept) greater that 0.01 ppm was found, so the linear correction can be reduced to applying a simple scaling factor. Thus, the fit was carried out as:

$$\delta_{\text{Other}}^{\text{QM}} = k_{\text{scaling}} \cdot \delta_{\text{B3LYP}}^{\text{QM}}, \tag{45}$$

where $\delta_{\text{B3LYP}}^{\text{QM}}$ is the chemical shift obtained at the B3LYP/6-311++G(d,p) level of theory, $\delta_{\text{Other}}^{\text{QM}}$ is the chemical shift obtained using another method and $k_{\text{scalinf}}$ is the fitted scaling constant.

It is evident that increasing the basis set size decreases the scaling factor relative to B3LYP/6-311++G(d,p). This is seen both for MP2 and HF calculations using series of Dunning's correlation consistent basis sets.

Increased inclusion of correlation energy was found to lower the scaling factor using a series of HF, MP2 CCSD and CCSD(T) methods and the same basis set. However, since the relatively small 6-311++G(d,p) basis set was used, no definite conclusion can be made based on this. The polarization consisten basis sets by Frank Jensen were used at the B3LYP level of theory. This lead to convergence of the scaling factor at a value of around 1.095. Attempts to use the augmented polarization consistent basis sets by Frank Jensen were carried out, but this lead to convergence problems and convergence was only obtained for two dimer calculations at the B3LYP/aug-pc-2 level of theory.

MP2 calculations using a series of Dunning's correlation consistent basis sets were carried out. In these calculations, however, the scaling factor diverged from 1.033 using cc-pVDZ up to 1.076 using cc-pVTZ, but down to 1.074 using cc-pVQZ. Since the series was not convergent, the scaling factor to be used was chosen as factor found at the highest level of theory, namely 1.074 at the MP2/cc-pVQZ level of theory. In the rest of the chapter on ring current effects, all calculated chemical shifts are obtained at the B3LYP/6-311++G(d,p) level of theory and scaled by this factor of 1.074.

| Method | $\sigma_{\mathrm{FMA}}$ | $\Delta\delta_{\mathrm{RC}}^1$ | $\Delta\delta_{\mathrm{RC}}^2$ | $\Delta\delta_{\mathrm{RC}}^3$ | $\Delta\delta_{\mathrm{RC}}^4$ | Scaling |
|---|---|---|---|---|---|---|
| B3LYP/6-311++G(d,p) | 27.64 | -0.72 | -0.43 | -0.21 | 0.15 | - |
| HF/6-311++G(d,p) | 27.76 | -0.80 | -0.46 | -0.23 | 0.16 | 1.103 |
| MP2/6-311++G(d,p) | 27.70 | -0.76 | -0.45 | -0.22 | 0.15 | 1.052 |
| CCSD/6-311++G(d,p) | 27.90 | -0.75 | -0.44 | -0.22 | 0.15 | 1.033 |
| CCSD(T)/6-311++G(d,p) | 27.85 | -0.73 | -0.43 | -0.21 | 0.15 | 1.012 |
| MP2/cc-pVDZ | 27.90 | -0.74 | -0.43 | -0.22 | 0.17 | 1.033 |
| MP2/cc-pVTZ | 27.16 | -0.77 | -0.46 | -0.23 | 0.17 | 1.076 |
| MP2/cc-pVQZ | 26.80 | -0.81 | -0.45 | -0.22 | 0.13 | 1.074 |
| HF/cc-pVDZ | 27.89 | -0.77 | -0.44 | -0.22 | 0.17 | 1.071 |
| HF/cc-pVTZ | 27.34 | -0.81 | -0.47 | -0.24 | 0.17 | 1.117 |
| HF/cc-pVQZ | 27.13 | -0.82 | -0.48 | -0.24 | 0.17 | 1.135 |
| HF/cc-pV5Z | 27.04 | -0.84 | -0.48 | -0.24 | 0.17 | 1.151 |
| B3LYP/pcS-0 | 28.88 | -0.75 | -0.42 | -0.21 | 0.16 | 1.042 |
| B3LYP/pcS-1 | 27.29 | -0.76 | -0.44 | -0.22 | 0.16 | 1.056 |
| B3LYP/pcS-2 | 26.91 | -0.80 | -0.47 | -0.24 | 0.14 | 1.087 |
| B3LYP/pcS-3 | 26.77 | -0.80 | -0.47 | -0.24 | 0.16 | 1.097 |
| B3LYP/pcS-4 | 26.76 | -0.80 | -0.47 | -0.25 | 0.16 | 1.095 |
| B3LYP/aug-pcS-0 | 28.62 | -0.71 | -0.44 | -0.23 | 0.22 | 1.069 |
| B3LYP/aug-pcS-1 | 26.93 | -0.72 | -0.43 | -0.22 | 0.17 | 1.012 |
| B3LYP/aug-pcS-2 | 26.74 | -0.85 | -0.51 | - | - | - |
| CCSD(T)/CBS | 26.64 | | | | | |

**Table 9:** For each method and basis set used in this section, the shielding constant of the FMA probe proton in vacuum is shown, as well as the chemical shift ring current interaction ($\Delta\delta_{\mathrm{RC}}^p$) for each of the four different conformations used. Further more the resulting scaling factor relative to data obtained at the B3LYP/6-311++G(d,p) level of theory is noted. The gas phase value of the proton chemical shift of FMA is noted for reference. All shielding constants and ring current chemical shift contributions are given in ppm.

### 5.3.3 The performance of Eq. 42

Here we show qualitatively, that Eq. 42 is efficient in removing noise from various intermolecular interaction in systems, where $\sigma_{Local}$ (see Eqn. 40-42) is significant. For the dimer systems of each amino acid, the linear correlation coefficient is calculated between quantum chemical data obtained via Eqn. 42 and the prediction by the three approximate model. When using the approach of Eq. 42, it is evident, that all the three approximate models are highly comparable in quality and correlate very well with QM data after removing noise. For reference, the modeling of the chemical shift contribution of a aromatic ring is also calculated without the use of a reference dimer, i.e. Eq. 44. In Fig. 14 a plot of the data obtained for the Point-dipole approximation is shown. Eq. 40 (Fig. 14b) shows a slightly better linear behavior than Eq. 44 (Fig. 14a). The use of Eq. 44 (Fig. 14a) does not ensure, that conformations with the dimer close to each other, will be free of other interactions, than the longer range ring current effect. It can be seen that dimers with a low amide proton chemical shift due to dimers being close deviate from the linear correlation of the rest of the data set. By the subtraction of the amide proton chemical shift in the olefinic reference system, however, these effects are removed. The effect is a minor increase in the linear correlation coefficient or $r$-value(Eqn. 55) from 0.991 to 0.993 for the point-dipole model. A schematic overview of the improvements can be found in table 10 for each ring type and corresponding geometric function, $G$, in the three ring current models.

|  | Point-Dipol | | Johnson-Bovey | | Haigh-Maillon | |
| --- | --- | --- | --- | --- | --- | --- |
| Side chain | Eq. 42 | Eq. 44 | Eq. 42 | Eq. 44 | Eq. 42 | Eq. 44 |
| Phenylalanine | 0.993 | 0.991 | -0.993 | -0.992 | 0.993 | 0.992 |
| Tyrosine | 0.990 | 0.957 | -0.990 | -0.957 | 0.990 | 0.958 |
| Histidine | 0.981 | 0.888 | -0.980 | -0.887 | 0.981 | 0.893 |
| Histidine$^{(+)}$ | 0.989 | 0.517 | -0.989 | -0.517 | 0.990 | 0.510 |
| Tryptophan* | 0.987 | N/A | 0.986 | N/A | 0.987 | N/A |

**Table 10:** The correlation between the geometric term for each approximate ring current interaction model and the chemical shift data obtained for each type of side chain. Note that since the $B$ and $i$ factors are not included, the Johnson-Bovey formalism differ by sign. *The data for tryptophan is obtained after fitting the realtive intensity factors for both rings in section 5.3.5 and is included for reference. The linear correlation of the data for the four other types of side chains is naturally conserved after fitting the linear scaling factors.

A clearer example of the benefits of Eq. 42, is the difficult modeling of the ring current interaction due to a charged histidine ring (approximated by an imidazolium ion). In this system strong electrostatic interactions are expected due to the positive charge, which is expected to have a large influence on the chemical shift. Also in this complicated case Eq. 42 proves extremely efficient in separating the ring current effect. In this case, we get a tremendous increase in correlation between calculated data and modeled (See Fig.14 c and d). In this case, the improvement in $r$-value is 0.52 to 0.99 for the Point-dipole model. Similar plots can be made for tyrosine and neutral histidine (see Appendix D). Neutral histidine (approximated by imidazole) and tyrosine (phenol) is expected to cause some deshielding, in cases where the lone pairs are interacting with the probe molecule, especially as weaker hydrogen bonds. Eqn. 42 proves very useful again, yielding modest improvements in $r$-value of 0.89 to 0.98 and 0.96 to 0.99 for the imidazole-NMA and phenol-NMA dimer systems, respectively. See Fig.34 in Appendix D. Note that due to the presence of two aromatic rings in tryptophan individual fitting of the relative ring current intensities for each ring is required in order to produce a similar plot. This is done in the next section. For this reason, the data for tryptophan in table 10 is obtained in section 5.3.5. See Appendix D for a scatter plot of the tryptophan data.

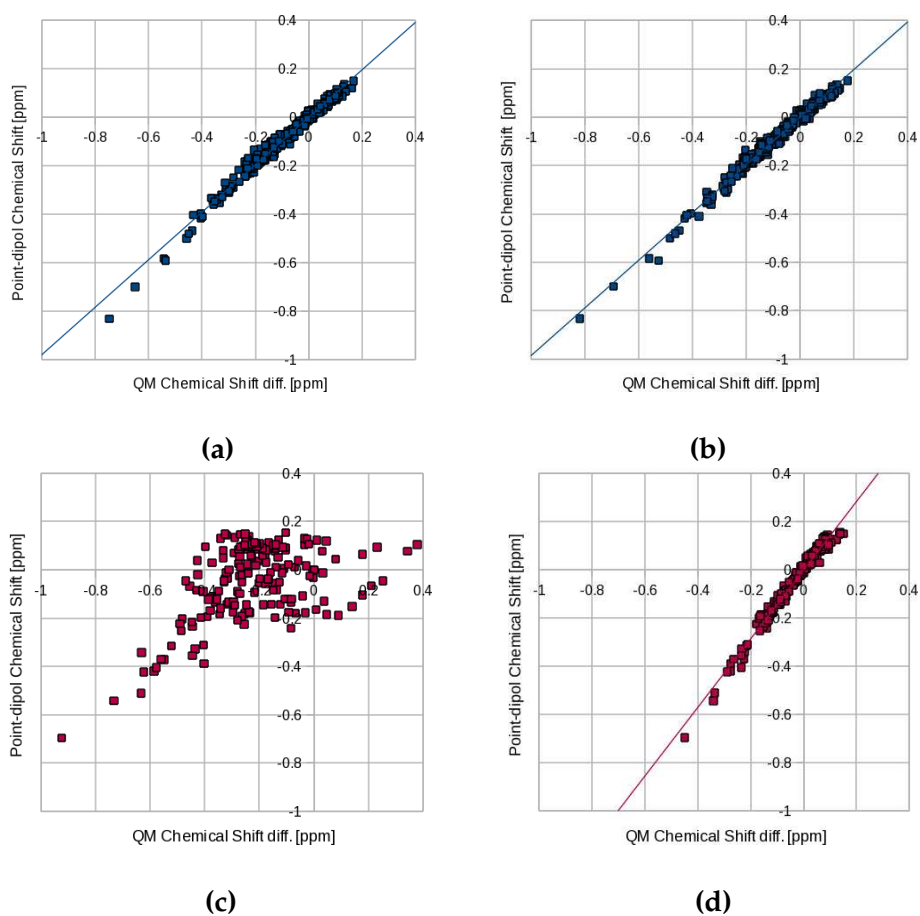

**Figure 14:** The X-axis describes the quantum chemical calculated chemical shift of a) and b) a benzene-NMA dimer, and c) and d)are a positive imidazolium-NMA dimer conformation, while the Y-axis is the chemical shift according the point-dipole ring current model by Pople, using an empirical $B$-value of 30.42 ppm Å$^{-3}$(see Eq. 31) as found in section 5.3.4. (a) and (c) The QC calculated data is derived using Eq. 44. (b) and (d) The QC calculated data is derived using Eq. 42, which increases the linearity of the data set. The relative $i$-factor for imidazolium in the Point-dipole model is found as the slope of the line. Similar plots are shown in Appendix D for the remaining three ring types.

### 5.3.4 Expressions for the $B$-factors

In the Point-dipole model we use the definition, that $i_{\text{Benzene}} = 1$, and fit the $B_{\text{PD}}$-factor using the chemical shifts obtained for all dimers to the corresponding $G_{\text{PD}}(\vec{r}, \theta)$ values using the following formula (similar to Eq. 31):

$$\Delta\delta^{\text{QM}} = B_{\text{PD}} \cdot G_{\text{PD}}(\vec{r}, \theta) \tag{46}$$

with $\Delta\delta^{\text{QM}}$ being the calculated chemical shifts of the amide protons using Eq. 42 and $G_{\text{PD}}(\vec{r}, \theta)$ being the geometric term of the benzene-FMA dimers. This gives a value of $B_{\text{PD}} = 30.42 \pm 0.16$ ppm $\text{Å}^{-3}$. The linear correlation of this fit is $r = 0.993$. See Fig. 15 for a scatter plot of the fitted data set.

In the litterature, the trend has been to use formally derived $B$-values in the Jonhson-Bovey formalism and scale the relative intensities accordingly [10]. Following this, the theoretically derived values of $B$ are used in the Jonhson-Bovey formalism in this work. The theoretical value of $B_{\text{JB}}$ (Eq. 36) evaluates to -3.79 ppm for 5-membered rings and -3.25 ppm for 6-membered rings To facilitate easy comparison to the ring current intensities to those found by Case [10], a $B_{\text{HM}}$-value of 5.455 ppm Å is adoped when fitting the relative intensities in the Haigh-Maillon model.

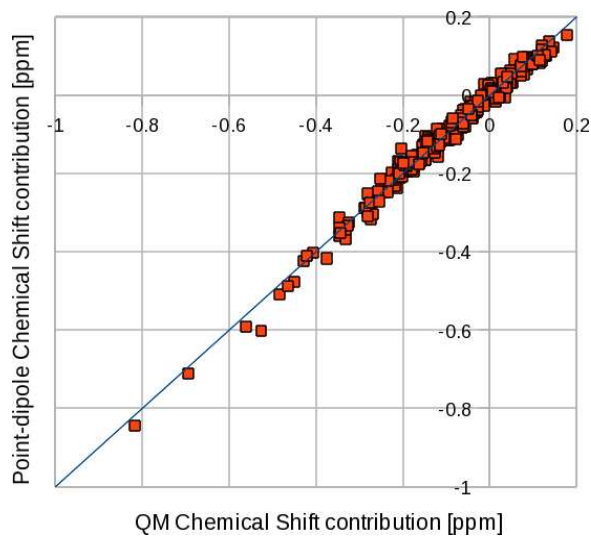

**Figure 15:** Correlation between the chemical shift predictions of the Point-dipole model and the chemical shifts obtained by Eq. 42 for a set of Benzene/NMA dimers using a best fit value of $B_{\text{PD}} = 30.42 \pm 0.16$ ppm $\text{Å}^{-3}$. The blue line represents the a best-fit between the two methods. The linear correlation between is 0.993.

### 5.3.5 Fitting he relative *i*-factors

Using the $B$-factor obtained in the previous subsection, the relative ring current intensities of all ring types in the three ring current models are obtained as best fit of $i$ when fitting Eq. 42 to Eq. 30.

The relative intensities of the two rings in tryptophan were trivially fit by using a two parameter fitting routine, although the contributions from the 5- and 6-membered ring was highly correlated. A table of the relative ring current intensites can be found in table 11, which also features a comparison to the $i$-factors found in other studies. A comparison is done to the values used in the `SHIFTX` program, `SHIFTS` program and to the values obtained by Case using DFT and measuring the ring current effect on methane hydrogen.

| Model | Point-dipole | Haigh-Mallion | | | | Johnson-Bovey | |
|---|---|---|---|---|---|---|---|
| Reference | This Work | SHIFTX | SHIFTS | Case | This Work | Case | This Work |
| PHE | 1.00 (0.02) | 1.05 (0.05) | 1.00 (0.05) | 1.46 (0.07) | 1.18 (0.03) | 1.27 (0.03) | 1.13 (0.02) |
| TYR | 0.81 (0.02) | 0.92 (0.02) | 0.84 (0.02) | 1.24 (0.06) | 0.93 (0.02) | 1.10 (0.04) | 0.91 (0.02) |
| HIS+ | 0.69 (0.02) | 0.43 (0.08) | 0.90 (0.06) | 1.35 (0.05) | 1.26 (0.03) | 1.40 (0.03) | 1.27 (0.03) |
| HIS | 0.68 (0.03) | 0.43 (0.08) | 0.90 (0.07) | 1.35 (0.06) | 1.22 (0.03) | 1.40 (0.04) | 1.25 (0.03) |
| TRP5 | 0.57 (0.03) | 0.90 (0.03) | 1.04 (0.03) | 1.32 (0.04) | 0.97 (0.02) | 1.02 (0.02) | 1.06 (0.02) |
| TRP6 | 1.02 | 1.04 | 1.02 | 1.24 | 1.18 | 1.27 | 1.18 |
| $B$-factor | 30.42 ppm Å$^{-3}$ | 5.13 ppm Å | 5.455 ppm Å | 5.455 ppm Å | 5.455 ppm Å | −3.25 ppm* −3.79 ppm* | −3.25 ppm* −3.79 ppm* |

**Table 11:** The relative ring current intensity factors of the different side chains, as found in this study, compared to the value of other studies. The RMSD associated with using the given intensity factor and $B$-value is given in paranthesis for each intensity factor. The RMSD is calculated over all dimer systems used in the fits to obtain intensity factors. The RMSD given for tryptophan, is the RMSD for a sum of both rings with the given intensities. *In the Johnson-Bovey model, values of -3.25 ppm and -3.79 ppm are used for six-and five-membered rings, respectively.

### 5.3.6   Comparison to already existing models

In this section, we compare the use of the JB, HM and PD models with the fitted parameters to already existing models. We compare our modelling to the model used in the SHIFTX program, the SHIFTS program and to the parameters in the Johnson-Bovey and Haigh-Maillon models obtained by Case. Benchmark systems is generated as polymer systems which mimic the total ring current effect of Protein G, by having aromatic rings at coordinates identical to those of the aromatic side chains of the protein. In this polymer system, NMA probes are placed with the amide group at coordinates corresponding to the coordinates of the amide group of the Protein G backbone. Again we use Eq. 42, with the reference system being the same system, but containing olefinic rings instead. An example, which is the model and reference systems corresponding to residue 13 in Protein G is shown i Fig. 16.

This allow us to calculate the linear correlation and RMSD between QM derived chemical shifts and the predictions of the ring current models and corresponding parameters used in various researches. See table 12 for an overview.

It is clear that neither of the point-dipole, Johnson-Bovey or Haigh-Mallion approximations are significantly more accurate than the other models. The difference in RMSD and linear correlation is also very invariant to the parameters. Hence the suggestion here is to use the simplest model, namely the point-dipole model, for which intensity parameters are presented here.

| Parameters | Model | RMSD [ppm] | $r$ |
|---|---|---|---|
| This work | Point-dipole | 0.12 | 0.963 |
| This work | Johnson-Bovey | 0.13 | 0.963 |
| Case (QM) | Johnson-Bovey | 0.12 | 0.962 |
| This work | Haigh-Mallion | 0.13 | 0.956 |
| Case (QM) | Haigh-Mallion | 0.11 | 0.956 |
| SHIFTX | Haigh-Mallion | 0.14 | 0.957 |
| SHIFTS | Haigh-Mallion | 0.14 | 0.957 |

**Table 12:** The RMSD and linear correlation between QM and approximated methods of calculating the total ring current effect in Protein G.

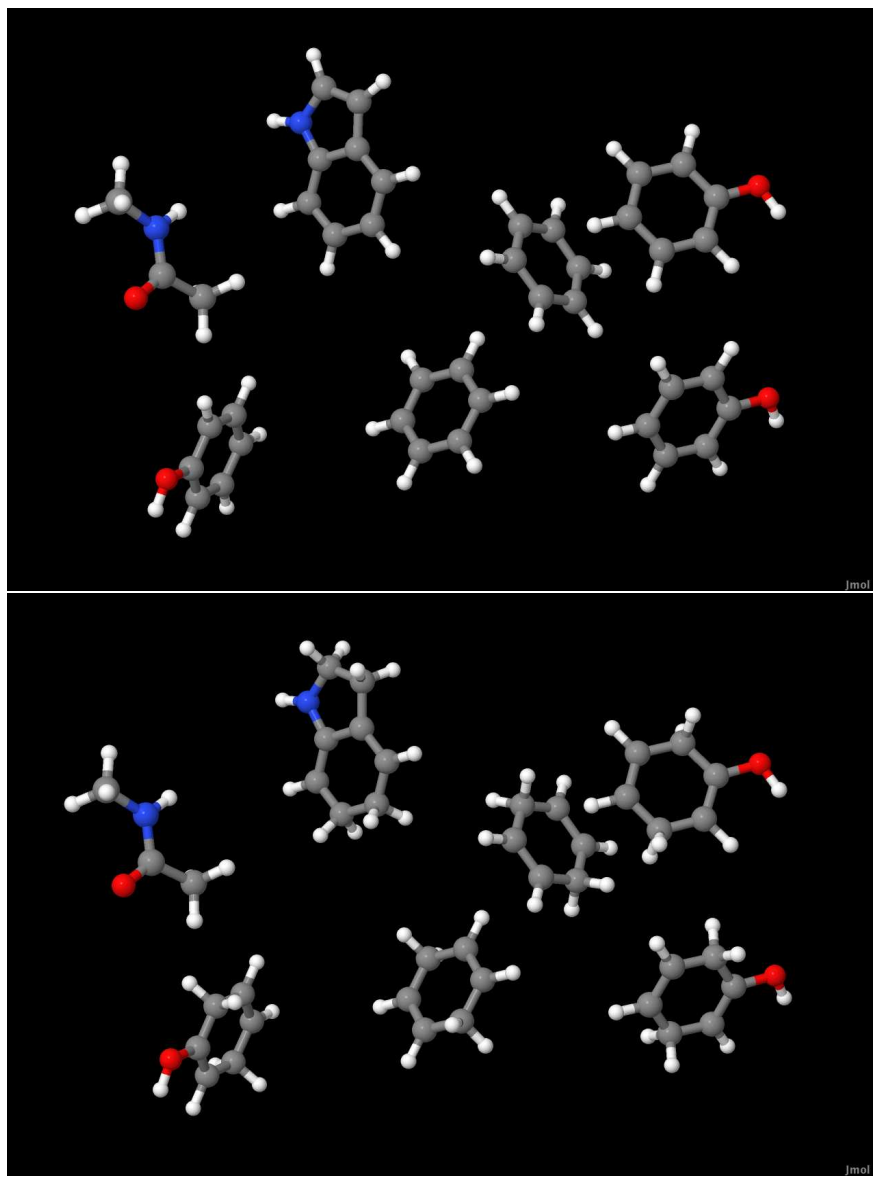

**Figure 16:** The systems corresponding to all aromatic rings and their relative positions to the amide group of residue 13 in Protein G. The top system is the system with aromatic rings, while the bottom system consists of olefinic analogues at the same coordinates as their aromatic counterparts. In both systems, the amide group of residue 13 is approximated by an NMA molecule.

# 6 The Padawan program

The method of calculating the chemical shift described in section 4 was programmed into a C++ program, which interfaces to the protein folding program `Phaistos`. The program itself is called `Padawan`. The aim of this section is to describe the function of `Padawan` and to compare the amide proton chemcial shift predictions of `Padawan` to the predictions of other available methods, as well as QM generated test data and experimental data.

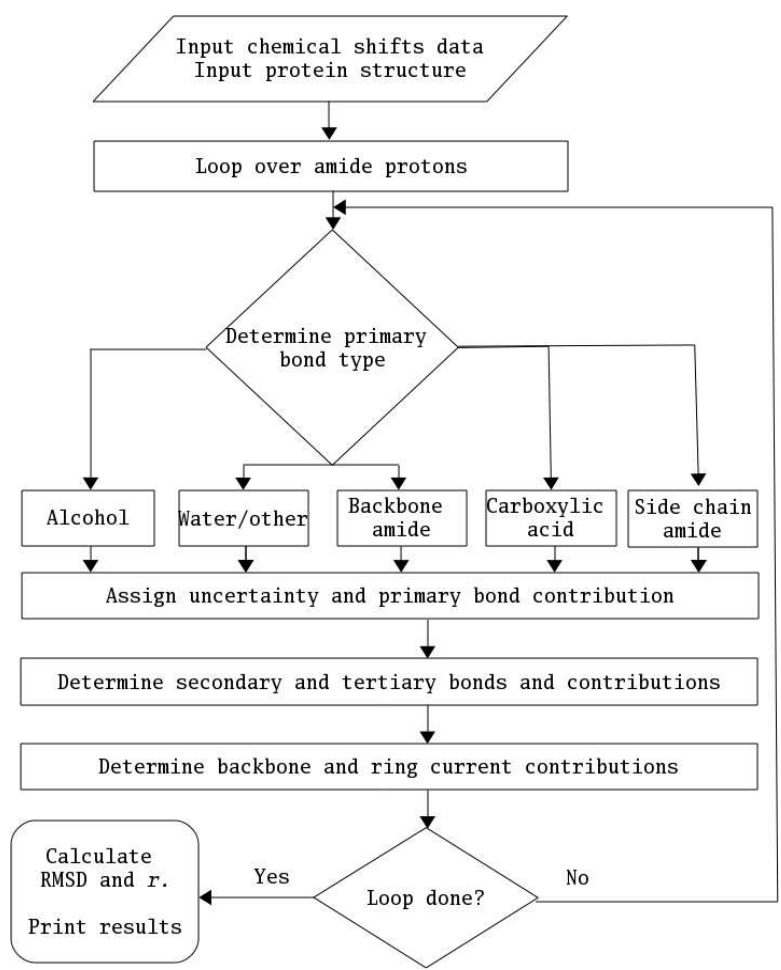

**Figure 17:** The internal workflow of `Padawan`. When an input structure and experimental chemical shifts are given, the programs loops over all amide protons in the protein and calculates the parameters of Eq. 20. At the end of the run, calculated and experimental chemical shifts are compared.

## 6.1   Function of Padawan

The `Padawan` program calculates the chemical shift of all amide protons in the protein using the method described in section 4 and returns the linear correlation and RMSD to an experimental set of data. A protein structure coordinate file and experimental chemical shifts must be supplied by the user. In the `Phaistos` interfaced version, the structure is generated by `Phaistos` and passed to `Padawan` which returns the pseudo energy based on experimental chemical shifts supplied by the user. A flow chart showing the concept of `Padawan` is displayed in Fig. 17.

### 6.1.1   Energy calculation and binning

It is easily seen in section 4, that the primary bond has great influence on the expected uncertainty of the total chemical shift. For instance, the approximation for the primary bonding term for non-bonded and solvent exposed amide protons is much a worse approximation than the similar term for amide-amide bonded protons. See also section 6.4.

| Primary bond type | $\sigma$ [ppm] |
|---|---|
| Backbone amide | 0.3 |
| Side chain amide | 0.5 |
| Carboxylic Acid | 0.8 |
| Alcohol | 0.8 |
| Solvent-exposed/other | 1.2 |

**Table 13:** The uncertainties ($\sigma$) for the amide protons, based on their primary bond type. The uncertainties are used when calculating the pseudo energy of a protein sample structure according to Eq. 16.

The energy expression derived in Eq. 16 allows us to specify different uncertainties for different data types when calculating the chemical shift pseudo energy of a structure. The choice was made to group the data into five bins, depending on the primary bond type. The five bins are (listed by primary bond type): 1) Backbone amide, 2) Side chain amide, 3) Side chain alcohol, 4) Side chain carboxylic acid, 5) Solvent exposed/non-bonded. The determination of the uncertainties are a bit of a guess-work since there is no obvious way of determining these. The decision was made to go with a uncertainty of 0.3 ppm in cases of bonds to backbone amides, since this is what was the RMSD obtained in the Parker, Houk and Jensen paper [50] for the method used on QM optimized structures. Since the same primary bond term is used for bonds both to back bone amides and side chain amides, the chemical shift contribution due to tertiary bonds are neglected for these amide protons, a slightly higher uncertainty of 0.5 ppm is chosen for these bonds. Since the bonds to side chain alcohols and carboxylic acids use similar bond terms, a term which is less thoroughly tested than the term due to Barfield, and because the chemical shift contribution from tertiary bonds is not included for these amide protons, an even higher uncertainty of 0.8 ppm is assigned for amide protons with bonds to these side chains. Non-bonded and solvent exposed amides are assigned with an uncertainty of 1.2 ppm to reflect the missing physical information in the chemical shift contribution due to these bonds. An overview is displayed in table 13.

### 6.1.2 Timings

The speed of evaluating the `Padawan` energy term in `Phaistos` is important, since a very large number of samples are required in a protein folding simulation. The average evaluation time of the Padawan energy term turned out to be very small, compared to the average Monte Carlo step time. The average time to calculate the energy for each energy function used in a protein folding simulation on the two structures the 2GB1 structure (56 residues) and 1ENH (54 residues) is displayed in Fig. 18. The time to evaluate the amide proton chemical shift based energy is 0.0016 s and 0.0014 s for 2GB1 and 1ENH, respectively. The average total time of all other energy terms in the same simulations is 0.053 s and 0.048 s, for the 2GB1 and 1ENH simulations, respectively. Thus it is computationally very cheap to include chemical shift based energy in the simulation.

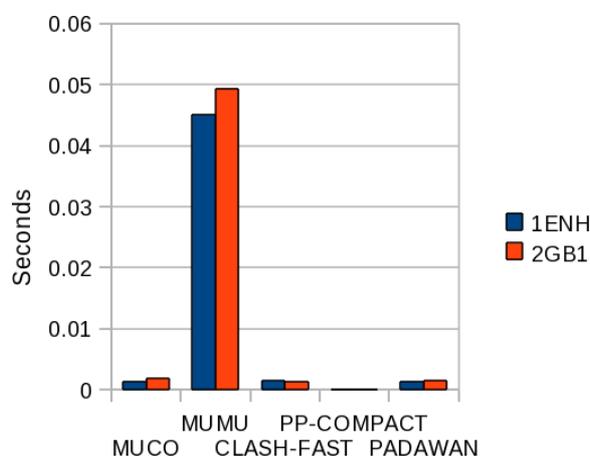

**Figure 18:** The timings of all energy function used in two protein folding simulations on the 2GB1 and 1ENH protein structures. See section 3.2 for a description of each energy term. The time is given in seconds. `Padawan` is the chemical shift based energy function.

### 6.1.3 Energy expression

In order to minimize the energy using the energy expression derived in Eq. 16, it is necessary to see that structures with small chemical errors also have low energy and vice versa. In Fig. 19, a graphic comparison between using the chemical shift energy and the chemical shift RMSD as a metric for the quality of a protein structure is displayed. Two sets of decoys were used (see section 8.3), each containing a series of structures of decreasing quality (measured as the $C^\alpha$-RMSD distance to the crystal structure) . The decoy sets were constructed from high quality crystal structures of Protein G (PDB code: 1IGD) and Pancreatic Trypsin inhibitor (PDB code: 5PTI). In the case of Protein G (Fig. 19A) the RMSD predicts that several decoys are better structures, since they have lower chemical shift RMSD than the native structure. When the weighting due to different uncertainties (depending on the primary bond type - see section 6.1.1) is introduced in the calculated energy, these erroneous predictions are filtered out. For the 5PTI decoy set (Fig. 19B), no structures are erroneously predicted better than the native crystal structure by the chemical shift RMSD. It is however seen, that where the chemical shift RMSD only shows a 50% increase for the worst-case decoy (compared to the chemical shift RMSD of the native structure), the variation in chemical shift energy is up to 300% for the worst-case decoy (compared to the chemical shift RMSD of the native structure).

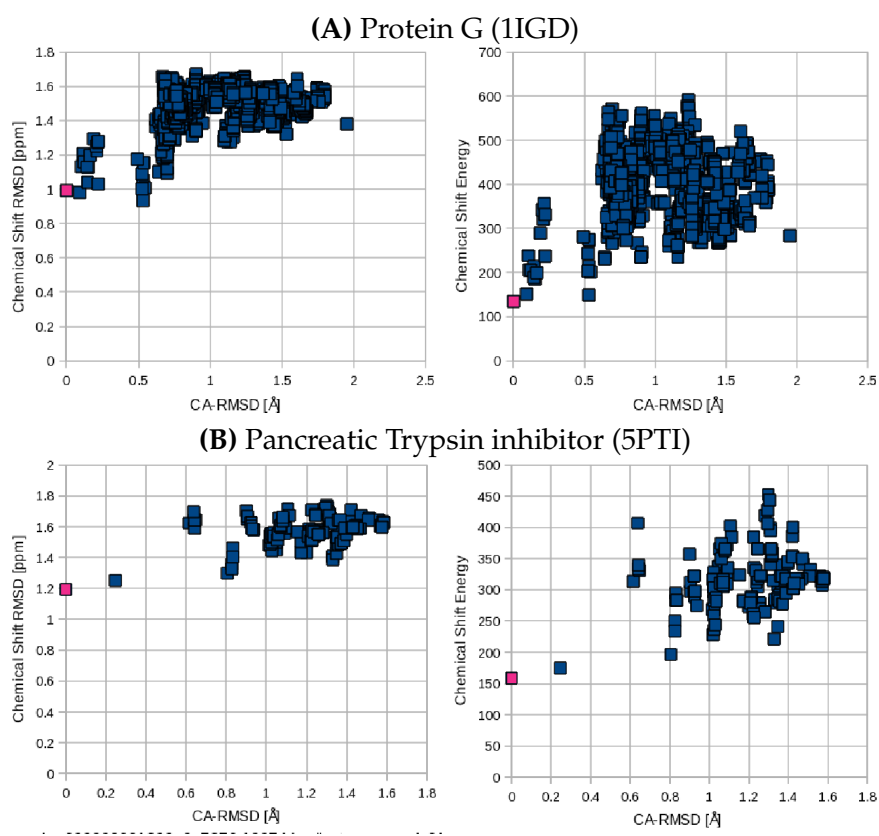

**Figure 19:** Two sets of decoys were used (see section 8.3), each containing a series of structures of decreasing quality. The decoy sets were constructed from high quality crystal structures of (A) Protein G (PDB code: 1IGD) and (A) Pancreatic Trypsin inhibitor (PDB code: 5PTI). The crystal structures are marked in red, while decoys are blue. The chemical shift based energy (left) and chemical shift RMSD (right) is displayed against the $C^\alpha$-RMSD distance to the crystal structure.

It is thus evident, that the weighted energy expression derived in Eq. 16 based on experimental chemical shifts is a viable metric for the protein structure quality.

## 6.2   Comparison to QM methods

In order to compare the accuracy of Padawan against QM methods, it was attempted to calculate the chemical shift of an entire protein by QM methods. The small structure of human parathyroid hormone, residues 1-34, was sufficiently small to allow chemical shifts to be calculated on the structure. A structure is avilable in the PDB database (PDB-entry: 1ET1), obtained at very high crystallographic resolution of 0.9 Å. No solvent model was used - this is addressed in the next subsection.

It was not possible to obtain convergence with a DFT NMR calculation on the system. The use of diffuse functions was also desired, but the use of these led to convergence failure. However, it was possible to carry out a calculation at the HF/6-31G(d) level of theory. The QM isotropic shielding is then plotted against the chemical shift predictions of `Padawan`, `SPARTA`, `SHIFTX`, `SHIFTS` and `PROSHIFT`. Ideally, the linear correlation will be close to -1 and the slope will also be close to -1, although deviance from these values are acceptable, due to the relatively low quality of HF/6-31G(d) shielding constants. For a comparison with experimental data, the ideal slope is 1 and the ideal linear correlation is also 1. Data is shown in Table 14.

|       |              | Padawan | SPARTA | SHIFTX | SHIFTS | PROSHIFT | Exp'tl | QM |
|-------|--------------|---------|--------|--------|--------|----------|--------|-------|
| $r$   | Experimental | 0.62    | 0.42   | 0.56   | 0.52   | 0.07     | 1      | -0.58 |
| $r$   | QM           | -0.94   | -0.64  | -0.45  | -0.86  | 0.03     | -0.58  | 1     |
| Slope | Experimental | 0.79    | 3.58   | 3.24   | 3.77   | 104.15   | 1      | -0.69 |
| Slope | QM           | -1.11   | -7.78  | -12.83 | -6.64  | 649.34   | -1.45  | 1     |
| RMSD  | Experimental | 0.61    | 0.41   | 0.40   | 0.42   | 0.31     |        |       |

**Table 14:** The chemical shifts of N-terminal fragment 1-34 of the human parathyroid hormone (PDB code: 1ET1) are calculated at the HF/6-31G(d) level of theory and compared to four different methods of chemical shift prediction.

Clearly, `Padawan` has by far the best agreement with the QM shielding constants. The linear correlation is -0.94 and the slope is -1.11. Scatter plots are shown on Fig. 20. The empirical based programs are all in very bad agreement with the QM data. This is presumably due to the very simple structural parameters used, which cause insensitivity to small errors in the crystal structure. `PROSHIFT` is especially insensitive to input structure. The explanation for this is likely to be the fact that `PROSHIFT` attempts to optimize hydrogen positions before calculating the chemical shifts, causing the calculated chemical shifts to be in better agreement with experimental data, but for a structure which is not the input structure.

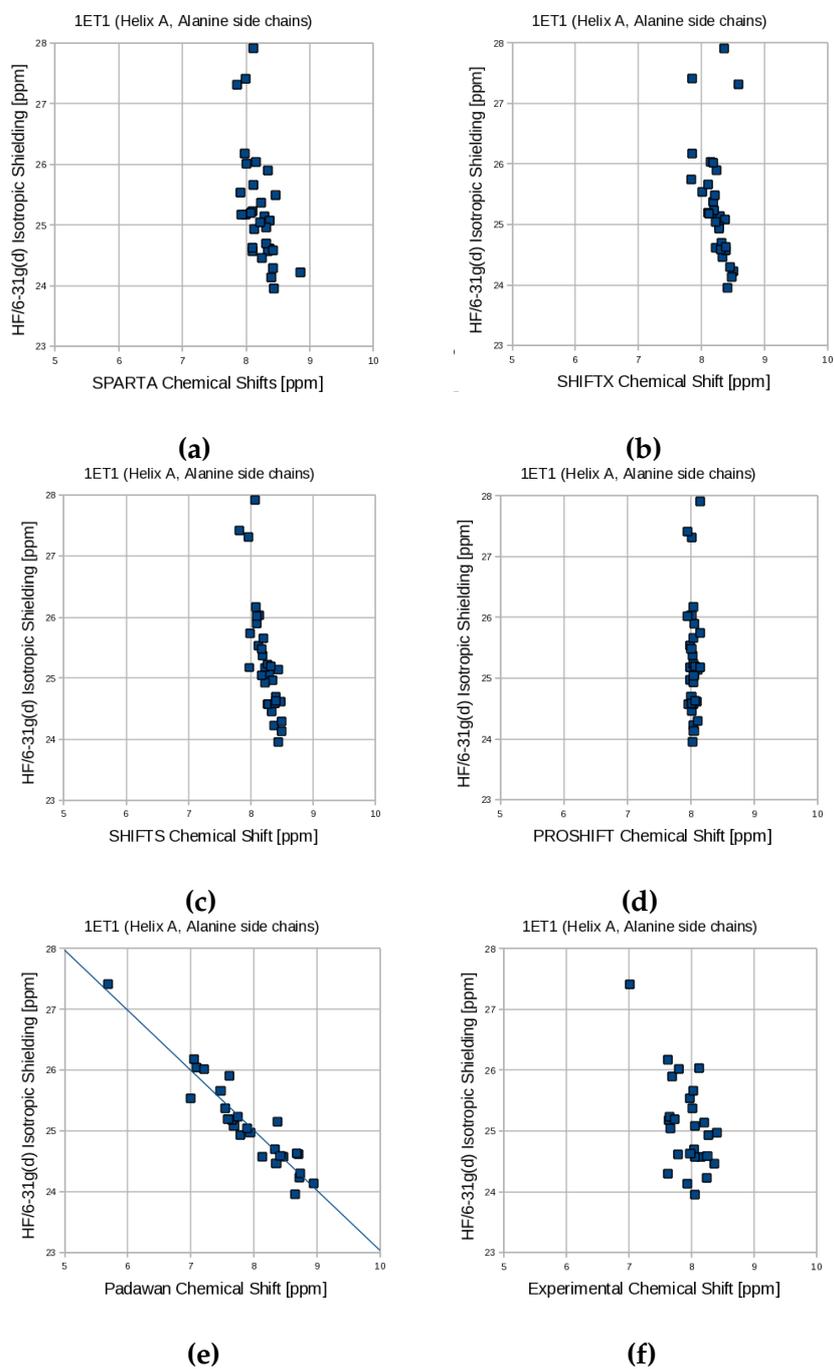

**Figure 20:** The correlation between the chemical shift of human parathyroid hormone (PDB code: 1ET1) calculated at the HF/6-31G(d) level of theory and five different methods of chemical shift prediction ((a) to (e)) The correlation to experimental data is given in (f).

Compared to experimental chemical shifts, all methods have comparable linear correlation coefficients, except `PROSHIFT`. However, the RMSD between predicted and experimental chemical shifts is significantly larger for `Padawan`, than for the empirical based methods. The QM calculated data is also found to be in somewhat good agreement with experimental data.

### 6.2.1 Solvent effects

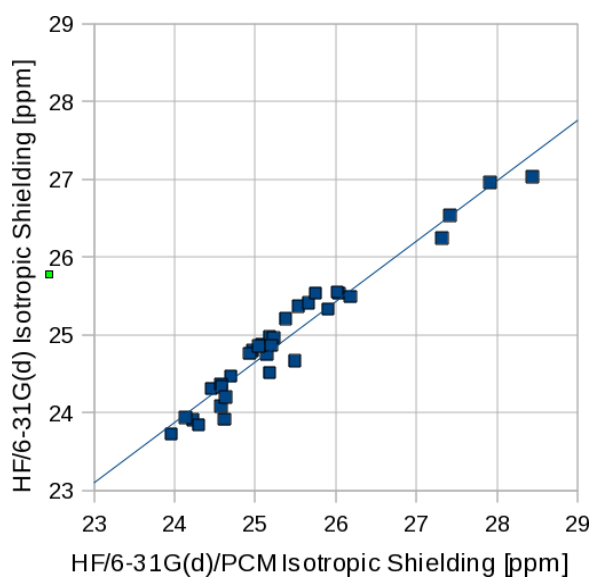

**Figure 21:** The correlation between backbone amide proton chemical shifts calculated at the HF/6-31G(d) level of theory in vacuum and with a PCM solvent model. The structure used is the N-terminal fragment 1-34 of the human parathyroid hormone (PDB code: 1ET1) with the side chains truncated as alanine (see text).

Here the inclusion of a solvent model is investigated. A polarizable continuum model (PCM) [41] is used to model a solvent around the protein. To ease the computational load, the side chains were truncated as alanine and protonated using the `PDB2PQR` web-interface [17]. For the artificial structure, the chemical shifts were calculated both in vacuum and with a PCM solvent model with a dielectric constant of 78.39 to model a water-like solvent.

Inclusion of a solvent causes an increase in the isotropic shielding of between 0 to 1 ppm. A high linear correlation of 0.97 is however found between solvent and vacuum chemical shifts. It is thus evident, that missing correlation between empirical based chemical shifts and QM chemical shifts obtained in vacuum, as found in section 6.2, cannot be attributed to lack of a solvent description, except perhaps for the few amide protons which are directly solvent exposed around the C-terminal in the end of the helix.

## 6.3 Experimental data

Here, the chemical shift predictions are tested on a test set of high resolution X-ray structures (see Appendix E) and their corresponding experimental chemical shifts. For a test set, the chemical shift RMSD and linear correlation between experimental data and predictions are calculated. `Padawan`, `SHIFTX`, `SHIFTS` and `SPARTA` are used to predict chemical shifts. See table 15 for an overview.

| | Padawan | | (amide only) | | SHIFTX | | SHIFTS | | SPARTA | |
|---|---|---|---|---|---|---|---|---|---|---|
| PDB | $r$ | RMSD | $r$ | RMSD | $r$ | RMSD | $r$ | RMSD | $r$ | RMSD |
| 1BRF | 0.56 | 1.26 | 0.88 | 1.06 | 0.99* | 0.11* | 0.82 | 0.43 | 0.82 | 0.40 |
| 1CEX | 0.57 | 1.19 | 0.79 | 0.98 | 0.99* | 0.14* | 0.71 | 0.41 | 0.80 | 0.36 |
| 1CY5 | 0.65 | 0.90 | 0.84 | 0.77 | 0.56 | 0.45 | 0.50 | 0.47 | 0.61 | 0.47 |
| 1ET1 | 0.62 | 0.75 | 0.62 | 0.71 | 0.56 | 0.40 | 0.52 | 0.42 | 0.42 | 0.41 |
| 1I27 | 0.47 | 1.13 | 0.82 | 0.88 | 0.68 | 0.43 | 0.44 | 0.55 | 0.74 | 0.39 |
| 1IFC | 0.60 | 1.03 | 0.78 | 0.90 | 0.89 | 0.29 | 0.71 | 0.47 | 0.81 | 0.39 |
| 1IGD | 0.67 | 0.99 | 0.84 | 0.67 | 0.80 | 0.41 | 0.66 | 0.51 | 0.80 | 0.45 |
| 1OGW | 0.63 | 1.19 | 0.71 | 1.06 | 0.65 | 0.48 | 0.63 | 0.52 | 0.83 | 0.40 |
| 1PLC | 0.43 | 1.32 | 0.67 | 1.11 | 0.64 | 0.57 | 0.56 | 0.60 | 0.92 | 0.34 |
| 1RGE | 0.65 | 1.39 | 0.76 | 1.33 | 0.78 | 0.54 | 0.63 | 0.71 | 0.80 | 0.60 |
| 1RUV | 0.59 | 1.17 | 0.78 | 1.01 | 0.67 | 0.56 | 0.53 | 0.56 | 0.88 | 0.36 |
| 3LZT | 0.44 | 1.22 | 0.60 | 1.00 | 0.61 | 0.52 | 0.43 | 0.61 | 0.68 | 0.48 |
| 5PTI | 0.70 | 1.17 | 0.87 | 1.02 | 0.99* | 0.13* | 0.78 | 0.49 | 0.80 | 0.50 |
| Average | 0.58 | 1.13 | 0.77 | 0.96 | 0.68 | 0.47 | 0.61 | 0.52 | 0.76 | 0.43 |

**Table 15:** Statistical data for backbone amide proton chemical shift prediction for a set of 14 protein structures. The linear correlation and RMSD between calculated and experimental chemical shifts are shown for the methods of `Padawan`, `SHIFTX`, `SHIFTS` and `SPARTA`. Furthermore, the predictions of Padawan for only amide-amide bonded protons is shown in the (amide only) coloumn. Three structures for which `SHIFTX` displays overfitting behavior are marked with an asterisk (*).

The predictions of `Padawan` is in worst agreement with experimental data compared to other methods, when high resolution X-ray structures are used. `SPARTA` generally has the highest accuracy in predicting experimental data. `SHIFTX` and `SHIFTS` are generally in better agreement with experimental data than `Padawan`, but worse than `SPARTA`.

For three structures, `SHIFTX` displayed clear signs of overfitting. The linear correlation between the calculated and experimental chemical shifts for the structures 1BRF, 1CEX and 5PTI is 0.99. A plausible explanation is the fact that the exact same chemical shifts data sets and protein structures are used in the `SHIFTX` training database [48]. Note that these are left out when calculating the average RMSD and $r$.

Using only predictions of amide protons in an amide-amide bonding conformation, the linear correlation to experimental data was found to be very good, and as high as 0.88 in one case. This indicates that the bonding term used to predict the chemical shifts of solvent exposed/non-bonded is not as accurate as the amide-amide bonding term.

The structure for which the highest accuracy of chemical shifts predictions is expected, is the bovine pancreatic trypsin inhipitor, PDB structure 5PTI, which is a joint X-ray and neutron diffraction structure. Neutron scattering is very accurate in determining hydrogen positions, since neutron radiation scatters at the nuclei. This is indeed also observed but is most pronounced for `Padawan` and `SHIFTS` (not considering `SHIFTX` in this case, for the reasons explained above).

## 6.4 Experimental data, for each primary bond type

From the protein test-set found in table 19, Appendix E, the linear correlation between `Padawan` chemical shift predictions and corresponding experimental values was calculated for each of the five bin types defined in section 6.1.1. Furthermore the RMSD and mean error is also calculated. The data is shown in table 16.

| Primary bond type | Counts | $r$ | RMSD [ppm] | Mean Error [ppm] |
|---|---|---|---|---|
| Solvent-exposed/other | 380 | 0.30 | 1.31 | -1.06 |
| Backbone amide | 718 | 0.73 | 1.08 | -0.46 |
| Carboxylic Acid | 60 | 0.36 | 1.41 | 0.68 |
| Alcohol | 42 | 0.42 | 1.75 | -1.57 |
| Side chain amide | 19 | 0.52 | 1.48 | -0.93 |
| All | 1221 | 0.56 | 1.21 | -0.64 |

**Table 16:** Statistical data correlating experimental chemical shifts to the predictions of `Padawan`, based on high quality X-ray structures from the data set in Appendix E.

The linear correlation is by far the best for the amide-amide primary bond type (0.73) and the lowest RMSD (1.08 ppm) and mean error (-0.46 ppm). Amide protons with primary bonds to side chains show a significantly smaller linear correlation (0.36-0.52). The chemical shift due to bonds to alcohol side chains and amide side chains are generally underestimated, while bonds to carboxylic acids are overestimated. This can be explained by the fact that aspartic acid and glutamic acid is always considered to be in the deprotonated state, which is not always true in real proteins. The increased underestimation in cases of bonds to side chain alcohols and amides, compared to bonds to backbone amides, can partially be explained by the missing inclusion of the tertiary bonding term. As expected, the amide protons which are solvent-exposed or have other primary bonds show only little correlation. However, the RMSD is better than for the amide protons with bonds to side chains.

These may seem as large errors compared to the empirical programs which also predicts chemical shifts. However, as it will be shown in the next section, these errors can be translated into small changes in 3-dimensional structure of the protein.

## 6.5   Chemical shift errors translated into structural parameters

Although `Padawan` gives worse chemical shifts prediction than competing prediction models, as compared to experimental data, given high resolution X-ray crystal structures as input stucutres, it is also clear from the exponential relationship between hydrogen bonding length and predicted chemical shifts (see Eq. 25), that chemical shifts predicted by Padawan are very sensitive to small errors in the used structures. This may not be the case for the competing model and is the subject of the next section.

For amide-amide bonded protons, it is possible to correlate the difference between the predicted chemical shifts and the experimental chemical shift to corresponding small change in hydrogen bonding distance. From the geometric term of Eq. 25 (the Barfield equation), it is possible to estimate a correction to the hydrogen bonding distance which cause the chemical shift error to become 0, if the $\rho$ and $\theta$ angles are approximated as fixed. If the difference between predicted and experimental chemical shifts can be attributed to structural errors, this hydrogen bond length correction must, in general, be very small for the hypothesis to physically make sense.

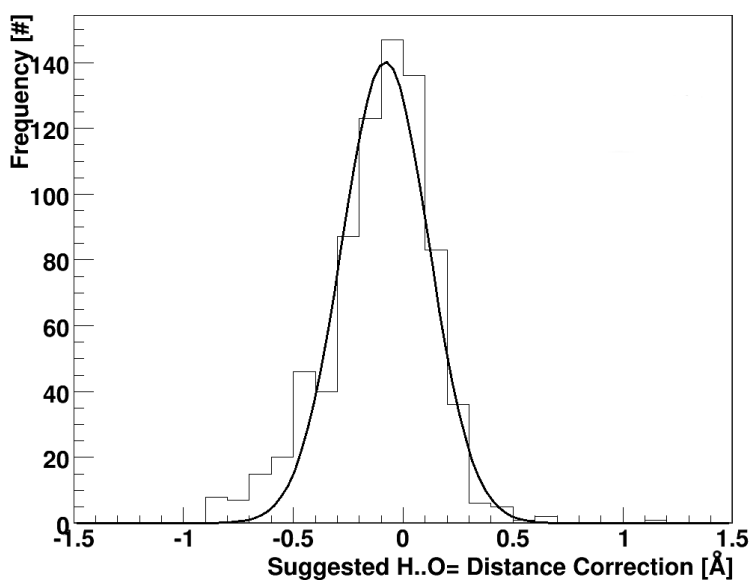

**Figure 22:** Via the Barfield equation (Eq. 25) a small hydrogen bond length correction is calculated (see text), which cause the difference between predicted and experimental data to be 0 ppm.

This correction is calculated for the 718 amide-amide bonded proton in the protein test set (see Appendix E). The standard deviation for this correction is 0.20 Å and the mean correction -0.08 Å. Thus for 68% of all chemical shifts in amide-amide bonding conformations 100% of the error can be attributed to a 0.2 Å or smaller correction in hydrogen bond distance. See Fig. 22 for a histogram of the suggested corrections.

### 6.6   Detecting structural errors: IL21

The effect that small errors in hydrogen bonding lengths will cause a large error in the predicted chemical shift is investigated here and a comparison between Padawan, SHIFTX, SHIFTS and SPARTA is carried out. A low quality NMR-structure is investigated, namely the human interleukin (PDB code: IL21). The first structure of the NMR-ensemble is considered as a case-study. In the structure, three amide-amide bonds with suspiciously short hydrogen bond lengths are discovered. The hydrogen bonds lengths from ILE17, LEU21 and LEU115 are 1.63 Å, 1.51 Å and 1.52 Å, respectively. All the named residues are found in $\alpha$-helices and the bond lengths are thus extremely uncommon (see for instance Fig. 4). It is thus expected to see, that the predicted chemical shifts of these amide protons are extremely overestimated compared to experimental data. Here it is investigated, whether these unusual bond lengths can be identified by looking at the chemical shifts. The predicted chemical shifts for the three amide protons from Padawan, SHIFTX, SHIFTS and SPARTA can be found in table 17. The RMSD and linear correlation, $r$, between experimental and predicted values is also found in table 17.

| Residue | HB distance [Å] | Experimental [ppm] | Padawan [ppm] | SHIFTX [ppm] | SHIFTS [ppm] | SPARTA [ppm] |
|---------|-----------------|--------------------|--------------|--------------|--------------|--------------|
| ILE17 | 1.63 | 7.84 | 11.38 | 8.72 | 8.83 | 8.74 |
| LEU21 | 1.51 | 8.66 | 12.51 | 9.22 | 9.15 | 9.35 |
| LEU115 | 1.52 | 7.76 | 12.69 | 8.67 | 8.58 | 9.09 |
| Total RMSD | - | - | 1.69 | 0.64 | 0.45 | 0.57 |
| Total $r$ | - | - | 0.10 | 0.20 | 0.11 | 0.26 |

**Table 17:** The chemical shift of four amide protons with very short hydrogen bond lengths as predicted by Padawan, SHIFTX, SHIFTS and SPARTA found in an NMR structure of human interleukin, PDB code: IL21. The experimental data is shown for reference.

The predictions based on all methods are overestimated, consistent with the observed short bond lengths. However, given the extremely short observed bond lengths and the exponential relationship between hydrogen bonding length and predicted chemical shifts, the predictions by SHIFTX,

SHIFTS and SPARTA are not large enough to be supported by the amide-amide hydrogen bonding model by Barfield [3]. The hydrogen bonding expressions used in these programs are thus too rough to properly take hydrogen bonding into account.

This is consistent with results obtained by Parker, Houk and Jensen [50], who found the chemical shift predictions of SHIFTX and SHIFTS to be unaffected by QM optimization of the hydrogen bonding network of local protein fragments, while the chemical shift predicted by QM was in better agreement with experimental data when the hydrogen bonding network was optimized (compared to a QM calculation on the crystal structure).

## 6.7 Decoy discrimination

This section focuses on the ability to discriminate between protein structures of varying quality based on chemical shift prediction. For this purpose chemical shifts are predicted on artificial protein structures, so called decoys. By adding increasing amounts of random geometric distortions to a high quality crystal structure, a decoy library is created which are very close to the crystal structure structre ($C^\alpha$-RMSD < 2 Å - see methodology section 8.3). Other decoy sets obtained from MD simulations which are much further from native structure was found in an online decoy collection[2]. The $C^\alpha$-RMSD to the native structure is measured for each decoy and a chemical shift prediction is carried out. If the method of chemical shift prediction is sensitive to the protein structure, a decoy will give worse chemical shift predictions (compared to experimental values) than the native structure, due to the random nature of the geometric distortions.

### 6.7.1 Tight decoy sets

The presented method is used to discriminate decoys and is compared to the programs SHIFTS, SHIFTX and SPARTA. Unfortunately the used tight (little variation in $C^\alpha$-RMSD) decoy sets supplied by Mikael Borg[Reference: personal communication] turned out to be the same structures, for which SHIFTX displayed overfitting behaviour.

Fig. 23 and Fig. 24 show plots of the chemical shift RMSD due to Padawan, SHIFTS, SHIFTX and SPARTA for each structure in the 5PTI and 1CEX decoy libraries, respectively. The 5PTI library varies over from 0 Å to 1.6 Å from the native structure, while the 1CEX library is extremely tight with 800 structures between 0 Å to 0.4 Å.

SPARTA wrongly predicts the crystal structure from which the decoys are generated to be in worse agreement with experimental data, for both the 5PTI and 1CEX libraries. SHIFTS show only very little variation in RMSD from the crystal structure to the worst decoy, although a clear correlation between $C^\alpha$-RMSD and chemical shift RMSD is found in both cases. `SHIFTX` displays very good correlation between $C^\alpha$-RMSD and chemical shift RMSD for both decoy sets, keeping in mind that `SHIFTX` displays overfitting behavior for both crystal structures. `Padawan` has a good correlation between $C^\alpha$-RMSD and chemical shift RMSD for both decoy sets. Only in a small range less than 0.1 Å $C^\alpha$-RMSD from the 1CEX crystal structure does `Padawan` fail. A similar decoy library generated from 1IGD was also investigated (not shown). For this library, all methods predicted the crystal structure to have the lowest RMSD of the decoy set.

---

[2]http://babylone.ulb.ac.be/decoys/index.php

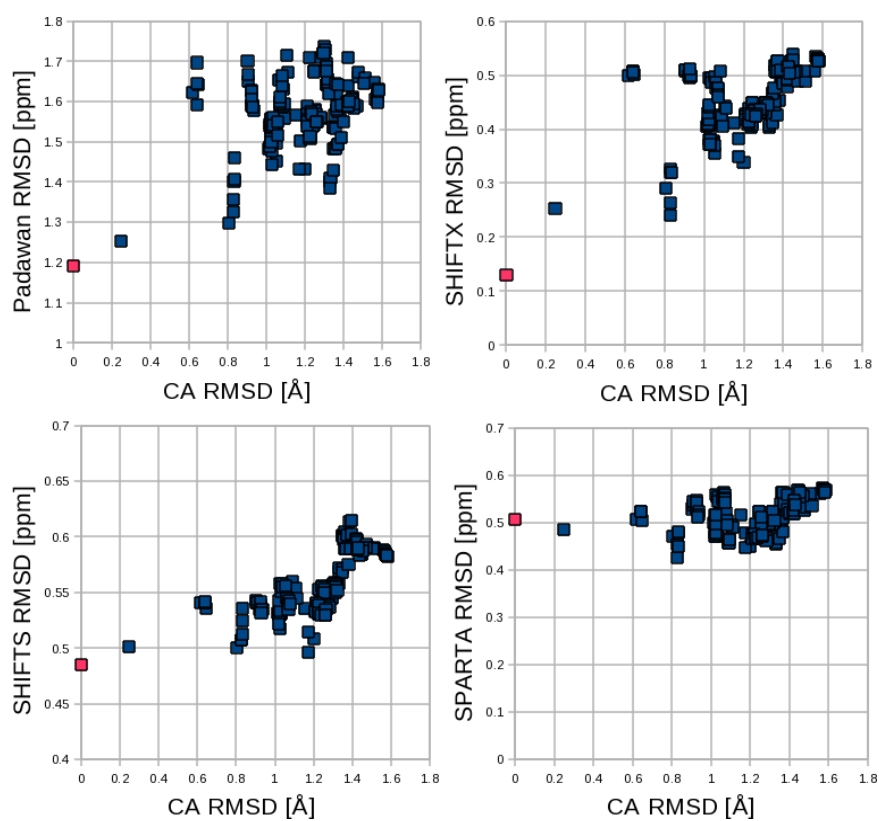

**Figure 23:** The correlation between $C^\alpha$-RMSD and chemical shift RMSD for a decoys library generated from the 5PTI crystal structure. The RMSD is calculated by `Padawan`, `SHIFTS`, `SHIFTX` and `SPARTA`. The native structure is displayed in red.

It is intersting to see, that `SPARTA`, which is the method employed to predic chemical shifts in the current state-of-the-art chemical shifts based protein stucture prediction methods, `CHESHIRE` [11] and `CS-ROSETTA` [57], turns out to have the worst perfomance in discriminating decoys from crystal structures.

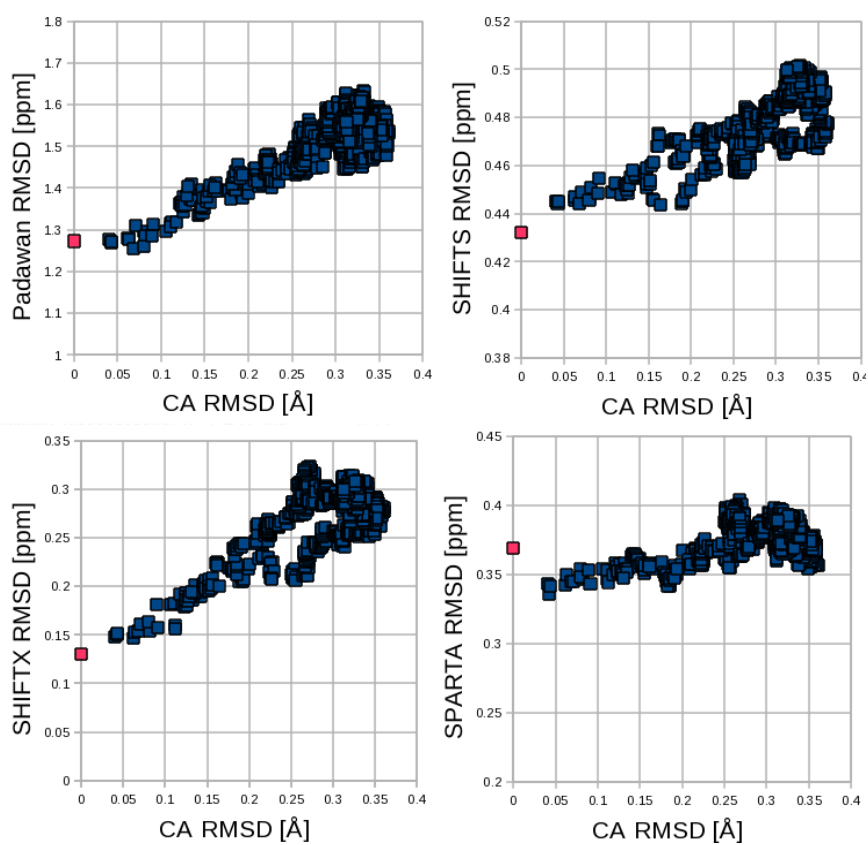

**Figure 24:** The correlation between C$^\alpha$-RMSD and chemical shift RMSD for a decoys library generated from the 1CEX crystal structure. The RMSD is calculated by `Padawan`, `SHIFTS`, `SHIFTX` and `SPARTA`. The native structure is displayed in red.

### 6.7.2 Loose decoy sets

The previous section concerned very tight decoy sets. Here decoy sets much further away from the native structure are considered. The decoy sets used in this section (see Fig. 25) are generated from the bovine pancreatic trypsin inhibitor (PDB code: 5PTI) (1) as snapshots from an MD simulation using an AMBER force field [53] at 300 K and (2) a `Phaistos` generated decoy set. These structures are between (1) 2.5 Å to 3 Å from the crystal structure and (2) 0.1 Å to 1.5 Å from the crystal structure.

The behaviour demonstrated here, is that for structures with a C$^\alpha$-RMSD of 1.5 Å from a high quality structure or worse, the chemical shift RMSD does not increase much, and is rarely found above 2 ppm. This was also observed in other decoys libraries (not shown). It is thus very difficult for `Padawan` to discriminate between decoys, when the C$^\alpha$-RMSD to the native structure is above 2 Å.

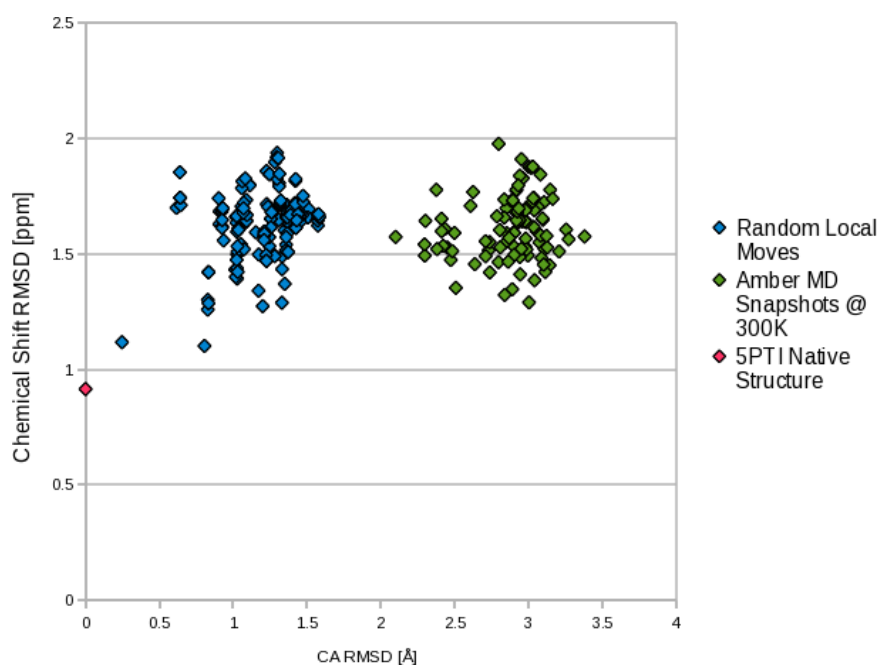

**Figure 25:** The correlation between C$^\alpha$-RMSD and chemical shift RMSD for a decoys library generated from the 5PTI crystal structure. The chemical shift RMSD is calculated from the prediction of `Padawan`. Blue structures are from a tight decoy library constructed by Mikael Borg[Reference: personal communication], while green structures represents snapshots from an MD simulation using an AMBER force field at 300 K. The native structure is displayed in red.

# 7   Results - folding and refinements

In this section, the methods developed in the previous sections are used to generate protein structures via `Phaistos` and the interface to `Padawan`. Both *de novo* folding simulations are carried out and a structure refinement is also carried out using chemical shift based energy.

## 7.1   Human parathyroid hormone (1ET1) folding

The easiest target for a *de novo* folding is a structure consisting of only $\alpha$-helix, since the Ramachandran allowed region is very limited, and more importantly the backbone hydrogen bonding network is principle completely defined by the secondary structure, if hydrogen bonds are well described by the energy functions. A protein consisting of only one $\alpha$-helix structure is the N-terminal fragment 1-34 of the human parathyroid hormone. Both chemical shifts and a 0.9 Å crystal structure is available for this fragment.

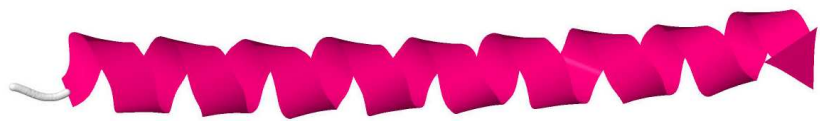

**Figure 26:** Cartoon structure of the $\alpha$-helical N-terminal fragment 1-34 of the human parathyroid hormone 1ET1.

Using a combination of the OPLS force field energy [28], the Clash-Fast energy term and amide proton chemical shifts energy, a folding simulation was carried out on the sequence of the human parathyroid hormone. The secondary structure assignment was taken from the a crystal structure, PDB-code: 1ET1. A similar simulation was carried out using only the OPLS force field energy and the Clash-Fast energy.

While using only the energy based on the OPLS force field, structures are sampled largely in a region of 1-4 Å from the crystal structure, but the energy shows no preference for structures closer to crystal structure. Introducing the chemical shift based energy, clearly creates a preference for structures around 1 Å from the crystal structure. See fig. 27.

The hydrogen bonding network as predicted by the OPLS force field alone did not describe a meaningful hydrogen bonding, however. It is expected to see backbone N-H and C=O bonds which are parallel to the $\alpha$-helical axis. However, the backbone carbonyl C=O bonds in the OPLS only simulation has a preference to sit at an angle to the $\alpha$-helical axis, making the critical hydrogen bonding network largely ill-defined. When amide proton chemical shifts are used in conjunction with OPLS, sampled structures with low energy have well defined hydrogen bonds.

The agreement between experimental chemical shifts and those predicted from the lowest energy structure in the chemical shift+OPLS folding simulation was an RMSD of 0.79 ppm and a linear correlation of $r = 0.44$. This can be compared to the crystal structure, for which the same values were an RMSD of 0.75 ppm and $r = 0.62$. The distance between the two structures were a $C^\alpha$-RMSD of 1.30 Å.

Based on these observations, using the amide proton chemical shifts from `Padawan` give physical meaningful hydrogen bonds in the structure, even in combination with energy functions which show no significant preference for these.

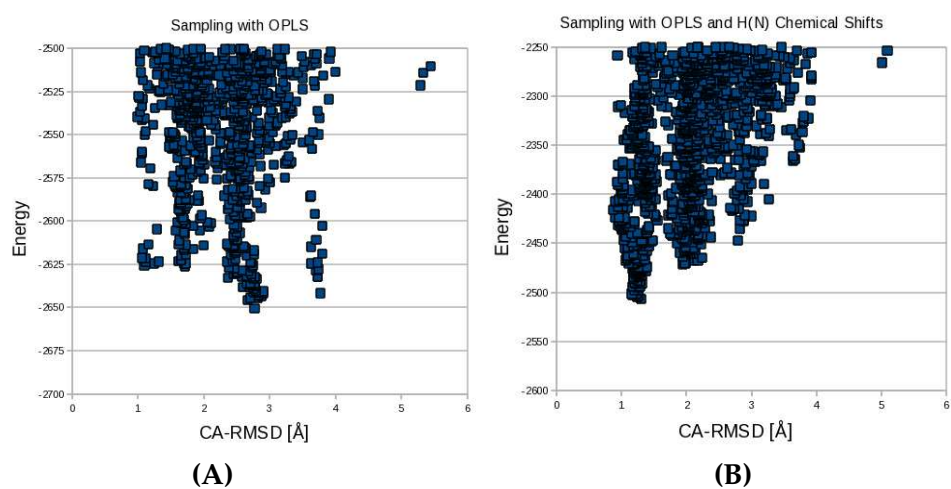

**Figure 27:** Energy as function of $C^\alpha$-RMSD to crystal structure for a *de novo* folding simulation of human parathyroid hormone residues 1-34. With only an OPLS force field (a), sampling shows no clear energy minimum in the sampled conformational space. Introduction of amide proton chemical shifts (b) cause lowest energy structures to be clustered around 1 Å from the crystal structure.

## 7.2   Engrailed homeodomain (1ENH) folding

An often used protein folding target is Drosophila melanogaster (fruit fly) engrailed homeodomain (PDB code: 1ENH), the secondary structure of which can best be described as a three helix bundle. The 1ENH structure is a 2.10 Å resolution crystal structure.

The engrailed homeodomain has previously been folded successfully using `Phaistos` [Reference: Mikael Borg, personal communication]. Here, chemical shifts data is added to a working folding protocol to test, whether adding amide proton chemical shifts energy gives a low-energy bias for structures closer to native, compared to a folding using the same parameters but without chemical shift energy. Two simulations were carried out for the amino acid sequence of Protein G. The energy functions used in the folding simulations were, PP-compactness, Clash, MuCo, MuMu. The Chemical shift based energy from `Padawan` was added to one simulation.

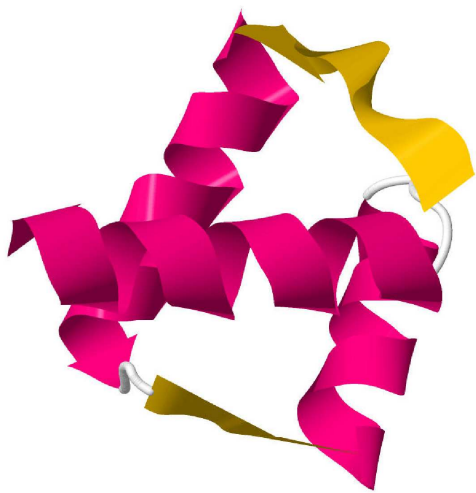

**Figure 28:** Cartoon of Drosophila melanogaster (fruit fly) engrailed homeodomain, PDB structure 1ET1.

In folding simulation using chemical shift based energy, a clear preference is seen for structures closer to the crystal structure, similarly to the 1ET1 folding simulation. The sampled structures with the lowest energy (and best agreement with experimental chemical shifts) are amongst the structures closest to the crystal structure. The lowest energy structures are clustered around 3.7 Å which is in good agreement with the crystal structure. See Fig. 29.

In a folding with the same parameters, but chemical shift energy turned off, the sampled structures with the lowest energy are not the ones which are closest to native. The lowest energy clustering happens at around 4.2 Å from native.

The agreement between experimental chemical shifts and those predicted from the lowest energy structure in the folding simulation including chemical shifts was an RMSD of 0.34 ppm and a linear correlation of $r = 0.45$. This can be compared to the NMR structure in the PDB database, for which the same values are an RMSD of 1.22 ppm and $r = 0.52$. The distance between the two structures were a $C^\alpha$-RMSD of 1.87 Å.

In this folding, an improvement in $C^\alpha$-RMSD of 0.5 Å to native is obtained only by including backbone amide proton chemical shifts. This further indicates, that backbone amide proton chemical shifts are efficacious in describing hydrogen bonding networks, especially for $\alpha$-helices. A cartoon of the lowest energy structure (obtained with chemical shifts) aligned with the crystal structure can be seen on Fig. 30. The $C^\alpha$-RMSD is 3.7 Å.

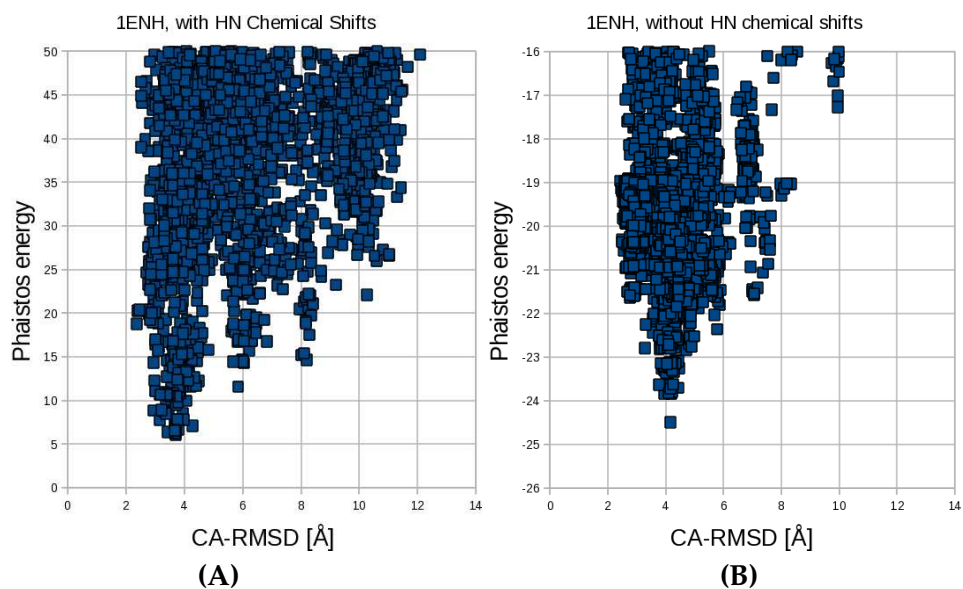

**Figure 29:** Energy as function of $C^\alpha$-RMSD to crystal structure for a *de novo* folding simulation of the Drosophila melanogaster engrailed homeodomain (PDB-code: 1ENH). The sampled structures in A) are generated using a Phaistos folding protocol as described in the text, but with added energy from amide proton chemical shifts. B) are samples using the only the standard Phaistos protocol.

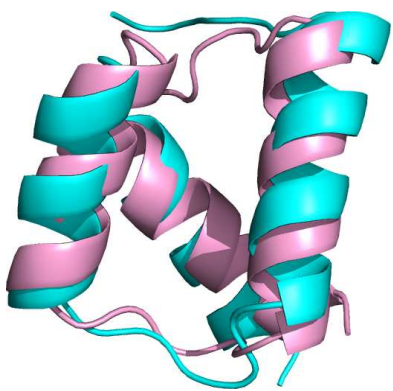

**Figure 30:** *De novo* folding of Drosophila melanogaster engrailed homeodomain (PDB code: 1ENH). A cartoon of the lowest energy structure (obtained with chemical shifts) aligned with the crystal structure can be seen on Fig. 30. The $C^\alpha$-RMSD is 3.7 Å

### 7.3 Protein G (2GB1) folding and refinement

Attempts were carried out to fold the Protein G structure using the same protocol used to fold the engrailed homeodomain. Despite a significant investment in CPU time, the converged structure could not be found close to the crystal structure. The runtime was 400 hours on 8 cores. Despite the heavy investment in CPU time, the folding failed, seemingly due to too little sampling. Presumably this is due to the difficulty of making the found $\beta$ strands come together, which can take lots of sampled structures (see Fig. 31). However, given the computational resources at hand, it was not possible to redo the folding with significantly longer run time (and thus enough samples to converge the simulation). Since a structure could not be found from a *de novo* folding of Protein G, the use of chemical shifts as structure refinement tool was tested on a 2GB1 decoy structure. A set of decoys was made using the same protocol as used in section 8.3. A 2.7 Å $C^\alpha$-RMSD structure was picked from the decoy set and a refinement was set up.

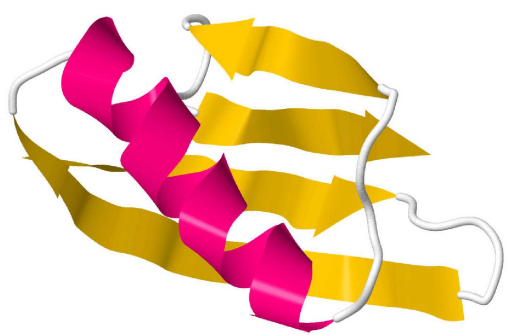

**Figure 31:** Cartoon structure of the Protein G structure 2GB1. Notice the four $\beta$-sheet strands which need to assemble in order to fold the structure correctly.

Using only chemical shifts energy and the Clash-Fast energy term, a simulation was started from the structure of the decoy 2.7 Å $C^\alpha$-RMSD decoy. In this case there was a clear transition towards the crystal structure. A large cluster centered around 1.75 Å $C^\alpha$-RMSD from the crystal structure had clearly the lowest energy found in the simulation (see Fig. 32) An improvement of around 1 Å $C^\alpha$-RMSD was thus obtained by only minimizing the backbone amide proton chemical shift errors.

The agreement between experimental chemical shifts and those predicted from the lowest energy structure in the folding simulation including chemical shifts was an RMSD of 0.24 ppm and a linear correlation of $r = 0.63$. This can be compared to the NMR structure in the PDB database, for which the same values are an RMSD of 1.49 ppm and $r = 0.26$. The distance between the two structures were a $C^\alpha$-RMSD of 1.87 Å.

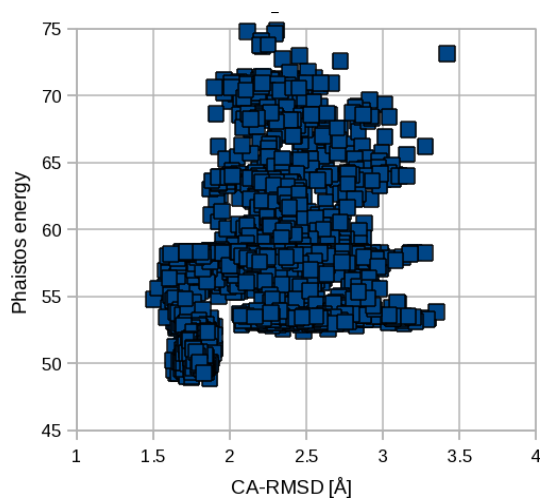

**Figure 32:** A chemical shift based structure refinement of a 2.7 Å $C^\alpha$-RMSD decoy of Protein G. A lowest energy cluster is centered around 1.75 Å, much closer than the decoy.

Other attempts to carry out *de novo* folding simulations were also carried out. For the structure of the artificial three helix bundle Protein A3D (PDB code: 2A3D) a folding simulation was also attempted. This protein is very similar to the engrailed homeodomain, but the predicted fold was not identical to the structure found in the PDB.

Several unsuccessful attempts to use chemical shift base energy to refine the structures of decoys created from the structure of bovine pancreatic trypsin inhibitor (PDB code: 5PTI) were also carried out. However, none of these could positively predict better structures than the decoy from which the refinement simulation was started.

# 8  Computational methodology

All calculations in this work are performed on a cluster consisting of three nodes with dual Intel Xeon X5450 CPUs, 32 GB RAM and 1.6 TB of disk space and four nodes with dual Intel Xeon X5560 CPUs, 32 GB RAM and 1.3 TB of disk space.

## 8.1  DFT calculation of proton chemical shifts

In chapter 4, we calculate the chemical shift of the amide proton, using the method suggested by Rablen, Pearlman and Finkbiner [56], who propose very convincing arguments for using a linear scaling technique to correlate DFT chemical to experimental values with chemical accuracy. Rablen, Pearlman and Finkbiner compare the accuracy of HF, DFT and MP2 NMR shielding calculations using both the GIAO and the IGAIM formulations as well as the basis set and input geometry dependence of the resulting chemical shielding. For a total of 80 different organic compounds, the isotropic shielding is calculated and compared to experimental data.

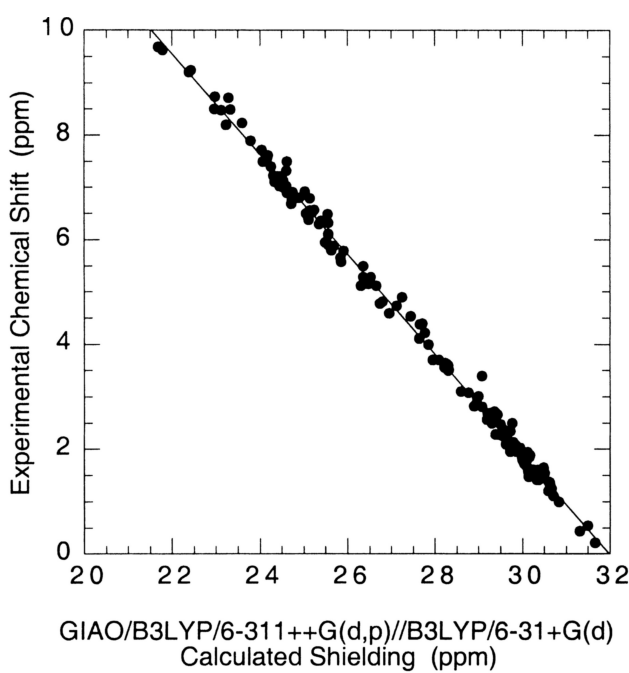

GIAO/B3LYP/6-311++G(d,p)//B3LYP/6-31+G(d)
Calculated Shielding  (ppm)

**Figure 33:** The correlation between experimental chemical shifts and shielding constants calculated at the GIAO/B3LYP/6-311++g(d,p)//B3LYP/6-31+g(d) level of theory. A best fit is found as $\delta_H = 30.60\,\text{ppm} - 0.957 \cdot \sigma_{\text{calc}}$

The computationally most efficient sweet-spot found in the study is the GIAO/B3LYP/6-311++g(d,p)//B3LYP/6-31+g(d) level of theory, at which the linear correlation to experimental data is as high as $r = 0.9981$. A scatter plot of the 105 data points is found in Fig. 33. A best fit relative to the experimental chemical shift relative to TMS in $CDCl_3$ is obtained as

$$\delta_H = 30.60 \text{ ppm} - 0.957 \cdot \sigma_{calc} \qquad (47)$$

where $\sigma_{calc}$ is the isotropic shielding calculated at the GIAO/B3LYP/6-311++G(d,p)//B3LYP/6-31+G(d). The RMSD of this fit is 0.15 ppm, which for DFT calculations on a set of 80 different molecules is remarkable, given the low computational expense of the calculations. Another method fit was made at the GIAO/B3LYP/6-311++G(d,p)//B3LYP/6-311++G(d,p) level of theory. In this case the relationship was almost identical:

$$\delta_H = 30.58 \text{ ppm} - 0.953 \cdot \sigma_{calc}, \qquad (48)$$

This approach was succesfully used by Molina and Jensen [43] to successfully predict proton chemical shifts in the active sites of chymotrypsin and $\alpha$-lytic protease. The proton chemical shifts of the 80 different compounds, many of which are chemically very different from an amide proton, is in a relatively wide range from 0 ppm to 10 ppm. Typically amide protons have chemical shifts in the range 7 ppm to 9 ppm, and we have no guarantee, that the best fit for the 80 compounds a large range of 10 ppm is a best fit for amide protons in more narrow range of a few ppm. And although the fit, which can be seen on Fig. 33 is very accurate, it still does not include an aqueous phase. Inclusion of aqueous solvent effect would lead to an increase in $\sigma_{Ref}$ to 30.93 ppm. Parker, Houk and Jensen, however, found that using a slightly lower offset of 30.30 ppm instead of 30.60 ppm leads to a much better agreement with experiments, presumably because the protein innards are not directly solvent exposed and all significant interactions of the amide proton are explicitly handled by the model systems used. Consistent with the Parker Houk and Jensen procedure, we use the following approximation of the chemical shift of amide protons:

$$\delta_{H^N} = 30.30 \text{ ppm} - 0.957 \cdot \sigma_{calc}, \qquad (49)$$

at the GIAO/B3LYP/6-311++G(d,p)//B3LYP/6-31+G(d) level of theory. When optimizing smaller molecules, it is easy to use a triple zeta quality basis set with diffuse and polarized functions on all atom. In this case, the approximation used is. Again we lower $\sigma_{Ref}$ by 0.30 ppm, consistent with Parker, Houk and Jensen:

$$\delta_{H^N} = 30.28 \text{ ppm} - 0.953 \cdot \sigma_{calc}, \qquad (50)$$

at the GIAO/B3LYP/6-311++G(d,p)//B3LYP/6-311++G(d,p) level of theory.

All DFT calculations are performed using the `Gaussian 03` package at default settings.

## 8.2   The calculation of ring current parameters

The dimers of an amide probe and an aromatic ring were generated from at data set of 21 former Proteins of the Month at the RCSB Protein Data Bank (PDB). This was done to ensure, that the conformations in the QM calculations were realistic conformations and no non-physical conformation were causing non-physical outliers. The structures used were (listed by PDB-code): 1F94, 1GK1, 1IGD, 1JYQ, 1JYR, 1JYU, 1Q3E, 1QJP, 1VJC, 1XA5, 1ZJK, 2ACO, 2B6C, 2D57, 2DRJ, 2ETL, 2F47, 2FZG, 2GOL, 2I4D and 2I4V. Since the structures were X-ray structure, no hydrogen atoms were present in the structures, and `PDB2PQR` [16] [17] was used to protonate the structures in order to obtain hydrogen atom posistions. The aromatic approximate units (benzene replacing the phenylalanine side chain; phenol, tyrosine; imidazole, histidine (neutral); imidalozium, histidine (positive); indole, tryptophan) and the amide probes (formamide and *N*-methylacetamide) were constructed in `Avogadro` and force field minimized using the MMFF94s force field. A set of olefinic analogues to the aromatic rings (see section XX) were constructed using the same method. The minimized structures were then minimized at the B3LYP/aug-cc-pVTZ level of theory in `Gaussian 03` [21]. From the protein structures, systems were identified, where the center of an aromatic ring was within 7 Åof an amide proton. For each of these systems, a dimer was created, with an aromatic approximate ring was put in place of the aromatic side chain, with the ring centers matching up and in the same plane. The given residue with the 7 Å   range was substituted by an NMA or an FMA molecule, with the nitrogen atoms at identical coordinates. Furthermore the N-H vector and the C-N-H planes were also aligned. Due to the rough substitution a few dimers had hydrogen from the two molecules within a very short distance. To avoid artifacts from these, all dimers with a shortest intermolecular distance of 3.4 Å   or less were discarded. 3.4 Åis roughly double the van der Waal radius of the largest atom (carbon) in the system [6]. NMR shielding constants were then calculated from the dimer systems. DFT calculations were carried out in Gaussian 03 at the B3LYP/6-311++G(d,p) level of theory. All Hartree-Fock (HF) calculations of NMR shielding constants were carried out in `Gaussian 03`. MP2/6-311++G(d,p) NMR calculations were carried out in `Gaussian 03`, while the calculation of MP2 shielding constants using Dunning's correlation consistent basis sets were carried out in `Turbomole` [1]. All calculations at the CCSD and CCSD(T) level of theory were carried out using `CFOUR` [63]

### 8.3 Decoy libraries

From a high quality crystal structure, protonated using `PDB2PQR`, the structure is loaded into `Phaistos` and a protein folding simulation is carried out. However, only the Clash-fast energy function is used, so the calculated energy is constant during the folding, leading to no preferences in the Monte-Carlo accepted structures, thus increasing the entropy of the structure. A PDB file containing the current sample structure was dumped each 100 Monte Carlo step. The $C^\alpha$-RMSD values are determined using the `RMSD` program included in the `Phaistos` package. The decoy libraries were constructed and supplied by Mikael Borg. Another decoy library constructed from bovine pancreatic trypsin inhibitor was taken from `http://babylone.ulb.ac.be/decoys/index.php`. This set consists of snapshots from a MD simulation using the AMBER force field at 300 K.

# 9 Summary and outlook

This thesis presents a new method of predicting backbone amide proton chemical shifts in proteins based on electronic structure calculation methods. The method has been implemented in the C++ program `Padawan`.

For a QM calculation at the HF/6-31G(d) level of theory on a 34 residue protein structure of the human parathyroid hormone (PDB code: 1ET1), the linear correlation between the QM calculated shielding constants and the chemical shifts predicted by `Padawan` was -0.94, with a slope close to -1. This linear correlation coefficient is the highest ever reported for a large data set of amide proton chemical shifts and closely matches the results obtained by Parker, Houk and Jensen [50].

For amide protons participating in hydrogen bonding to other backbone amide protons, the predictions of `Padawan` are in good agreement with experimental data (when high quality crystal structures are used). It is thus strongly implied, that electronic structure based methods are capable of parametrizing the chemical shifts of an atom in the easily polarized backbone amide group. QM based approaches should thus serve as viable method for parametrization of the chemical shifts of all atoms in the protein backbone. This is a requirement for the electronic structure based chemical shift predictions necessary to alleviate the current need for more accurate protein chemical shift predictions.

However, it is clear, that the chemical shift predictions of solvent exposed amide protons in the presented model is generally in very bad agreement with experimental data, due to the extremely crude approximation for the chemical shift contribution due to solvent exposure. It is thus the suggestion of the presented work to use another solvent model than the one presented here. A solvent model based on QM/MM and MD approaches containing explicit water has previously been shown to improve shielding constants and is an approach that needs to be explored. [68]

For the first time the three most widely used models of ring-current chemical shift interactions, the point-dipole model due to Pople, the Haigh-Mallion model and the Johnson-Bovey model has been compared by QM methods with large basis sets. This thesis presents a new and general method of quantifying ring current effects.

The first set of intensity parameters for the point-dipole model, derived using QM method is also presented. The point-dipole model is much simpler than the Haigh-Mallion and the Johnson-Bovey, but with demonstrated comparable accuracy, and the presented intensity parameters are thus a useful contribution to the field of QM based chemical shift prediction

The presented work offers a computationally very fast method of pre-dicting chemical shifts in proteins which was interfaced to the protein structure determination software framework, `Phaistos`. Despite the sparse information contained in the amide proton chemical shifts, it was in one case possible to refine a decoy structure of Protein G with a $C^\alpha$-RMSD of 2.7 Å to the crystal structure to a $C^\alpha$-RMSD distance of 1.87 Å for the lowest energy structure, based solely on optimizing the amide proton chemical shifts as predicted by `Padawan`. This is strongly indicates that electronic structure based methods are potentially very useful in the determination of very accurate protein structures. In two protein folding simulations, the inclusion of a chemical shift based energy term was demonstrated to bias the lowest energy structures towards a crystal structure.

A version of `Padawan` will be released under the terms of the GNU General Public License.

# 10 Appendices

## 10.1 Appendix A

This section contains the definitions of various statistical observables, as they are used in this work. For a Gaussian probability distribution, we use to following definitions: [64] [13] The arithmetic mean of a sample population is by definition:

$$\bar{x} = \frac{1}{N} \sum_{i=1}^{N} x_i \qquad (51)$$

Two definitions of the standard deviation exists. The difference is the in the normalization factor of the sum of squared deviations of the samples with respect to the arithmetic mean. See below. In this work, Eq. 52 is referred to as the *root mean square deviation* (*RMSD*), while the definition of Eq. 53 is used when talking about the standard deviation.

$$\sigma_x^{\text{RMSD}} = \sqrt{\frac{1}{N} \sum_{i=1}^{N} (x_i - \bar{x})^2} \qquad (52)$$

$$\sigma_x^{\text{St.Dev.}} = \sqrt{\frac{1}{N-1} \sum_{i=1}^{N} (x_i - \bar{x})^2} \qquad (53)$$

The extent to which a sample population $(x_1, y_1), \ldots, (x_N, y_N)$ supports a linear relation between $x$ and $y$ is measured by the sample *linear correlation coefficient* also known as the Pearson product-moment correlation coefficient, and referred to, in this work, as the *r*-value. In order to define the *r*-value, we briefly define the covariance as between the sample population $(x_1, y_1), \ldots, (x_N, y_N)$ as:

$$\sigma_{xy} = \frac{1}{N-1} \sum_{i=1}^{N} (x_i - \bar{x})(y_i - \bar{y}) \qquad (54)$$

This allow us to write the *r*-value as a function of the covariance and the standard deviations of $x$ and $y$:

$$\begin{aligned} r = \frac{\sigma_{xy}}{\sigma_x \sigma_y} &= \frac{\frac{1}{N-1} \sum_{i=1}^{N} (x_i - \bar{x})(y_i - \bar{y})}{\sqrt{\frac{1}{N-1} \sum_{i=1}^{N} (x_i - \bar{x})^2} \sqrt{\frac{1}{N-1} \sum_{i=1}^{N} (y_i - \bar{y})^2}} \\ &= \frac{\sum (x_i - \bar{x})(y_i - \bar{y})}{\sqrt{\sum (x_i - \bar{x})^2 \sum (y_i - \bar{y})^2}} \end{aligned} \qquad (55)$$

Please take note, that the linear correlation are unaffected by any linear scaling correction to the $x$- or $y$-axes, while the RMSD is not.

## 10.2 Appendix B

$N$ = 10

$a$ = 4.8361

$b$ = [0.0256  0.4592  0.1984  −0.1214  −0.1250  0.1929  −0.0457  −0.0202  −0.0151  −0.0189]

$c$ = [0.0912  0.1343  −0.0091  0.1169  −0.0259  0.0722  0.0015  −0.0137  −0.0174  0.0092]

$$
d = \begin{bmatrix}
-0.3898 & 0.0364 & -0.0770 & -0.0175 & 0.1608 & -0.0005 & 0.0963 & 0.0116 & 0.0619 \\
0.0227 & 0.6617 & 0.0079 & 0.2755 & 0.0268 & 0.1250 & -0.0085 & 0.0254 \\
0.3494 & 0.0076 & -0.0237 & -0.0553 & -0.1020 & 0.0067 & -0.0794 \\
0.0223 & 0.1552 & -0.0001 & 0.0719 & -0.0035 & 0.0366 \\
0.0215 & -0.0927 & -0.0558 & 0.0137 & 0.0444 \\
-0.0290 & -0.0438 & -0.0337 & 0.0356 \\
0.0425 & 0.0305 & 0.0694 \\
0.0151 & -0.0633 \\
-0.0339
\end{bmatrix}
$$

## 10.3   Appendix C

| Residue | $\omega$ [°] | $r_\omega$ [Å] | $\phi$ [°] | $\psi$ [°] | $\Delta\delta_{\text{PHJ}}^{\text{QM}}$ [ppm] | $\Delta\delta_{\text{BB}}^{\text{CC}}$ [ppm] (scaled) | $\Delta\delta_{\text{BB}}^{\text{QM}}$ [ppm] |
|---|---|---|---|---|---|---|---|
| 6 | 18.73 | 2.54 | 175.22 | 146.52 | 5.5 | 5.60 | 5.20 |
| 7 | 12.01 | 2.83 | -87.53 | 122.13 | 5.5 | 5.15 | 4.92 |
| 8 | 21.21 | 2.52 | -104.35 | 148.02 | 5.5 | 5.33 | 5.31 |
| 9 | 14.78 | 2.48 | -118.36 | 145.27 | 5.5 | 5.57 | 5.29 |
| 10 | -2.84 | 2.59 | -119.83 | 125.01 | 5.5 | 5.36 | 5.14 |
| 11 | 4.43 | 2.54 | -97.89 | 122.77 | 5.5 | 5.21 | 5.20 |
| 22 | 8.67 | 2.30 | -142.41 | 163.53 | 5.5 | 6.12 | 5.87 |
| 23 | -0.42 | 2.15 | -151.80 | 158.33 | 6.5 | 6.10 | 6.08 |
| 24 | 10.18 | 2.53 | -105.78 | 138.17 | 5.5 | 5.35 | 5.24 |
| 25 | 3.25 | 2.31 | -156.49 | 160.36 | 5.5 | 6.15 | 5.89 |
| 26 | -131.44 | 4.32 | -73.77 | -22.98 | 4.5 | 4.69 | 4.61 |
| 27 | -2.68 | 2.18 | -155.10 | 168.08 | 5.5 | 6.25 | 6.35 |
| 28 | -116.38 | 4.40 | -61.92 | -34.24 | 4.5 | 4.68 | 4.49 |
| 29 | -159.87 | 4.30 | -65.48 | -36.62 | 4.5 | 4.67 | 5.04 |
| 30 | -164.12 | 4.26 | -67.49 | -41.72 | 4.5 | 4.65 | 4.82 |
| 31 | -167.43 | 4.24 | -62.53 | -44.15 | 4.5 | 4.64 | 4.76 |
| 32 | -163.32 | 4.30 | -61.99 | -45.03 | 4.5 | 4.64 | 4.87 |
| 33 | -161.76 | 4.29 | -60.92 | -42.56 | 4.5 | 4.65 | 4.64 |
| 34 | -161.49 | 4.29 | -62.86 | -43.88 | 4.5 | 4.65 | 4.58 |
| 35 | -155.55 | 4.26 | -70.27 | -39.63 | 4.5 | 4.66 | 4.73 |
| 36 | -162.14 | 4.24 | -63.15 | -38.50 | 4.5 | 4.67 | 4.66 |
| 37 | -171.14 | 4.25 | -63.66 | -46.75 | 4.5 | 4.63 | 4.66 |
| 38 | -166.72 | 4.27 | -60.32 | -46.45 | 4.5 | 4.63 | 4.60 |
| 39 | -168.11 | 4.26 | -62.58 | -45.65 | 4.5 | 4.64 | 4.52 |
| 40 | -164.59 | 4.29 | -63.58 | -45.35 | 4.5 | 4.64 | 4.85 |
| 49 | 4.41 | 2.25 | -130.57 | 161.14 | 5.5 | 5.92 | 5.85 |
| 50 | -6.78 | 2.39 | -133.23 | 132.37 | 5.5 | 5.54 | 5.47 |
| 51 | -7.20 | 2.77 | -111.40 | 110.46 | 5.5 | 5.13 | 4.99 |
| 52 | -115.26 | 4.41 | -65.37 | -20.28 | 4.5 | 4.71 | 5.05 |
| 53 | -146.16 | 4.34 | -69.98 | -34.42 | 4.5 | 4.67 | 5.05 |
| 54 | -120.88 | 4.02 | -111.48 | 1.19 | 4.5 | 4.46 | 4.47 |
| **55** | **122.80** | **4.44** | **54.89** | **40.77** | **4.5** | **5.02** | **5.76** |
| 56 | 15.46 | 2.78 | -122.36 | 130.75 | 5.5 | 5.46 | 5.28 |
| 57 | 15.77 | 2.40 | -102.92 | 149.50 | 5.5 | 5.30 | 5.34 |
| 58 | -0.22 | 2.37 | -131.19 | 139.12 | 5.5 | 5.65 | 5.64 |

**Table 18:** Table of geometric parameters of optimized backbone geometries taken from Protein G (PDB code: 1IGD).

## 10.4    Appendix D

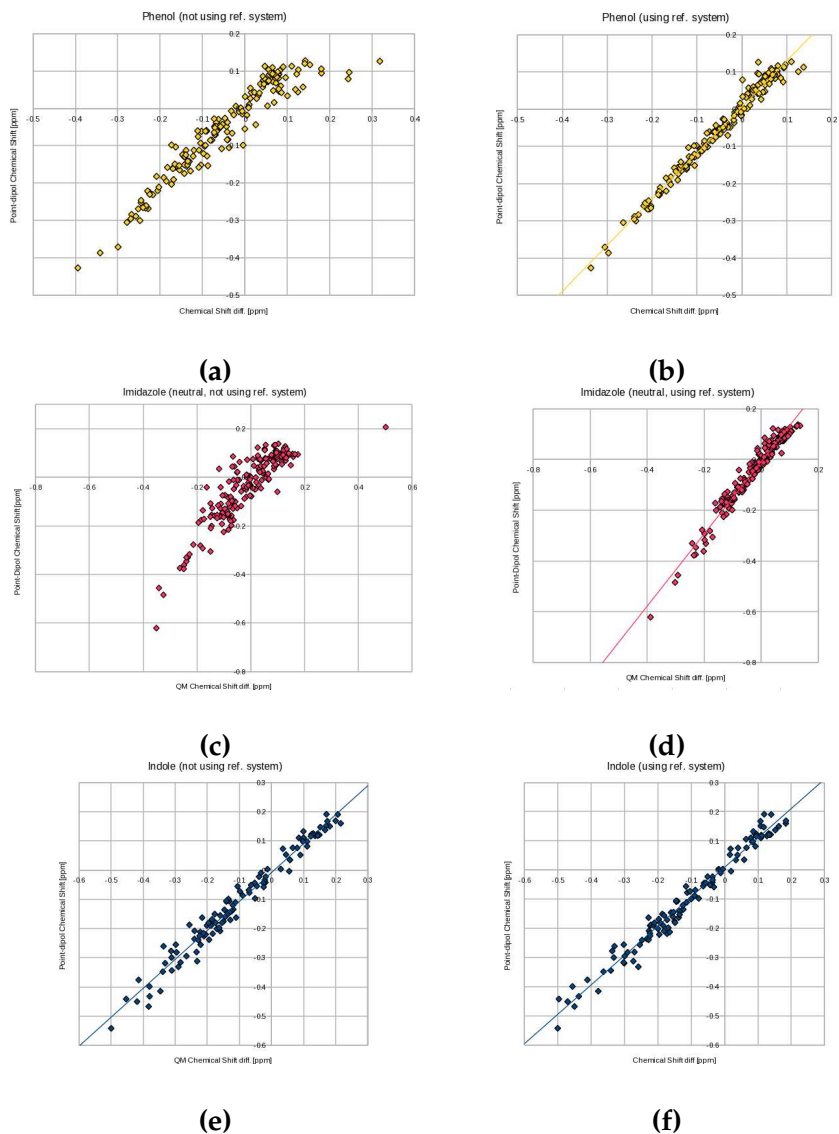

**Figure 34:** The X-axis describes the B3LYP/6-311++G(d,p)/GIAO//B3LYP/aug-cc-pVTZ calculated chemical shift of a dimers between an NMA molecule and the the molecule in the headline, while the Y-axis is the chemical shift according the point-dipole ring current model by Pople. The QM data in (a), (c) and (e) is derived using Eq. 44, while the data in (b), (d) and (f) is derived using Eq. 42, which increases the linearity of the data set.

## 10.5  Appendix E

A set of high resolution X-ray crystal structures, for which $^1$H chemical shifts are available, is constructed. To ensure high quality of the used structures, only structures better with a criystalographic resolution of 1.35 Å or better was used. The protein structures were taken from the RCSB Protein Data Bank (PDB), while matching chemical shifts were found in the Biological Magnetic Resonance Data Bank (BMRB).

| PDB id | Resolution [Å] | BMRB id | Size [residues] | Hydrogen |
|--------|----------------|---------|-----------------|----------|
| 1BRF | 0.95 | 5601 | 53 | PDB2PQR |
| 1CEX | 1.00 | 4101 | 214 | X-ray |
| 1CY5 | 1.30 | 4661 | 97 | PDB2PQR |
| 1D4T | 1.10 | 5211 | 104 | PDB2PQR |
| 1ET1 | 0.90 | 3427 | 34 | PDB2PQR |
| 1F41 | 1.30 | 4637 | 127 | PDB2PQR |
| 1I27 | 1.02 | 5685 | 82 | PDB2PQR |
| 1IFC | 1.19 | 15082 | 132 | PDB2PQR |
| 1IGD | 1.10 | 2575 | 61 | PDB2PQR |
| 1OGW | 1.32 | 6457 | 76 | PDB2PQR |
| 1PLC | 1.33 | 4922 | 99 | X-ray |
| 1RGE | 1.15 | 4259 | 96 | PDB2PQR |
| 1RUV | 1.25 | 4031 | 125 | X-ray |
| 3LZT | 0.93 | 4562 | 129 | PDB2PQR |
| 5PTI | 1.00 | 5359 | 58 | neutron |

**Table 19:** Brief description of the proteins used in this study. For each PDB entry, the crystallographic resolution of the X-ray structure is listed. Note that only structures of high resolution (better than 1.35 Å) are used. The NMR data source is the Biological Magnetic Resonance Data Bank (BMRB). The number of residues is listed for each protein, as well as the protonation method. "PDB2PQR" denotes the use of the PDB2PQR 1.6 as the protonation method, while "X-ray" means that the hydrogen coordinates are determined by X-ray crystallography. "Neutron" denotes the use of joint X-ray and neutron scattering experiments to determine hydrogen positions.

## 10.6   Appendix F

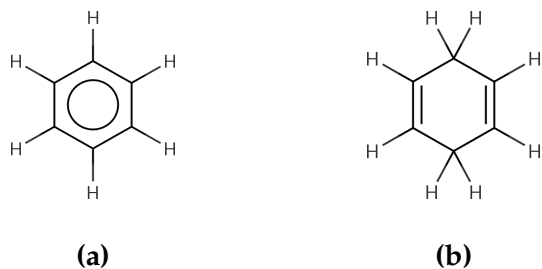

**(a)**                    **(b)**

**Figure 35:** The phenylalanine analogues used in section 5: (a)benzene and (b)1,4-cyclohexadiene

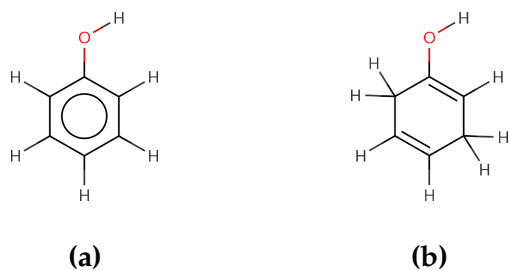

**(a)**                    **(b)**

**Figure 36:** The tyrosine analogues used in section 5: (a)phenol and (b)cyclohexa-1,4-diene-1-ol

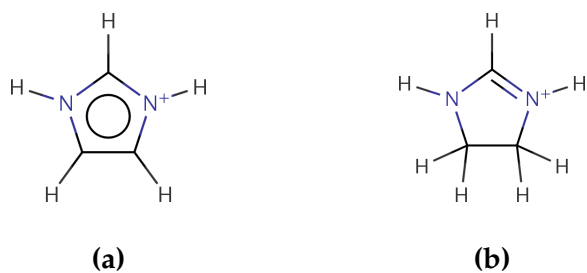

**(a)**                    **(b)**

**Figure 37:** The charged histidine analogues used in section 5: (a)imidazolium and (b)4,5-dihydroimidazolium

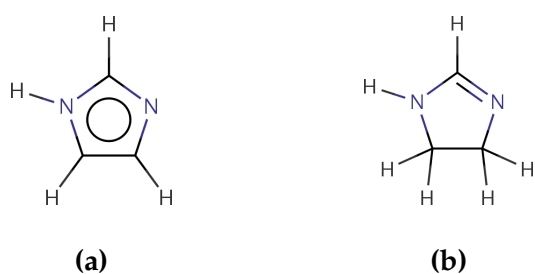

**(a)**          **(b)**

**Figure 38:** The neutral histidine analogues used in section 5: (a)imidazole and (b)4,5-dihydroimidazole

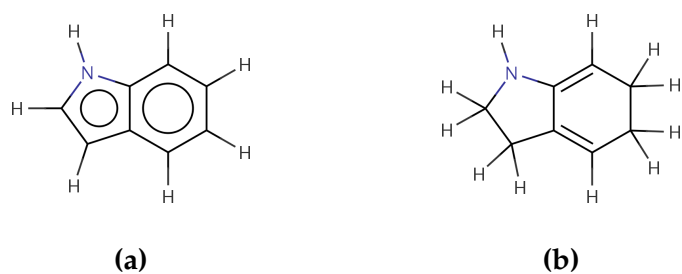

**(a)**                    **(b)**

**Figure 39:** The tryptophan analogues used in section 5: (a)indole and (b)2,3,5,6-tetrahydroindole